\documentclass[12pt,twoside]{report}

\usepackage[T1]{fontenc}
\usepackage[cp1250]{inputenc}
\usepackage{amsmath}
\usepackage{amsthm}
\usepackage{amssymb,graphics,pstricks}
\usepackage{amssymb}

\def\thetitle{Regeneration and Fixed-Width Analysis of Markov
Chain Monte Carlo Algorithms}
\def\theauthor{Krzysztof Łatuszyński}
\def\theday{
18}
\def\themonth{
February}
\def\theyear{
2008}
\def\thesupervisor{
dr~hab.~Wojciech Niemiro}

\def\thedate{\themonth{} \theday, \theyear}
\def\themonthyear{\themonth{} \theyear}

\author{\theauthor}
\title{\thetitle}

\def\titlepages{\newpage
\thispagestyle{empty}
\begin{centering}
\Large
Warsaw University\\
\large
Faculty of Mathematics, Informatics and Mechanics\\
\vspace{3.75cm}
\theauthor\\
\vspace{0.85cm} \LARGE
\thetitle\\
\vspace{0.3cm} \normalsize
\textit{PhD dissertation}\\
\vspace{5.6cm}
\begin{flushright}
Supervisor\\
\vspace{0.25cm}
\thesupervisor\\
\vspace{0.75cm}
Institute of Applied Mathematics and Mechanics\\
Warsaw University\\
\end{flushright}
\vfill
\themonthyear\\
\end{centering}
\newpage\thispagestyle{empty}
\hfill \par \vfill \noindent Author's declaration:\par\noindent
aware of legal responsibility I hereby declare that I have written
this dissertation myself and all the contents of the dissertation
have been obtained by legal means.
\par
\vspace{0.35cm} \hfill \par \noindent
\begin{tabular}[t]{p{0.3\textwidth}p{0.2\textwidth}p{0.4\textwidth}}
\thedate&&\dotfill\\
\hfill\scriptsize \textit{date} \hfill \phantom{.} &&
\hfill\scriptsize \textit{\theauthor} \hfill \phantom{.}
\end{tabular}\par
\vspace{1.8cm} \hfill \par \noindent Supervisor's declaration:
\par\noindent
the dissertation is ready to be reviewed
\par
\vspace{0.35cm} \hfill \par \noindent
\begin{tabular}[t]{p{0.3\textwidth}p{0.2\textwidth}p{0.4\textwidth}}
\thedate&&\dotfill\\
\hfill\scriptsize \textit{date} \hfill \phantom{.} &&
\hfill\scriptsize \textit{\thesupervisor} \hfill \phantom{.}
\end{tabular}
\newpage
}

\theoremstyle{plain}
\newtheorem{thm}{Theorem}[section]
\newtheorem{assu}[thm]{Assumption}
\newtheorem{alg}[thm]{Algorithm}
\newtheorem{prop}[thm]{Proposition}
\newtheorem{lemma}[thm]{Lemma}
\newtheorem{cor}[thm]{Corollary}
\theoremstyle{definition}
\newtheorem{ex}[thm]{Example}
\newtheorem{defi}[thm]{Definition}
\theoremstyle{remark}
\newtheorem{remark}[thm]{Remark}
\newtheorem{concluding remark}[thm]{Concluding Remark}

\def\be#1\ee{\begin{equation}#1\end{equation}}
\newcommand{\ba}{\begin{eqnarray} }
\newcommand{\ea}{\end{eqnarray} }
\def\bt#1\et{\begin{theo}#1\end{theo}}
\def\bl#1\el{\begin{lema}#1\end{lema}}
\def\bp#1\ep{\begin{prop}#1\end{prop}}
\def\bd#1\ed{\begin{defi}#1\end{defi}}

\def\stany{{\cal{X}}}
\def\borel{{\cal{B}(\stany)}}
\def\ptalfa{\check{\alpha}}
\def\1c{\mathbb{I}_C(x)}
\def\lancuch{(X_n)_{n\gs 0}}
\def\borelplus{\mathcal{B}^{+}(\stany)}

\def\ccX{{\cal X}}
\def\va{\varepsilon}
\def\ra{\rightarrow}

\def\R{{\mathbb R}}

\def\Var{\mathrm{Var}}

\def\ccX{\mathcal{X}}

\def\ls{\leqslant}
\def\gs{\geqslant}

\begin{document}
\titlepages

\large{Abstract}

\bigskip
\normalsize In the thesis we take the split chain approach to
analyzing Markov chains and use it to establish fixed-width
results for estimators obtained via Markov chain Monte Carlo
procedures (MCMC). Theoretical results include necessary and
sufficient conditions in terms of regeneration for central limit
theorems for ergodic Markov chains and a regenerative proof of a
CLT version for uniformly ergodic Markov chains with $E_{\pi}f^2<
\infty.$ To obtain asymptotic confidence intervals for MCMC
estimators, strongly consistent estimators of the asymptotic
variance are essential. We relax assumptions required to obtain
such estimators. Moreover, under a drift condition, nonasymptotic
fixed-width results for MCMC estimators for a general state space
setting (not necessarily compact) and not necessarily bounded
target function $f$ are obtained. The last chapter is devoted to
the idea of adaptive Monte Carlo simulation and provides
convergence results and law of large numbers for adaptive
procedures under path-stability condition for transition kernels.

\bigskip


\bigskip


\newpage

\large{Streszczenie}

\bigskip
\normalsize W pracy przedstawione są rezultaty dotyczące estymacji
stałoprecyzyjnej dla algorytmów Monte Carlo opartych na łańcuchach
Markowa (MCMC). Podstawową techniką w analizie łańcuchów Markowa i
związanych z nimi procedur MCMC, jest łańcuch rozszczepiony i
regeneracja, co prowadzi do koniecznego i dostatecznego warunku w
terminach regeneracji dla centralnego twierdzenia granicznego dla
ergodycznych łańcuchów Markowa. Dodatkowym rezultatem jest
regeneracyjny dowód CTG dla jednostajnie ergodycznych łańcuchów
Markowa przy założeniu $E_{\pi} f^2< \infty.$ Aby otrzymać
asymptotyczne przedziały ufności za pomocą algorytmów MCMC
konieczna jest m.in. mocno zgodna estymacja wariancji
asymptotycznej. Osłabiamy znane założenia wymagane do konstrukcji
takich estymatorów. Przy założeniu warunku dryfu, ale bez założeń
o ograniczoności funkcji podcałkowej $f$ i zwartości przestrzeni
stanów, otrzymujemy nieasymptotyczną estymację stałoprecyzyjną.
Ostatni rozdział poświęcony jest procedurom adaptacyjnym, a
uzyskane tam wyniki dotyczące zbieżności i prawa wielkich liczb
zakładają stabilność operatorów przejścia względem trajektorii.

\bigskip

\textbf{Słowa kluczowe: łańcuch Markowa, MCMC, adaptacyjne Monte
Carlo, łańcuch rozszczepiony, regeneracja, warunek dryfu,
$(\varepsilon-\alpha)-$aproksymacja, przedziały ufności,
asymptotyczne przedziały ufności, centralne twierdzenie graniczne,
prawo wielkich liczb}

\bigskip

\textbf{Klasyfikacja tematyczna wg. AMS: 60J10, 60J05, 60F15,
60F05}

\tableofcontents

   \chapter{Introduction} \label{chapter: introduction}

In this chapter we give some background for results presented in
later chapters and introduce main ideas behind the thesis in an
informal way. Therefore mathematical rigour will not always be our
priority here. We start with defining the problem addressed by
Markov chain Monte Carlo methods in Section \ref{Section MCMC} and
proceed to describing typical sampling schemes and MCMC algorithms
(the Metropolis algorithm and the Gibbs sampler) in Section
\ref{Section algorithms}. 
Section \ref{Section - results} provides an overview of the
results of the thesis.

\section{Markov Chain Monte Carlo} \label{Section MCMC}

Let $\stany$ be a region in a possibly high-dimensio\-nal space,
and let $f$ be a real valued function on $\stany.$ Moreover
consider a probability distribution $\pi$ with density $p$ with
respect to some standard measure $dx,$ usually either Lebesque or
counting measure, i.e. $\pi(dx) = p(x) dx.$
 An essential part of many problems in Bayesian
inference, statistical physics and combinatorial enumeration is
the computation of analytically intractable integral
\begin{equation}
I = E_{\pi} f = \pi f = \int_{\stany} f(x) \pi (dx),
\label{integral}
\end{equation}
where $p$ and thus $\pi$ is known up to a normalizing constant and
direct simulation from $\pi$ is not feasible (see e.g.
\cite{RoCa}, \cite{Liu}). The common approach to this problem is
to simulate an ergodic Markov chain $(X_{n})_{n\geq 0}$, using a
transition kernel $P$, with stationary distribution $\pi$, which
ensures the convergence in distribution of $X_{n}$ to a random
variable from $\pi$. Thus, for a "large enough" $t$, $X_{n}$ for
$n \geq t$ can be considered as having distribution approximately
equal to $\pi$. Since a simple and powerful algorithm for
constructing such a Markov chain has been introduced in 1953 by
Metropolis et al. in the very seminal paper \cite{Metrop}, various
sampling schemes and approximation strategies for estimating the
unknown value of $I$ have been developed and analyzed
(\cite{NiePo}, \cite{Liu}, \cite{RoCa}). The method is referred to
as Markov chain Monte Carlo (MCMC).

To avoid problems with integrating functions with respect to
probability distributions with unknown normalizing constants,
Bayesian statisticians used to restrict attention to conjugate
priors (see e.g. \cite{Robert Bayesian Choice}). This concept,
although technically appealing, deprives the bayesian approach of
flexibility which is one of its main strengths. Also, when
building complex models with many parameters (as in the example of
Section \ref{Section - Bayesian Example}), even using conjugate
priors usually leads to intractable multidimensional posterior
distributions.

The invention of MCMC has transformed dramatically Bayesian
inference since it allows practitioners to sample from complicated
posterior distributions and to integrate functions with respect to
these distributions. Thus Bayesian inference became a feasible and
powerful approach for practitioners and now receives immense
attention from the statistics community (\cite{RobRos
survey},\cite{RoCa}).

In addition to their importance for applications, MCMC algorithms
rise numerous questions related to Markov chains and probability.
It is crucial to understand the nature and speed of convergence of
the distribution of $X_n$ to $\pi$ as $n \to \infty.$

\section{Sampling Schemes and MCMC Algorithms} \label{Section
algorithms}

Before we proceed to the description of MCMC algorithms let us
recall the independent Monte Carlo solution to the problem in
(\ref{integral}) when simulating from $\pi$ is feasible. In this
case one takes i.i.d. random variables $X_i, \dots, X_n \sim \pi$
and estimates $I$ by
\begin{equation}
\hat{I}_n = \frac{1}{n}\sum_{i=1}^n f(X_i). \label{estimator -
standard MC}
\end{equation}
\begin{remark} Basic properties of the independent Monte Carlo estimation
are very easy to obtain.
\begin{itemize}
\item If $I$ exists then
$\hat{I}_n$ is its unbiased and (by the weak law of large numbers)
consistent estimate.
\item Furthermore, if $\pi f^2 < \infty,$ then by the classical
Central Limit Theorem $$\sqrt{n}(\hat{I}-I) \stackrel{d}{\to} N(0,
\pi f^2 - (\pi f)^2).$$
\item Confidence intervals for $I$ can be obtained e.g. by the Chebyshev
inequality
 $$P(|\hat{I}_n-I|\geq \varepsilon) \leq \frac{\pi f^2 -
(\pi f)^2}{n\varepsilon^2},$$
 provided that the variance $\pi f^2
- (\pi f)^2$ can be bounded a priori.
\item Asymptotic confidence intervals can be derived from the CLT,
$$P(|\hat{I}_n-I| \geq \varepsilon) \lesssim
2-2\Phi \left(\frac{\sqrt{n} \varepsilon}{\sqrt{\pi f^2 - (\pi
f)^2}}\right),$$ and effectively computed using a consistent
estimate or an upper bound of $\pi f^2 - (\pi f)^2.$
\end{itemize} \end{remark}
Assume now the MCMC setting, where no efficient procedure for
sampling independent random variables from $\pi$ is available. Let
$(X_{n})_{n\geq 0}$ be an ergodic Markov chain on $\stany$ with
transition kernel $P$ and stationary limiting distribution $\pi.$
Let $\pi_0$ denote the initial distribution of the chain, i.e.
$X_0 \sim \pi_0.$ The distribution of $X_t$ is $\pi_t=\pi_0 P^t
\to \pi,$ but $X_0, X_1, \dots$ are dependent random variables and
(\ref{estimator - standard MC}) is no longer an obvious and easy
to analyze estimator. There are several possible strategies (cf.
\cite{Geyer practical}, \cite{NiePo}, \cite{chan yue - jrssb -
asymptotic}, \cite{Liu}, \cite{RoCa}).
\begin{itemize}
\item \textit{Estimation Along one Walk.} Use average along a single trajectory
of the underlying Markov
chain and discard the initial part to reduce bias. In this case
the estimate is of the form
\begin{equation}\label{est along one walk}
\hat{I}_{t,n}=\frac{1}{n}\sum_{i=t}^{t+n-1}f(X_i)\end{equation}
and $t$ is called the burn-in time.
\item \textit{Estimation Along one Walk with Spacing.}
Discard the initial part of a single trajectory to reduce bias and
then take every $s-$th observation to reduce correlation. In this
case the estimate is of the form
\begin{equation}\label{est along one walk with spacing}
\hat{I}_{t,n,s}=\frac{1}{n}\sum_{i=t}^{t+n-1}f(X_{is})\end{equation}
and $s$ is called the spacing parameter.
\item \textit{Multiple Run.} Use average over final states of multiple independent runs
of the chain. Thus we need first to simulate say $n$ trajectories
of length say $t$:
\begin{eqnarray} X_0^{(1)},X_1^{(1)} &\dots, &X_t^{(1)}, \nonumber \\
\vdots \qquad \nonumber \\
X_0^{(n)},X_1^{(n)} &\dots, &X_t^{(n)}, \nonumber
\end{eqnarray}
and for an estimate we take
\begin{equation}\hat{I}_{t,n}=\frac{1}{n}\sum_{m=1}^{n}f(X_{t}^{(m)}),
\label{est multiple run} \end{equation} where $m$ numbers the
independent runs of the chain and $t$ should be large enough to
reduce bias.
\item \textit{Median of Averages.} Use median of multiple independent
shorter runs. Here we simulate
\begin{itemize}
\item Simulate $m$ independent runs of length $t+n$ of the
underlying Markov chain, $$X_{0}^{(k)}, \dots, X_{t+n-1}^{(k)},
\quad k=1,\dots, m.$$
\item Calculate $m$ estimates of $I,$ each based on a single run,
$$\hat{I}_{k}=\hat{I}_{t,n}^{(k)}=\frac{1}{n}\sum_{i=t}^{t+n-1}
f(X_{i}^{(k)}), \quad k=1,\dots,m.$$
\item For the final estimate take
$$\hat{I}=\textup{med}\{\hat{I}_1,\dots,\hat{I}_m\}.$$
\end{itemize}
\end{itemize}
The one walk estimators are harder to analyze since both $X_t,
\dots X_{t+n-1}$ and $X_{ts}, \dots X_{(t+n-1)s}$ are not
independent, whereas $X_t^{(1)}, \dots, X_t^{(m)}$ are. Yet one
walk strategies are believed to be more efficient and are usually
the practitioners' choice. Some precise results comparing the
first three estimators under certain assumptions are available and
confirm the practitioners' intuition. We refer to them later.

For each choice of estimation strategy additional questions arise,
since one has to decide how to chose parameters $t, n$ or $t,n,s$
or $t,n,m$ respectively, that assure ''good quality of
estimation''. This choice must clearly depend on how one defines
the desired ''quality of estimation''.

Moreover, we see from the above that MCMC requires a Markov chain
on $\stany$ which is easily run on a computer, and which has $\pi$
as its stationary limiting distribution. It may be a bit
surprising that there exist reasonably general recipes for
constructing such a chain that converges to $\pi$ in most settings
of practical interest.

\subsection{The Metropolis Algorithm} \label{section algs
metropolis}

The Metropolis algorithm has been introduced by Metropolis et al.
in \cite{Metrop}. Let $Q$ be a transition kernel of any other
Markov chain that is easily simulated on a computer. Recall that
$\pi(\cdot)$ has a density $\pi(dx)= p(x) dx,$ with possibly
unknown normalizing constant. Let also $Q(x, \cdot)$ have a
density $Q(x, dy)=q(x, y) dy.$ These densities are taken with
respect to some $\sigma-$finite reference measure $dx$, which
typically is the Lebesgue measure on $R^d$, however other settings
are possible, including counting measures on discrete state
spaces.

The Metropolis algorithm proceeds as follows.

\begin{enumerate}
\item Draw $X_0$ from an initial distribution $\pi_0$ (typically
$\pi_0=\delta_{x_0}$ for some $x_0 \in \stany$).
\item Given $X_n$ draw a proposal $Y_{n+1}$ from $Q(X_n,\cdot).$
\item Set $$X_{n+1} = \left\{\begin{array}{lll} Y_{n+1} &
\textrm{with probability} & \alpha (X_n,Y_{n+1}), \\ X_n &
\textrm{with probability} & 1-\alpha (X_n, Y_{n+1}), \end{array}
\right.$$ where $$\alpha (x,y):= \min \left\{1,
\frac{p(y)q(y,x)}{p(x)q(x,y)}\right\}$$ (Also, set $\alpha
(x,y)=1$ whenever $p(x)q(x,y)=0.$)
\item Replace $n$ by $n+1$ and go to 2.
\end{enumerate}

Note that one only has to compute the ratio of densities
$p(y)/p(x),$ and hence the unknown normalizing constant for $\pi$
in the acceptance probability $\alpha(x,y)$ simplifies and one
does not need to know it to run the chain.

Choosing the proposal density is another question that arises when
implementing the Metropolis algorithm and different ways of doing
it lead to different classes of algorithms. Typical classes
include
(see e.g. \cite{RobRos survey}
)
\begin{itemize}
\item \textit{Symmetric Metropolis Algorithm.} In this case
$q(x,y)=q(y,x)$ and hence
$\alpha(x,y)=\min\{1,\frac{\pi(y)}{\pi(x)}\}.$
\item \textit{Random Walk Metropolis-Hastings.} In this case
$q(x,y)=q(y-x).$
\item \textit{Independence Sampler.} In this case the proposal
does not depend on $x,$ i.e. $q(x,y) = q(y).$
\item \textit{Langevin Algorithm.} Where $Q(X_n, \cdot)=N(X_n
+(\delta/2) \nabla \log \pi(X_n), \delta)$ for some $\delta >0.$
\end{itemize}

\subsection{The Gibbs Sampler} \label{section algs gibbs}

The Gibbs Sampler is suitable in a setting where $\stany$ is a
product space. For simplicity we suppose in this section that
$\stany$ is an open subset of $R^d,$ and write $x=(x_1, \dots,
x_d).$

The $i-$th component $P_i$ of the Gibbs sampler $P$ replaces $x_i$
by a draw from the conditional distribution $\pi(x_i|x_1, \dots,
x_{i-1}, x_{i+1}, \dots, x_d).$

To state it more formally let, similarly as in \cite{RobRos
survey}, $$S_{x,i,a,b}= \{y \in \stany; y_j=x_j \textrm{ for } j
\neq i, \textrm{ and } a \leq y_i \leq b \}.$$

And \begin{equation}\label{Gibbs component} P_i(x, S_{x,i,a,b})=
\frac{\int_a^b p (x_1, \dots, x_{i-1},t,x_{i+1},\dots, x_d)
dt}{\int_{-\infty}^{\infty} p (x_1, \dots,
x_{i-1},t,x_{i+1},\dots, x_d) dt}.\end{equation}

Now the \textit{deterministic scan Gibbs sampler} uses the
transition kernel \begin{equation} \label{deterministic Gibbs
sampler} P=P_1P_2\cdots P_d,\end{equation} i.e. updates the
coordinates of $X_n$ in a systematic way, one after another, with
draws from full conditional distributions.

On the other hand the \textit{random scan Gibbs sampler} choses a
coordinate uniformly at random and performs its update, i.e. it
uses the transition kernel
\begin{equation} \label{random Gibbs sampler}
P=\frac{1}{d}\sum_{i=1}^d P_i.\end{equation}

In the example of Section \ref{sec:example HREM} drawing from
conditional distributions will be straightforward and in fact this
is often the case for bayesian posterior distributions. However,
if this step is infeasible, then instead of using $P_i$ as defined
in (\ref{Gibbs component}), one performs one step of a Metropolis
algorithm designed to update $i-$th coordinate. Such a procedure
is then called \textit{Metropolis within Gibbs algorithm.}

\section{Overview of the Results}
\label{Section - results}

Existing literature on Markov chains and their applications to
Markov chain Monte Carlo procedures is to large extent focused on
obtaining bounds on convergence rates to the stationary
distribution (\cite{Bax}, \cite{DMR}, \cite{Jones Hobert Gibbs for
rand eff mod}, \cite{Roberts Tweedie}, \cite{Rosenthal drift}) and
on asymptotical results for MCMC estimators
(\cite{JonesNaranCaffoNeath}, \cite{KipnisVaradhan},
\cite{MeynTweedie}). However, when analyzing MCMC estimators,
results on the rate of convergence to the stationary distribution
allow only to keep bias in control and do not translate in a
straightforward way into bounds on the mean square error or
confidence intervals. Moreover, asymptotic results may turn out
useless in practice and may even be misleading (\cite{RobRos
survey}).

\medskip

The main goal of this thesis is to obtain fixed-width results for
an estimator, say $\hat{I},$ based on an MCMC algorithm. In
particular we strive for the $(\varepsilon-\alpha)-$approximation,
i.e.
\begin{equation} \label{eps-alpha intro}
P(|\hat{I} - I| \geq \varepsilon) \leq \alpha,
\end{equation}
where $\varepsilon$ is the desired quality of estimation and
$\alpha$ is the confidence level.

In analyzing Markov chains and estimators based on MCMC procedures
we take the regenerative approach based on the split chain. The
split chain construction allows to divide the Markov chain
trajectory into independent or $1-$dependent blocks and turns out
to be an extremely powerful technique with wide range of
applications. The approach has been introduced independently in
\cite{Athreya Ney} and \cite{Nummelin paper} and immensely
developed in \cite{Nummelin book} and \cite{MeynTweedie}. We give
the basics of the approach in Chapter \ref{Section - Markov
chains}.

\medskip

Results related to (\ref{eps-alpha intro}) are known in literature
for discrete state space $\stany$ and bounded function $f$
(\cite{Aldous}, \cite{Gillman}, \cite{Perron}). For general state
space $\stany,$ and uniformly ergodic Markov chains (which in
practice implies that $\stany$ is compact) and bounded function
$f,$ exponential inequalities are available (due to \cite{Glynn}
and an improved result due to \cite{MeynKonto}) thus
$(\varepsilon-\alpha)-$approximation can be easily deduced.

For a general, not necessarily compact, state space $\stany$ (or
equivalently, not uniformly ergodic chains) and unbounded function
$f$ (which is e.g. the case when computing bayesian estimators for
a quadratic loss function) no nonasymptotic results of type
(\ref{eps-alpha intro}) are available. Fixed-width estimation is
performed by deriving asymptotic confidence intervals based on
$$\hat{I}_n = \frac{1}{n} \sum_{i=0}^{n-1}f(X_i).$$ This
construction requires two steps. First requirement is that a
central limit theorem must hold, i.e.
\begin{equation}\label{ctg overview}
\frac{\hat{I}_n - I}{\sqrt{n}} \stackrel{d}{\longrightarrow}
N(0,\sigma_f^2),
\end{equation}
where $\sigma_f^2 < \infty$ is the asymptotic variance. The second
step is to obtain a strongly consistent estimator
$\hat{\sigma}^2_f$ of $\sigma_f^2.$ Recent paper
\cite{JonesNaranCaffoNeath} presents the state of the art approach
to the problem.

\medskip

Results of Chapter \ref{chapter: CLT} and Chapter \ref{Chapter
fixed-width asymptotics} are related to this methodology.

In Chapter \ref{chapter: CLT}, based on \cite{clt ecp}, a
necessary and sufficient condition in terms of regeneration for a
central limit theorem for functionals of ergodic Markov chains (as
defined in (\ref{ctg overview}) have been obtained. It turns out,
that the CLT holds if and only if excursions between regenerations
are square integrable. An additional result of Chapter
\ref{chapter: CLT} is a solution to the open problem posed in
\cite{RobRos survey}, i.e. a regeneration proof of a CLT for
uniformly ergodic Markov chains with $E_{\pi} f^2 < \infty.$

Chapter \ref{Chapter fixed-width asymptotics}, based on \cite{BeLa
JASA}, is devoted to relaxing assumptions for strongly consistent
estimators of $\sigma_f^2.$ Results of Chapter \ref{Chapter
fixed-width asymptotics} improve the methodology of
\cite{JonesNaranCaffoNeath}.

\medskip

In Chapter \ref{CHAPTER Fixed-Width Nonasymptotic Results under
Drift Condition} nonasymptotic results of type (\ref{eps-alpha
intro}) are obtained for noncompact state space $\stany$ and
without assuming boundedness of the target function $f.$

More precisely, the goal of this chapter is to analyze estimation
along one walk
\begin{equation}\label{estym t n}
\hat{I}_{t,n} = \frac{1}{n}\sum_{i=t}^{t+n-1}f(X_i)
\end{equation}
of the unknown value $I$ under the following drift condition
towards a small set.
\begin{itemize}
\item[(A.1)]{Small set.} There exist $C\in \borel,$ $\tilde{\beta}>0$ and a
probability measure $\nu$ on $(\stany, \borel)$ such that for all
$x\in C$ and $A \in \borel$ $$P(x,A)\geq \tilde{\beta} \nu (A).$$
\item[(A.2)]{Drift.} There exist a function $V:\stany \to
[1,\infty)$ and constants $\lambda<1$ and $K<\infty$ satisfying
$$PV(x) \leq \left\{\begin{array}{lcc}
\lambda V(x), & \text{if} & x\notin C, \\ K, & \text{if} & x \in
C.
\end{array}
\right.$$
\item[(A.3)]{Aperiodicity.} There exists $\beta>0$ such that
$\tilde{\beta} \nu(C) \geq \beta.$
\end{itemize}
Under this assumption we provide explicit lower bounds on the
\textit{burn-in} time $t$ and the length of simulation $n$ that
guarantee $(\varepsilon-\alpha)-$approximation. These bounds
depend only and explicitly on the estimation parameters
$\varepsilon$ and $\alpha,$ drift parameters $\tilde{\beta},
\beta, \lambda, K$ and the the $V-$norm of the target function
$f,$ i.e. $|f^2|_V = \sup_{x} f^2(x)/V(x).$

Moreover we analyze also estimation by the \textit{median of
averages} introduced in the previous section. It turns out that
for small $\alpha$ sharper bounds on the total simulation cost
needed for $(\varepsilon-\alpha)$-approximation are available in
this case by a simple exponential inequality.

The results of Chapter \ref{CHAPTER Fixed-Width Nonasymptotic
Results under Drift Condition} have been applied for Gibbs
samplers for a Hierarchical Random Effects Model of practical
interest enabling nonasymptotic fixed-width analysis of this
model. In particular this extends the results form \cite{Jones
Hobert Gibbs for rand eff mod}, where burn in bounds in terms of
total variation norm have been established for this model.

\medskip

Chapter \ref{CHAPTER Convergence Results for Adaptive Monte Carlo}
deals with a slightly different topic, namely adaptive procedures.
The idea is to modify the transition kernel based on the
information collected during the simulation. This usually leads to
a stochastic process that are not Markov chains any more and are
less tractable theoretically. On the other hand, an adaptive
procedure at time $n$ as allowed to make use of an additional
information: the sample trajectory up to time $n.$ Clearly the
class of stochastic processes used for simulation is bigger. Thus
a smart use of the idea may lead to improvements in estimation
quality. Simulations confirm this expectations and numerical
examples for numerous specific algorithms outperform classical
procedures \cite{RobRos Adapt Examples}, \cite{Nott}. An important
example of the application of adaptive schemes is the Metropolis
algorithm with multivariate normal proposal. In this case
adaptation allows for automated choice of the covariance matrix
for the proposal distribution \cite{aro}. Theoretical results on
convergence and quality of estimation for adaptive procedures are
very modest so far. Typical conditions that allow for
investigation of convergence are called \textit{diminishing
adaptation} will be provided in Chapter \ref{CHAPTER Convergence
Results for Adaptive Monte Carlo}. Time stability conditions for
transition kernels assumed in (\cite{aro}, \cite{Nott}) fit into
the diminishing adaptation framework. Intuitively time stability
means that the adaptive process approaches a time homogeneous
Markov chain.

In Chapter \ref{CHAPTER Convergence Results for Adaptive Monte
Carlo} we prove two results a convergence rate theorem and a law
of large numbers for adaptive schemes. For both results we assume
a \textit{path stability condition} for transition kernels which
is weaker then the \textit{time stability} condition, assumed in
\cite{aro} to prove similar results. The \textit{path stablity
condition} results from \textit{time stability} condition by the
triangle inequality and intuitively means that the adaptive
process approaches a time in-homogeneous Markov chain.

   \chapter{Some Markov Chains}
\label{Section - Markov chains}

In this chapter we give some basic definitions and facts about
stationarity and ergodicity of Markov chains that justify the
Metropolis algorithm and the Gibbs sampler of Section \ref{Section
algorithms} and provide grounds for the MCMC methodology. Next we
outline the regeneration construction and the split chain and
introduce typical objects and tools useful in for analyzing
regenerative chains. Systematic, applications driven development
of Markov chains theory via regeneration can be found e.g. in
\cite{MeynTweedie} and \cite{Nummelin book} that constitute an
immense body of work. Hence we we do not attempt a systematic
treatment of the Markov chain theory here and this chapter, based
on \cite{MeynTweedie}, \cite{Nummelin book}, \cite{RobRos survey}
and \cite{Numm MC MCMCists} is nothing more then a place for
notions and tools frequently used in later chapters.

\section{Stationarity and Ergodicity} \label{sec: chap chains:
stationarity and ergodicity}

Although majority of the results we describe carry over to the
setting where $\stany$ is a general set and $\borel$ is a
countably generated $\sigma-$algebra (see e.g.
\cite{MeynTweedie}), in our applications driven development we
believe Polish spaces offer more then sufficient generality and a
grat deal of ''comfort''. Thus, if not stated otherwise, the state
space $\stany$ shall be a Polish and $\borel$ shall denote the
Borel $\sigma-$algebra on $\stany$. A transition
kernel $P$ 
on $(\stany, \borel)$ is a map $P: \stany \times \borel \to
[0,1],$ such that
\begin{itemize}
\item for any fixed $A \in \borel$ the function $P(\cdot, A)$ is
measurable,
\item for any fixed $x \in \stany$ the function $P(x, \cdot)$ is a
probability measure on $(\stany, \borel).$
\end{itemize}

For a probability measure $\mu$ and a transition kernel $Q$, by
$\mu Q$ we denote a probability measure defined by $$\mu
Q(\cdot):=\int_{\stany} Q(x,\cdot)\mu (dx),$$ furthermore if $g$
is a real-valued measurable function on $\stany$ let
$$Qg(x):=\int_{\stany} g(y)Q(x,dy)$$ and $$\mu
g:=\int_{\stany} g(x)\mu(dx).$$ We will also use $E_{\mu}g$ for
$\mu g,$ especially if $\mu=\delta_x$ we will write $E_x g.$ For
transition kernels $Q_1$ and $Q_2$, $Q_1Q_2$ is also a transition
kernel defined by $$Q_1Q_2(x,\cdot):=\int_{\stany}
Q_2(y,\cdot)Q_1(x,dy).$$

Let $(X_n)_{n\geq0}$ denote a time homogeneous Markov chain on
$\stany$ evolving according to the transition kernel $P,$ i.e.
such that $\mathcal{L}(X_{n+1}|X_n)=P(X_n,\cdot).$ By $\pi_0$
denote the distribution of $X_0,$ i.e. the initial distribution of
the chain. Then, using the above notation the distribution of
$X_n$ is $\pi_n= \pi_0 P^n.$ In particular, if $\pi = \delta_x,$
then $X_n$ is distributed as $\pi_n=\delta_x P^n=P^n(x, \cdot).$
Clearly the behavior of $\pi_n$ is of our vital interest.

We say that a probability distribution $\pi$ is stationary for
$P,$ if $\pi P=\pi.$ A crucial notion related to stationarity via
Proposition \ref{prop - rev-stat} is reversibility.
\begin{defi} A Markov chain on a state space $\stany$ with transition kernel
$P$ is reversible with respect to a probability distribution $\pi$
on $\stany$, if
$$\int_A
P(x,B)\pi(dx)=\int_B P(y,A) \pi(dy), \qquad \textrm{for all } A, B
\in \borel $$ we shall write equivalently
$$\pi(dx)P(x, dy)=\pi(dy)P(y,dx), \quad \textrm{for all} \quad
x,y \in \stany.$$
\end{defi}
\begin{prop} \label{prop - rev-stat} If a Markov chain with transition
kernel $P$ is
reversible with respect to $\pi,$ then $\pi$ is stationary for
$P.$ \end{prop}
\begin{proof} \begin{eqnarray} \pi P (A)&=&\int_{\stany}
P(x,A)\pi(dx)=\int_A P(y, \stany) \pi(dy) = \int_A
\pi(dy)=\pi(A).\nonumber \end{eqnarray} \end{proof}
It is straightforward to check that the acceptance probability
$\alpha(x,y)$ of the Metropolis algorithm of Section \ref{section
algs metropolis} makes the procedure reversible with respect to
$\pi$ and thus it has $\pi$ as its stationary distribution.

Also the $i-$th component $P_i$ of the Gibbs sampler of Section
\ref{section algs gibbs} is a special case of the Metropolis
algorithm (with $\alpha(x,y)=1$) and hence $\pi$ is stationary for
$P_i.$ This implies that the random scan Gibbs sampler is
reversible and has $\pi$ as its stationary distribution. The
deterministic scan Gibbs sampler usually is not reversible,
however since $\pi$ is stationary for each $P_i,$ it is also
stationary for $P.$

Obviously stationarity is not enough for the applications in
question since it does not even imply $\pi_n \to \pi$ (see
\cite{RobRos survey} for examples), not to mention justifying any
of the estimation schemes (\ref{est along one walk}-\ref{est
multiple run}). One needs some more assumptions and notions to
investigate convergence of $\pi_n$ to $\pi$ and properties of
estimation strategies of previous sections.

In particular the total variation distance is a very common tool
to evaluate distance between two probability measures $\mu_1$ and
$\mu_2$ and is defined as
\begin{equation}\label{total var dist}
\|\mu_1 - \mu_2\|_{tv}=  \sup_{A \in \borel} |\mu_1(A)-\mu_2(A)|.
\end{equation}
We shall distinguish between the two following types of
convergence to $\pi.$
\begin{eqnarray}\label{conv def pi a.e.} \lim_{n\ra
\infty}\|P^n(x,\cdot)-\pi\|_{tv}=0, &  \mbox{for} &
\mbox{$\pi-$almost every }\;x\in \stany, \\
\label{conv def all x} \lim_{n\ra
\infty}\|P^n(x,\cdot)-\pi\|_{tv}=0, &\mbox{for} & \mbox{all
}\;x\in \stany. \end{eqnarray}
$\phi-$irreducibility and aperiodicity are properties that
guarantee convergence in (\ref{conv def pi a.e.}).
\begin{defi} A Markov chain $(X)_{n\geq 0}$ with transition kernel $P$
is $\phi-$irreducible if there exists a non-zero $\sigma-$finite
measure $\phi$ on $\stany$ such that for all $A\subseteq \stany$
with $\phi(A) > 0,$ and for all $x \in \stany,$ there exists a
positive integer $n=n(x,A)$ such that $P^n(x,A)>0.$
\end{defi}
\begin{defi}  A Markov chain $(X)_{n\geq 0}$ with transition kernel
$P$ and stationary distribution $\pi$ is periodic with period $d
\geq 2$ if there exist disjoint subsets $\stany_0, \dots,
\stany_{d-1} \subseteq \stany$ such that $\pi(\stany_1)>0$ and for
all $0\leq i \leq d-1,$ and for all $x \in \stany_i,$ $P(x,
\stany_{i+1 \mod d})=1.$ And $d$ is maximal for the property.
Otherwise the chain is called aperiodic.\end{defi}
\begin{thm} \label{thm introduction convergence}
If a Markov chain $(X)_{n\geq 0}$ with transition kernel $P$ and
stationary distribution $\pi$ on a state space $\stany$ is
$\phi-$irreducible and aperiodic, then (\ref{conv def pi a.e.})
holds. \newline Moreover, if a function $f:\stany \to \mathbb{R}$
is such that $\pi(|f|) < \infty,$ then a strong law of large
numbers holds in the following sense
\begin{equation} \frac{1}{n}
\sum_{i=0}^{n-1}f(X_i) \to \pi f, \quad \textrm{as} \quad n \to
\infty, \qquad \textrm{w.p. } 1.\label{eqn thm intr conv
2}\end{equation}
\end{thm}
The foregoing convergence result is one of many possible
formulations. A proof of the first part can be found in
\cite{RobRos survey} Section 4.6 and the strong law of large
numbers part results e.g. from Theorem 17.0.1 of
\cite{MeynTweedie}.
Theorem \ref{thm introduction convergence} is widely applicable to
MCMC algorithms. The Metropolis algorithm and the Gibbs samplers
of Section \ref{Section algorithms} are designed precisely so that
$\pi$ is stationary. Also, it is usually straightforward to verify
that the chain is aperiodic and $\phi-$irreducible with e.g.
$\phi$ being the Lebesgue measure or $\phi = \pi.$

The following example due to C. Geyer (cf. \cite{RobRos survey})
provides a simple Markov chain that exhibits a ''bad'' behavior on
a null set.
\begin{ex}\label{example: pi a.e. convergence}
Let $\stany = \{1,2,\dots\}$ and define transition probabilities
by $P(1, \{1\}) =1,$ and for $x \geq 2,$ let $P(x, \{1\}) = 1/x^2$
and $P(x, \{x+1\}) = 1-1/x^2.$ Then the chain is aperiodic and
$\pi = \delta_1$ is the invariant distribution. The chain is also
$\pi-$irreducible. However, if $X_0 = x \geq 2,$ then $P(X_n = x+n
\textrm{ for all } n) > 0,$ and $\|P^n(x,\cdot) - \pi(\cdot)\|
\nrightarrow 0.$ Thus the convergence holds only for $x=1$ which
in this case is $\pi-$a.e. $x \in \stany.$
\end{ex}

To guarantee convergence for all $x \in \stany,$ as in (\ref{conv
def all x}) one needs to assume slightly more, namely Harris
recurrence.

\begin{defi}[Harris Recurrence] A Markov chain $\lancuch$ with
transition kernel $P$ and stationary probability measure $\pi$ is
Harris recurrent if for all $A \in \borel,$ such that $\pi(A)>0,$
and all $x \in \stany,$ the chain started at $x$ will eventually
reach $A$ with probability 1, i.e. $P(\exists n : X_n \in A | X_0
=x) =1.$\end{defi}

\begin{thm} \label{thm convergence equiv Harris rec} Ergodicity as
defined in (\ref{conv def all x}) in equivalent to Harris
recurrence and aperiodicity.
\end{thm}

The foregoing Theorem \ref{thm convergence equiv Harris rec}
results form Proposition 6.3 in \cite{Nummelin book}. Harris
recurrent and aperiodic chains are often referred to as
\textit{Harris ergodic.}

The speed of convergence in (\ref{conv def pi a.e.}) or (\ref{conv
def all x}) is another natural criterion for classifying chains.
Geometrically ergodic and uniformly ergodic chains are of
particular interest.

\begin{defi}[Uniform Ergodicity and Geometric Ergodicity]
\label{defi geom and unif ergodicity} We say that a Markov chain
$(X_n)_{n\geq 0}$ with transition kernel $P$ and stationary
distribution $\pi$ is
\begin{itemize}
\item \textit{geometrically ergodic,} if
$\|P^n(x,\cdot)-\pi(\cdot)\|_{tv} \leq M(x) \rho^n, $ for some
$\rho < 1$ and $M(x) < \infty \quad \pi-$almost everywhere,
\item \textit{uniformly ergodic,} if
$\|P^n(x,\cdot)-\pi(\cdot)\|_{tv} \leq M \rho^n, $ for some $\rho
< 1$ and $M < \infty,$
\end{itemize}
\end{defi}

The difference between geometric ergodicity and uniform ergodicity
is that $M$ may depend on the initial state $x.$ Obviously, if a
chain is geometrically ergodic and $M(x)$ is a bounded function,
then the chain is also uniformly ergodic. In particular, if the
state space is finite, then every geometrically ergodic Markov
chain is uniformly ergodic. (And from the standard theory of
discrete state space Markov chains we know that every ergodic
chain is uniformly ergodic.) Verifying uniform or geometric
ergodicity is in general nontrivial and we will refer to it later.
An interesting result for the algorithms presented in Chapter
\ref{chapter: introduction} is for example that a symmetric
random-walk Metropolis algorithm is geometrically ergodic if and
only if $\pi$ has finite exponential moments, as shown in
\cite{Mengersen Tweedie}.

Since in the sequel we deal with integrals of unbounded functions
$f$ with respect to probability measures, the very common total
variation distance defined by (\ref{total var dist}) is in this
case inappropriate for measuring distances between probability
measures and we need to introduce the $V-$norm and $V-$norm
distance.

Let $V:\stany\to [1,\infty)$ be a measurable function. For
measurable function $g:\stany \to R$ define its \emph{V-norm} as
\[ |g|_{V}:=\sup_{x \in
\stany}\frac{|g(x)|}{V(x)}. \]
 To evaluate the distance between two probability measures
$\mu_{1}$ and $\mu_{2}$ we use the \emph{V-norm distance}, defined
for probability measures $\mu_{1}$ and $\mu_{2}$ as
\[ \|\mu_{1}-\mu_{2}\|_{V} := \sup_{|g| \leq V}
\left|\mu_1g-\mu_2g
\right|.\]
Note that for $V\equiv 1$ the $V-$norm distance $||\cdot||_{V}$
amounts to the total variation distance, i.e.
$\|\mu_{1}-\mu_{2}\|_{V}=2\sup_{A\in \borel}
|\mu_{1}(A)-\mu_{2}(A)|=2||\mu_{1}-\mu_{2}||_{tv}.$
 Finally for two transition kernels $Q_1$ and $Q_2$ the \emph{V-norm
 distance} between $Q_1$ and $Q_2$ is defined by
 \[|||Q_1-Q_2|||_{V}:=\big|\|Q_1(x,\cdot)-Q_2(x,\cdot)\|_{V}\big|_{V}=
 \sup_{x\in
 \stany}\frac{\|Q_1(x,\cdot)-Q_2(x,\cdot)\|_{V}}{V(x)}.
 \]
 For a probability distribution $\mu,$ define a transition
 kernel $\mu(x,\cdot):=\mu(\cdot),$ to allow for writing
 $|||Q-\mu|||_{V}$ and $|||\mu_1-\mu_2|||_{V}.$
 Define also the following Banach space
 \[B_V:=\{f:f:\stany\to R, |f|_{V}<\infty\}.\]
 Now if $|||Q_1-Q_2|||_{V}<\infty,$ then $Q_1-Q_2$ is a bounded operator
 from $B_V$ to itself, and $|||Q_1-Q_2|||_{V}$ is its operator norm.
 See \cite{MeynTweedie} Chapter 16 for details.

Now we are in a position to introduce the $V-$uniform ergodicity.
\begin{defi}[$V-$uniform ergodicity] \label{defi V-unif ergodicity}
We say that a Markov chain $(X_n)_{n\geq 0}$ with transition
kernel $P$ and stationary distribution $\pi$ is $V-$uniformly
ergodic, if \begin{equation} |||P^n-\pi |||_{V} \to 0, \quad
\textrm{as} \quad n\to \infty. \label{defi V-unif ergodicity -
eqn} \end{equation}
Moreover, since $||| \cdot |||_{V}$ is an operator norm (\ref{defi
V-unif ergodicity - eqn}) is equivalent to
\begin{equation} |||P^n-\pi |||_{V} \leq M\rho^n, \quad
\textrm{for some} \quad M< \infty \; \textrm{and} \; \rho < 1.
\label{defi V-unif ergodicity - eqn geom} \end{equation}\end{defi}

\section{Small Sets and the Split Chain} \label{section: preliminaries}

The regeneration construction has been invented independently by
\cite{Nummelin paper} and \cite{Athreya Ney} and is now a very
celebrated technique. The development of this approach resulted in
intuitive and rather simple proofs of most results about Markov
chains and enabled better understanding and rapid progress of the
theory. In this section we provide the basics of the regeneration
and split chain construction needed for the following chapters.
Systematic development of the theory can be found in
\cite{Nummelin book} and \cite{MeynTweedie} which we exploit here.

We begin with the following definition of an atom.
\begin{defi}[Atom]
A set $B \in \borel$ is called an atom for a Markov chain
$(X)_{n\geq0}$ with transition kernel $P$ if there exists a
probability  measure $\nu$ on $\borel,$ such that for all $x \in
B,$
$$P(x, \cdot)=\nu(\cdot).$$ If the Markov chain is
$\psi-$irreducible and $\psi(B)>0$ then $B$ is called an
accessible atom.
\end{defi}

A single point $x \in \stany$ is always an atom. For a discrete
state space irreducible Markov chain every single point is an
accessible atom. Much of the discrete state space theory is
developed by studying Markov chain tours between consecutive
visits to a distinguished atom $c \in \stany.$ On a general state
space accessible atoms typically do not exist. However such atoms
can be artificially constructed. First we provide a general
version of a minorization condition that enables this
construction.

\begin{defi}[Minorization Condition - a general version]
\label{def: chains: minorization general} Let $s:
\stany \to [0,1]$ be a function for which $E_{\pi} s > 0$ and
there exists an $m>0$ and such a probability measure $\nu_m$ on
$\borel,$ that for all $x \in \stany,$
\begin{equation} \label{eqn:
small_function} P^m(x,\cdot) \geq s(x) \nu_m(\cdot).\end{equation}
\end{defi}

However, a special case of this condition with $s(x) = \varepsilon
\1c$ usually turns out to be as powerful as the general version
and is often more suitable to work with.

\begin{defi}[Small Set] \label{defi: small set}
A set $C \in \borel$ is
$\nu_m-$small, if there exist $m>0,$ $\varepsilon > 0,$ and a
probability measure $\nu_m$ on $\borel,$ such that for all $x \in
C,$ \begin{equation} \label{eqn: small_1} P^m(x,\cdot)\geq
\varepsilon \nu_m(\cdot).\end{equation}
\end{defi}

\begin{remark} \label{remark: small sets exist}
Theorem 5.2.2 of \cite{MeynTweedie} states that any
$\psi-$irreducible Markov chain is well-endowed with small sets
$C$ of positive measure $\psi$ and such that $\nu_m(C) > 0$. Since
ergodic Markov chains are $\pi-$irreducible, for an ergodic chain
a small set $C$ with $\pi(C)>0$ and $\nu_m(C)>0$ always exists.
\end{remark}

Definition \ref{defi: small set} and Remark \ref{remark: small
sets exist} imply the following minorization condition.

\begin{defi}[Minorization Condition]
For some $\varepsilon>0,$ some $C$ such that $\psi(C)>0,$ and some
probability measure $\nu_m$ with $\nu_m(C) = 1$ we have for all $x
\in C,$
\begin{equation} \label{eqn: small} P^m(x,\cdot)\geq \varepsilon
\nu_m(\cdot).\end{equation}
\end{defi}

The minorization condition (\ref{eqn: small}) allows for
constructing the split chain for $(X_n)_{n\geq 0}$ which is the
central object of the approach (see Section 17.3 of
\cite{MeynTweedie} for a detailed description). Let
$(X_{nm})_{n\geq 0}$ be the $m-$skeleton of $(X_n)_{n\geq 0},$
i.e. a Markov chain evolving according to the $m-$step transition
kernel $P^m.$ The minorization condition allows to write $P^m$ as
a mixture of two distributions:
 \begin{equation} \label{mieszanka} P^m(x,\cdot)=\varepsilon
\1c \nu_m(\cdot)+ [1- \varepsilon \1c]R(x,\cdot),\end{equation}
where $R(x,\cdot)=[1- \varepsilon \1c]^{-1}[P(x,\cdot)-
\varepsilon \1c \nu_m(\cdot)].$ Now let $(X_{nm},Y_n)_{n\geq 0}$
be the split chain of the $m-$skeleton i.e. let the random
variable $Y_n \in \{0,1\}$ be the level of the split $m-$skeleton
at time $nm.$ The split chain $(X_{nm},Y_n)_{n\geq 0}$ is a Markov
chain that obeys the following transition rule $\check{P}.$
\begin{eqnarray}
\check{P}(Y_n=1, X_{(n+1)m}\in dy | Y_{n-1},
X_{nm}=x)&=&\varepsilon \1c
\nu_m(dy) \label{skeleton split 1} \\
\check{P}(Y_n=0, X_{(n+1)m}\in dy | Y_{n-1},
X_{nm}=x)&=&(1-\varepsilon \1c) R(x,dy), \quad \quad
\label{skeleton split 2}
\end{eqnarray}
and $Y_n$ can be interpreted as a coin toss indicating whether
$X_{(n+1)m}$ given $X_{nm}=x$ should be drawn from $\nu_m(\cdot)$
- with probability $\varepsilon \1c$ - or from $R(x,\cdot)$ - with
probability $1-\varepsilon \1c.$

Obviously $(X_{nm}, Y_n)_{n\geq 0},$ i.e. the split chain of the
$m-$skeleton is a Markov chain and the crucial observation follows
from the Bayes rule, namely the set $\check{\alpha}:= C \times
\{1\}$ is an accessible atom for this chain.

One obtains the split chain $(X_k,Y_n)_{k\geq 0, n\geq 0}$ of the
initial Markov chain $(X_n)_{n\geq 0}$ by defining appropriate
conditional probabilities. To this end let
$X_0^{nm}=\{X_0,\dots,X_{nm-1}\}$ and
$Y_0^n=\{Y_0,\dots,Y_{n-1}\}.$
\begin{eqnarray}
\label{split 1} \check{P}(Y_n=1, X_{nm+1} \in dx_1, \dots,
X_{(n+1)m-1}\in dx_{m-1}, X_{(n+1)m}\in dy | \qquad \qquad \\
|Y_0^{n}, X_0^{nm};X_{nm}=x)=\frac{\varepsilon \1c
\nu_m(dy)}{P^m(x,dy)} P(x,dx_1) \cdots P(x_{m-1},dy),
\nonumber \\
\label{split 0}\check{P}(Y_n=0, X_{nm+1} \in dx_1, \dots,
X_{(n+1)m-1}\in dx_{m-1}, X_{(n+1)m}\in dy | \qquad \qquad \\
|Y_0^{n},
X_0^{nm};X_{nm}=x)=\frac{(1-\varepsilon\1c)R(x,dy)}{P^m(x,dy)}
P(x,dx_1) \cdots P(x_{m-1},dy), \nonumber
\end{eqnarray}
where $\frac{\nu_m(dy)}{P^m(x,dy)}$ and
$\frac{R(x,dy)}{P^m(x,dy)}$ are Radon-Nykodym derivatives. Note
that the marginal distribution of $(X_k)_{k\geq 0}$ in the split
chain is that of the underlying Markov chain with transition
kernel $P.$

An important characterization of the invariant measure obtained
via the splitting technique is a generalization of the Kac's
Theorem, namely Theorem \ref{thm: invariant measure}, which is the
key conclusion of Chapter 10 in \cite{MeynTweedie}. Let
$$U(x, A):= \sum_{n=1}^{\infty}P^n(x,A) = E_x \bigg(
\sum_{n=1}^{\infty} \mathbb{I}_A(X_n)\bigg)$$
 and for a measure $\psi$
define $$\mathcal{B}^{+}(\stany):=\{A \in \borel: \psi(A)>0\}.$$
\begin{defi}[Recurrent Chains] A chain $(X_n)_{n\geq 0}$ with a
transition kernel $P$ is called recurrent if it is
$\psi-$irreducible and $U(x,A) = \infty$ for any $x \in \stany$
and every $A \in \mathcal{B}^{+}(\stany).$\end{defi}
\begin{remark} Recurrence is a weaker condition then Harris
recurrence, in particular the Markov chain defined in Example
\ref{example: pi a.e. convergence} is recurrent but not Harris
recurrent.
\end{remark}
Moreover, for a set $A \in \stany$ define its hitting time
$\tau_A$ as
$$ \tau_A:= \min\{n\geq 1 : X_n \in A \}.$$
\begin{thm} \label{thm: invariant measure} Let the Markov chain $\lancuch$
be recurrent. Then there exists an unique (up to constant
multiples) invariant measure $\pi_u.$ This measure $\pi_u$ has the
following representation for any $A \in \borelplus$
\begin{equation} \label{thm eqn: invariant measure}
\pi_u(B)= \int_{A} E_x \left[ \sum_{n=1}^{\tau_A} \mathbb{I}_B
(X_n)\right] \pi_u(dx), \quad B \in \borel.\end{equation}
Moreover, the measure $\pi_u$ is finite if there exists a small
set $C$ such that $$\sup_{x \in C} E_x[\tau_C]<\infty.$$
\end{thm}

To take advantage of the splitting technique for analyzing Markov
chains and functionals of Markov chains we need a bit more
formalism. For a measure $\lambda$ on $(\stany,\borel)$ let
$\lambda^*$ denote the measure on $\stany\times\{0,1\}$ (with
product $\sigma-$algebra) defined by $\lambda^*(B\times
\{1\})=\varepsilon \lambda (B\cap C)$ and $\lambda^*(B\times
\{0\})=(1-\varepsilon) \lambda (B\cap C)+\lambda (B \cap C^c).$ In
the sequel we shall use $\nu_m^*$ for which $\nu_m^*(B\times
\{1\})= \varepsilon \nu_m(B)$ and $\nu_m^*(B \times \{0\}) =
(1-\varepsilon)\nu_m(B)$ due to the fact that $\nu_m(C)=1.$
\smallskip

\noindent Now integrate (\ref{split 1}) over $x_1, \dots, x_{m-1}$
and then over $y.$ This yields
\begin{equation} \label{eqn: do bayesa 1}
 \check{P}(Y_n=1, X_{(n+1)m} \in dy | Y_0^n, X_0^{nm}; X_{nm}=x)
 = \varepsilon \1c \nu_m(dy),\end{equation}
and
 \begin{equation} \label{eqn: do bayesa 2}
\check{P}(Y_n=1 | Y_0^n, X_0^{nm}; X_{nm}=x)
 = \varepsilon \1c.\end{equation}
From the Bayes rule we obtain
 \begin{equation} \label{eqn: z bayesa} \check{P}(X_{(n+1)m}
\in dy | Y_0^n, X_0^{nm}; Y_n=1, X_{nm}=x)
 = \nu_m(dy),\end{equation}
and the crucial observation due to Meyn and Tweedie, emphasized
here as Lemma \ref{lemma:conditional indep} follows.
\begin{lemma} \label{lemma:conditional indep}
Conditional on $\{Y_n=1\},$ the pre$-nm$ process $\{X_k,Y_i:k\ls
nm, i\ls n\}$ and the post$-(n+1)m$ process $\{X_k,Y_i:k\gs
(n+1)m, i\gs n+1\}$ are independent. Moreover, the post$-(n+1)m$
process has the same distribution as $\{X_k, Y_i: k\gs 0, i\gs
0\}$ with $\nu_m^*$ for the initial distribution of $(X_0,Y_0).$
\end{lemma}
Next, let $\sigma_{\check{\alpha}}(n)$ denote entrance times of
the split chain to the set $\check{\alpha}=C\times \{1\},$ i.e.
$$
\sigma_{\check{\alpha}}(0)  =  \min\{k\gs 0: Y_k=1\}, \quad
\sigma_{\check{\alpha}}(n)  =  \min\{k > \sigma(n-1): Y_k=1\},
\;\; n \gs 1,
$$
whereas hitting times $\tau_{\check{\alpha}}(n)$ are defined as
follows:
$$
\tau_{\check{\alpha}}(1)  =  \min\{k\gs 1: Y_k=1\},\quad
\tau_{\check{\alpha}}(n)  =  \min\{k > \tau_{\check{\alpha}}(n-1):
Y_k=1\}, \;\; n \gs 2.
$$
In view of Lemma \ref{lemma:conditional indep} it should be
intuitively clear that the following tours
$$\big\{\{X_{(\sigma_{\check{\alpha}}(n)+1)m},
X_{(\sigma_{\check{\alpha}}(n)+1)m +1}, \dots,
X_{(\sigma_{\check{\alpha}}(n+1)+1)m-1}\}, n=0,1,\dots \big\}$$
that start whenever $X_k \sim \nu_m$ are of crucial importance. In
fact in the next chapter they will turn out to be much more
tractable then the crude chain $\lancuch$ on $\stany$.

Since we are interested in functionals of the Markov chain
$\lancuch,$ for a real-valued function, say $g,$ on $\stany,$ we
define here also
\begin{equation} \label{defi eqn si of g} s_i=s_i(g)=
\sum_{j=m(\sigma_{\check{\alpha}}(i)+1)}^{m(\sigma_{\check{\alpha}
}(i+1)+1)-1}g(X_j)= \sum_{j= \sigma_{\check{\alpha}}(i)+1
}^{\sigma_{\check{\alpha}}(i+1)} Z_j(g),\end{equation}
where
\begin{equation} \label{defi eqn Zi of g}
Z_j(g)=\sum_{k=0}^{m-1} g(X_{jm+k}).
\end{equation}

\begin{remark} Clearly, one can construct the split chain based
on the more general minorization condition (\ref{eqn:
small_function}) instead of (\ref{eqn: small}). We chose
(\ref{eqn: small}) for simplicity. However, we use the split chain
construction based on (\ref{eqn: small_function}) in Chapter
\ref{Chapter fixed-width asymptotics}.
\end{remark}

   \chapter{A Complete Characterisation of $\sqrt{n}-$CLTs for
Ergodic Markov Chains via Regeneration} \label{chapter: CLT}

Central limit theorems for functionals of general state space
Markov chains are of crucial importance in sensible implementation
of Markov chain Monte Carlo algorithms as well as of vital
theoretical interest. Different approaches to proving this type of
results under diverse assumptions led to a large variety of CTL
versions. However due to the recent development of the
regeneration theory of Markov chains, many classical CLTs can be
reproved using this intuitive probabilistic approach, avoiding
technicalities of original proofs. In this paper we provide an if
and only if characterization of $\sqrt{n}-$CLTs for ergodic Markov
chains via regeneration and then use the result to solve the open
problem posed in \cite{RobRos survey}. We then discuss the
difference between one-step and multiple-step small set condition.

Results of this chapter are based on paper \cite{clt ecp} and are
joint work with Witold Bednorz and Rafał Latała.

\section{CLTs for Markov Chains}

Let $(X_n)_{n\gs 0}$ be a time homogeneous, ergodic Markov chain
on a measurable space $(\stany,\borel)$, with transition kernel
$P$ and a unique stationary measure $\pi$ on $\stany.$ We remark
that here ergodicity means that
\be\label{wb1} \lim_{n\ra \infty}\|P^n(x,\cdot)-\pi\|_{tv}=0,\quad
\;\;\mbox{for all}\;x\in \stany, \ee where $\|\cdot\|_{tv}$
denotes the total variation distance. The process $(X_n)_{n\gs 0}$
may start from any initial distribution $\pi_0$. Let $g$ be a real
valued Borel function on $\ccX$, square integrable against the
stationary measure $\pi$. We denote by $\bar{g}$ its centered
version, namely $\bar{g}=g-\int g d\pi$ and for simplicity
$S_n:=\sum_{i=0}^{n-1}\bar{g}(X_i)$. We say that a $\sqrt{n}-$CLT
holds for $(X_n)_{n\gs 0}$ and $g$ if
\begin{equation} \label{sqrt n CLTs
1} S_n / \sqrt{n} \stackrel{d}{\longrightarrow} N(0,\sigma^2_{g}),
\qquad \textrm{as} \quad n \to \infty,
\end{equation}
where $\sigma^2_{g}<\infty$.

Central limit theorems as defined by condition (\ref{sqrt n CLTs
1}) are crucial for assessing the quality of Markov chain Monte
Carlo estimation as we demonstrate in Chapter \ref{Chapter
fixed-width asymptotics} (c.f. \cite{JonesNaranCaffoNeath} and
\cite{Geyer practical}) and are also of independent theoretical
interest. Thus a large body of work on CLTs for functionals of
Markov chains exists and a variety of results have been
established under different assumptions and with different
approaches to proofs (see \cite{Jones_CLT_surv} for a review).

First we aim to provide a general result, namely Theorem
\ref{Thm1}, that gives a necessary and sufficient condition for
$\sqrt{n}$-CLTs for ergodic chains (which is a generalization of
the well known Theorem 17.3.6 \cite{MeynTweedie}). Assume for a
moment that there exists an accessible atom $\alpha \in \borel,$
i.e. such a set $\alpha$ that $\pi(\alpha)>0$ and there exists a
probability measure $\nu$ on $\borel,$ such that $P(x, A) =
\nu(A)$ for all $x \in \alpha.$ Let $\tau_{\alpha}$ be the first
hitting time for $\alpha.$ In this simplistic case we can rephrase
our Theorem \ref{Thm1} as follows:
\begin{thm}\label{thm: CLT 1}
Suppose that $(X_n)_{n\gs 0}$ is ergodic and possess an accessible
atom $\alpha$, then the $\sqrt{n}-$CLT holds  if and only if
\begin{equation}\label{cond: CTG 1}
E_{\alpha}\bigg[\bigg(\sum_{k=1}^{\tau_{\alpha}}\bar{g}(X_{k})\bigg)^2\bigg]
< \infty.
\end{equation}
Furthermore we have the following formula for the variance
$$\sigma_g^2 = \pi(\alpha)E_{\alpha}\bigg[\bigg(
\sum_{k=1}^{\tau_{\alpha}}\bar{g}(X_{k})\bigg)^2\bigg].$$
\end{thm}
We discuss briefly the relation between two classical CLT
formulations for geometrically ergodic and uniformly ergodic
Markov chains (recall Definition \ref{defi geom and unif
ergodicity}). Recently the following CLT provided by \cite{Ibrag
Linn geometr} has been reproved in \cite{RobRos survey} using the
intuitive regeneration approach and avoiding technicalities of the
original proof (however see Section \ref{section: problems} for a
commentary).
\begin{thm} \label{CLT geom}
If a Markov chain $(X_n)_{n\gs 0}$ with stationary distribution
$\pi$ is geometrically ergodic, then a $\sqrt{n}-$CLT holds for
$(X_n)_{n\gs 0}$ and $g$ whenever $\pi (|g|^{2+\delta})<\infty$
for some $\delta > 0$. Moreover $\sigma_g^2:= \int_{\stany}
\bar{g}^2 d\pi + 2\int_{\stany}\sum_{n=1}^{\infty}
\bar{g}(X_0)\bar{g}(X_n)d\pi$.
\end{thm}
\begin{remark}Note that for reversible chains the condition $\pi
(|g|^{2+\delta})<\infty$ for some $\delta
> 0$ in Theorem \ref{CLT geom} can be weakened to $\pi (g^2)<\infty$
as proved in \cite{RobRos hybrid}, however this is not possible
for the general case, see \cite{Bradley} or \cite{Haggstrom} for
counterexamples. \end{remark} Roberts and Rosenthal posed an open
problem, whether the following CLT version for uniformly ergodic
Markov chains due to \cite{Cogburn uniform} can also be reproved
using direct regeneration arguments.
\begin{thm} \label{CLT uniform}
If a Markov chain $(X_n)_{n\gs 0}$ with stationary distribution
$\pi$ is uniformly ergodic, then a $\sqrt{n}-$CLT holds for
$(X_n)_{n\gs 0}$ and $g$ whenever $\pi (g^2)<\infty.$ Moreover
$\sigma_g^2:= \int_{\stany} \bar{g}^2d\pi +
2\int_{\stany}\sum_{n=1}^{\infty} \bar{g}(X_0)\bar{g}(X_n)d\pi$.
\end{thm}
The aim of this chapter is to prove Theorem \ref{Thm1} and show
how to derive from this general framework the regeneration proof
of Theorem \ref{CLT uniform}. The outline of the chapter is as
follows. In Section \ref{section: tools} we provide some
preliminary results which may also be of independent interest. In
Section \ref{section: CLT} we detail the proof of Theorem
\ref{Thm1}, and derive \mbox{Theorem \ref{CLT uniform}} as a
corollary in Section \ref{section: CLT uniform}. Section
\ref{section: problems} comprises a discussion of some
difficulties of the regeneration approach.

\section{Tools and Preliminary Results}\label{section: tools}

Recall the split chain construction of the previous chapter and
the notation therein. In particular $s_i,$ defined by (\ref{defi
eqn si of g}) will be of our vital interest.

In this section we take $\bar{g},$ the centered version of $g,$
and analyze the sequence $s_i(\bar{g})$, $i\gs 0$. The basic
result we often refer to is Theorem 17.3.1 in \cite{MeynTweedie},
which states that $(s_i)_{i\gs 0}$ is a sequence of $1$-dependent,
identically distributed r.v.'s with $\check{E} s_i=0$. In our
approach we use the following decomposition:
$s_i=\underline{s}_i+\overline{s}_i$, where
\begin{eqnarray}
\underline{s}_i:&=&
\sum^{\sigma_{\check{\alpha}}(i+1)-1}_{j=\sigma_{\check{\alpha}}(i)+1}
Z_j(\bar{g})- \check{E}_{\pi^{\ast}_0}
\bigg[\sum^{\sigma_{\check{\alpha}}(i+1)-1}_{j=\sigma_{\check{\alpha}}(i)+1}
Z_j(\bar{g})\bigg], \label{clt: eqn: underline s} \\
\overline{s}_i: &= & Z_{\sigma_{\check{\alpha}}(i+1)}(\bar{g})-
\check{E}_{\pi^{\ast}_0}\bigg[Z_{\sigma_{\check{\alpha}}(i+1)}(\bar{g})\bigg].
 \label{clt: eqn: overline s} \end{eqnarray}
A look into the proof of Lemma \ref{lemma: overline s} later in
this section clarifies that $\underline{s}_i$ and $\overline{s}_i$
are well defined.
\begin{lemma}\label{lemma: underline s}
The sequence $(\underline{s}_i)_{i\gs 0}$ consists of i.i.d.
random variables.
\end{lemma}
\begin{proof} First note that $\underline{s}_i$ is a function of
$\{X_{(\sigma_{\check{\alpha}}(i)+1)m},
X_{(\sigma_{\check{\alpha}}(i)+1)m +1}, \dots \}$ and that
$Y_{\sigma_{\check{\alpha}}(i)}=1,$ hence by Lemma
\ref{lemma:conditional indep} $\underline{s}_0, \underline{s}_1,
\underline{s}_2, \dots$ are identically distributed. Now focus on
$\underline{s}_i, \underline{s}_{i+k}$ and
$Y_{\sigma_{\check{\alpha}}(i+k)}$ for some $k \gs 1.$ Obviously
$Y_{\sigma_{\check{\alpha}}(i+k)}=1.$ Moreover $\underline{s}_i$
is a function of the pre$-\sigma_{\check{\alpha}}(i+k)m$ process
and $\underline{s}_{i+k}$ is a function of the
post$-(\sigma_{\check{\alpha}}(i+k)+1)m$ process. Thus
$\underline{s}_i$ and $\underline{s}_{i+k}$ are independent again
by Lemma \ref{lemma:conditional indep} and for $A_i, A_{i+k},$
Borel subsets of $R,$ we have
$$\check{P}_{\pi^{\ast}_0}(\{\underline{s}_i \in A_i\} \cap \{\underline{s}_{i+k}
\in A_{i+k}\})=\check{P}_{\pi^{\ast}_0}(\{\underline{s}_i \in
A_i\})\check{P}(\{\underline{s}_{i+k} \in A_{i+k}\}).$$
 Let $0 \ls i_1 < i_2 < \dots < i_l.$ By the same pre- and
 post- process reasoning we obtain for $A_{i_1}, \dots, A_{i_l}$
 Borel subsets of $R$ that
 $$
 \check{P}_{\pi^{\ast}_0}(\{\underline{s}_{i_1} \in A_{i_1}\}
\cap \dots \cap \{\underline{s}_{i_l} \in A_{i_l}\})  = \qquad
\qquad \qquad \qquad \qquad \qquad \qquad \qquad \qquad$$
$$\qquad
\qquad \qquad = \check{P}_{\pi^{\ast}_0}(\{\underline{s}_{i_1} \in
A_{i_1}\} \cap \dots \cap \{\underline{s}_{i_{l-1}} \in
A_{i_{l-1}}\}) \cdot
\check{P}_{\pi^{\ast}_0}(\{\underline{s}_{i_l} \in A_{i_l}\}),
$$
and the proof is complete by induction.
\end{proof}
Now we turn to prove the following lemma, which generalizes the
conclusions drawn in \cite{Hob Rob} for uniformly ergodic Markov
chains.
\begin{lemma} \label{lemma: pi truncated}
Let the Markov chain $\lancuch$ be recurrent (and
$(X_{nm})_{n\gs0}$ be recurrent) and let the minorization
condition (\ref{eqn: small}) hold with $\pi(C)>0.$ Then
\begin{equation}\label{eqn: pi truncated}
\mathcal{L}(X_{\tau_{\check{\alpha}}(1)} | \{X_0,Y_0\} \in
\check{\alpha})=\mathcal{L}(X_{\sigma_{\check{\alpha}}(0)} |
\{X_0,Y_0\} \sim \nu_m^*) = \pi_C(\cdot),\end{equation} where
$\pi_C(\cdot)$ is a probability measure proportional to $\pi$
truncated to $C,$ that is $\pi_C(B)=\pi(C)^{-1}\pi(B \cap C).$
\end{lemma}
\begin{proof} The first equation in (\ref{eqn: pi truncated}) is a
straightforward consequence of the split chain construction. To
prove the second one we use Theorem \ref{thm: invariant measure}
for the split $m-$skeleton with $A=\check{\alpha}.$ Thus
$\tau_A=\tau_{\check{\alpha}}(1)$ and $\check{\pi}:=\pi^*$ is the
invariant measure for the split $m-$skeleton. Let $C \supseteq B
\in \borel,$ and compute
\begin{eqnarray}
\varepsilon \pi(B) & = & \check{\pi}(B \times \{1\})=
\int_{\check{\alpha}}\check{E}_{x,y}\left[\sum_{n=1}^{\tau_{\check{\alpha}}(1)}
\mathbb{I}_{B\times \{1\}} (X_{nm},Y_n)\right] \check{\pi}(dx,dy) \nonumber\\
& = & \check{\pi}(\check{\alpha})
\check{E}_{\nu_m^*}\left[\sum_{n=0}^{\sigma_{\check{\alpha}}(0)}\mathbb{I}_{B
\times \{1\}}(X_{nm},Y_n)\right]= \check{\pi}(\check{\alpha})
\check{E}_{\nu_m^*}
\mathbb{I}_{B}(X_{\sigma_{\check{\alpha}}(0)}). \nonumber
\end{eqnarray}
This implies proportionality and the proof is complete.
\end{proof}
\begin{lemma} \label{lemma: overline s}
$\check{E}_{\pi^{\ast}_0}\overline{s}_i^2 \leq \frac{m^2 \pi
\bar{g}^2}{\varepsilon \pi(C)} < \infty$ and
$(\overline{s}_i)_{i\gs 0}$ are $1$-dependent identically
distributed r.v.'s.
\end{lemma}
\begin{proof} Recall that $\overline{s}_i = \sum_{k=0}^{m-1}
\bar{g}(X_{\sigma_{\check{\alpha}}(i+1) m
+k})-\check{E}_{\pi^{\ast}_0}\left(\sum_{k=0}^{m-1}
\bar{g}(X_{\sigma_{\check{\alpha}}(i+1) m +k})\right)$ and is a
function of the random variable
\begin{equation} \label{eqn: ogonek}
\{X_{\sigma_{\check{\alpha}}(i+1) m}, \dots,
X_{\sigma_{\check{\alpha}}(i+1) m +m-1}\}.
\end{equation}
By $\mu_i(\cdot)$ denote the distribution of (\ref{eqn: ogonek})
on $\stany^m.$ We will show that $\mu_i$ does not depend on $i$.
From (\ref{split 1}), (\ref{eqn: do bayesa 2}) and the Bayes rule,
for $x \in C,$ we obtain
\begin{equation}\label{eqn: do distr overline s}
\check{P}\Big(X_{nm+1} \in dx_1, \dots, X_{(n+1)m-1}\in dx_{m-1},
X_{(n+1)m}\in dy \Big| \qquad \qquad \qquad
\end{equation}
\begin{equation}\nonumber
\qquad \qquad \Big| Y_0^{n}, X_0^{nm};Y_n=1, X_{nm}=x\Big) =
\frac{\nu_m(dy)}{P^m(x,dy)} P(x,dx_1) \cdots P(x_{m-1},dy).
\end{equation}
Lemma \ref{lemma: pi truncated} together with (\ref{eqn: do distr
overline s}) yields
\begin{equation}
\check{P}\Big(X_{nm}\in dx, X_{nm+1} \in dx_1, \dots,
X_{(n+1)m-1}\in dx_{m-1}, X_{(n+1)m}\in dy \Big| \quad \label{eqn:
distr overline s}\end{equation} \begin{equation} \Big|Y_0^{n},
X_0^{nm};Y_n=1;\sigma_{\check{\alpha}}(0)<n\Big) \;= \;
\pi_C(dx)\frac{\nu_m(dy)}{P^m(x,dy)} P(x,dx_1) \cdots
P(x_{m-1},dy). \nonumber
\end{equation}
Note that $\frac{\nu_m(dy)}{P^m(x,dy)}$ is just a Radon-Nykodym
derivative and thus (\ref{eqn: distr overline s}) is a well
defined measure on $\stany^{m+1}$, say $\mu(\cdot)$. It remains to
notice, that $\mu_i(A)=\mu(A\times \stany)$ for any Borel
$A\subset \stany^m$. Thus $\mu_i$, $i\gs 0$ are identical and
hence $\overline{s}_i$, $i\gs 0$ have the same distribution. Due
to Lemma \ref{lemma:conditional indep} we obtain that
$\overline{s}_i$, $i\gs 0$ are $1$-dependent. To prove
$\check{E}_{\pi^{\ast}_0}\overline{s}_i^2<\infty$, we first note
that $\frac{\nu_m(dy)}{P^m(x,dy)} \ls 1/\varepsilon$ and also
$\pi_C(\cdot) \ls \frac{1}{\pi(C)}\pi(\cdot).$ Hence
$$\mu_i(A) = \mu(A \times \stany) \ls \frac{1}{\varepsilon
\pi(C)} \mu_{\textrm{chain}}(A),$$ where $\mu_{\textrm{chain}}$ is
defined by $\pi(dx)P(x,dx_1)\dots P(x_{m-2},dx_{m-1}).$ Thus
$$\left|\check{E}_{\pi^{\ast}_0}\left(\sum_{k=0}^{m-1}
\bar{g}(X_{\sigma_{\check{\alpha}}(i+1) m +k})\right)\right| \leq
\frac{m \pi |\bar{g}|}{\varepsilon \pi(C)} <\infty.$$ Now let
$\tilde{s}_i = \sum_{k=0}^{m-1}
\bar{g}(X_{\sigma_{\check{\alpha}}(i+1) m +k})$ and proceed
\begin{eqnarray}
\check{E}_{\pi^{\ast}_0}\overline{s}_i^2  & \ls &
\check{E}_{\pi^{\ast}_0}\tilde{s}_i^2  \ls \frac{1}{\varepsilon
\pi(C)} \mu_{\textrm{chain}}\tilde{s}_i^2 = \frac{1}{\varepsilon
\pi(C)} E_{\pi}\left( \sum_{k=0}^{m-1}\bar{g}(X_k) \right)^2
\nonumber \\
& \ls & \nonumber \frac{m}{\varepsilon \pi(C)} E_{\pi}\left[
\sum_{k=0}^{m-1}\bar{g}^2(X_k) \right] \ls \frac{m^2 \pi
\bar{g}^2}{\varepsilon \pi(C)}.
\end{eqnarray}
\end{proof}
We need a result which gives the connection between stochastic
boundedness and the existence of the second moment of
$\underline{s}_i$. We state it in a general form.
\begin{thm}\label{thm: rafal}
Let $(X_n)_{n\gs 0}$ be a sequence of independent identically
distributed random variables and $S_n=\sum^{n-1}_{k=0}X_k$.
Suppose that $(\tau_n)$ is a  sequence of positive, integer valued
r.v.'s such that $\tau_n/n\rightarrow a\in (0,\infty)$ in
probability when $n\rightarrow\infty$ and  the sequence
$(n^{-1/2}S_{\tau_n})$ is stochastically bounded. Then $E
X_0^2<\infty$ and $E X_0=0$.
\end{thm}
The proof of Theorem \ref{thm: rafal} is based on the following
lemmas.
\begin{lemma}\label{lem: lemat2}
Let $\delta\in (0,1)$ and $t_0:=\sup\{t>0\colon \sup_{0\ls k\ls
n}P(|S_k|\gs t)\gs \delta\}$. Then $P(|S_{10n}|\gs 4t_0)\gs
(1-\delta)(\delta/4)^{20}$ and $P(\sup_{k\ls n}|S_k|\ls 3t_0)\gs
1-3\delta$.
\end{lemma}
\begin{proof}
By the definition of $t_0$ there exists $0\ls n_0\ls n$ such that
$P(|S_{n_0}|\gs t_0)\gs \delta$. Then either $P(|S_{n}|\gs
t_0/2)\gs \delta/2$ or $P(|S_n|\gs t_0/2)<\delta/2$ and
consequently
\begin{eqnarray}
P(|S_{n-n_0}|\gs t_0/2)&=&P(|S_{n}-S_{n_0}|\gs t_0/2)\nonumber \\
&\gs &P(|S_{n_0}|\gs t_0)-P(|S_{n}|\gs t_0/2)\gs \delta/2.
\nonumber \end{eqnarray}
Thus there exists $n/2\ls n_1\ls n$ such
that $P(|S_{n_1}|\gs t_0/2)\gs \delta/2$. Let $10n=an_1+b$ with
$0\ls b<n_1$, then $10\ls a\ls 20$,
\begin{eqnarray}
P(|S_{an_1}|\gs 5t_0) & \gs & P(S_{an_1}\gs at_0/2)+P(S_{an_1}\ls
-at_0/2) \nonumber
\\
&\gs & (P(S_{n_1}\gs t_0/2))^a+(P(S_{n_1}\ls -t_0/2))^a \gs
(\delta/4)^{a}, \nonumber
\end{eqnarray}
hence
\begin{eqnarray}
P\big(|S_{10n}|\gs 4t_0\big) & \gs & P\big(|S_{an_1}|\gs 5t_0\big)
P\big(|S_{10n}-S_{an_1}|\ls t_0\big) \nonumber \\ & \gs &
(\delta/4)^a(1-\delta)\gs (1-\delta)(\delta/4)^{20}.\nonumber
\end{eqnarray}
Finally by the Levy-Octaviani inequality we obtain
$$
P\Big(\sup_{k\ls n}|S_k|> 3t_0\Big) \; \ls \; 3\sup_{k\ls
n}P\big(|S_k|>t_0\big) \; \ls \; 3\delta.
$$
\end{proof}
\begin{lemma}\label{lem: lemat3}
\label{two} Let $c^2<\Var(X_1)$, then for sufficiently large $n$,
$P(|S_n|\gs c\sqrt{n}/4)\gs 1/16$.
\end{lemma}
\begin{proof}
Let $(X_i')$ be an independent copy of $(X_i)$ and
$S_k'=\sum_{i=1}^nX_i'$. Moreover let $(\va_i)$ be a sequence of
independent symmetric $\pm 1$ r.v.'s, independent of $(X_i)$ and
$(X_i')$. For any reals $(a_i)$ we get by the Paley-Zygmund
inequality,
\begin{eqnarray} P\bigg(\big|\sum_{i=1}^{n}a_i\va_i \big|
 \gs  \frac{1}{2}\Big(\sum_i a_i^2\Big)^{1/2}\bigg)
&=&P\bigg(\Big|\sum_{i=1}^{n}a_i\va_i\Big|^2 \gs
\frac{1}{4}E\Big|\sum_{i=1}^{n}a_i\va_i\Big|^2\bigg) \nonumber \\
& \gs & \Big(1-\frac{1}{4}\Big)^2
\frac{\big(E|\sum_{i=1}^{n}a_i\va_i|^2\big)^2}{E|\sum_{i=1}^{n}a_i\va_i|^4}
\gs \frac{3}{16}. \nonumber \end{eqnarray}
Hence
\begin{eqnarray}
P\Big(|S_n-S_n'|\gs \frac{c}{2}\sqrt{n}\Big) & = &
P\Big(|\sum_{i=1}^n\va_i(X_i-X_i')|\gs \frac{c}{2}\sqrt{n}\Big)
\nonumber \\ & \gs &
\frac{3}{16}P\Big(\sum_{i=1}^{n}(X_i-X_i')^2\gs c^2n\Big)\gs
\frac{1}{8} \nonumber \end{eqnarray}
for sufficiently large $n$ by the Weak LLN. Thus
\begin{eqnarray}
\frac{1}{8} \ls P\Big(|S_n-S_n'|\gs \frac{c}{2}\sqrt{n}\Big) & \ls
& P\Big(|S_n|\gs \frac{c}{4}\sqrt{n}\Big)+ P\Big(|S_n'|\gs
\frac{c}{4}\sqrt{n}\Big) \nonumber \\ & \ls & 2P\Big(|S_n|\gs
\frac{c}{4}\sqrt{n}\Big).\nonumber
\end{eqnarray}
\end{proof}
\begin{cor}\label{cor: wniosek1}
Let $c^2<\Var(X_1)$, then for sufficiently large $n$,
$$P(\inf_{10n\ls k\ls 11n}|S_k|\gs \frac{1}{4}c\sqrt{n})\gs
2^{-121}.$$
\end{cor}
\begin{proof}
Let $t_0$ be as in Lemma \ref{lem: lemat2} for $\delta=1/16$, then
\begin{eqnarray}
P\Big(\inf_{10n\ls k\ls 11n}|S_k|\gs t_0\Big) & \gs &
P\Big(|S_{10n}|\gs 4t_0, \sup_{10n\ls k\ls 11n}|S_k-S_{10n}|\ls
3t_0\Big) \nonumber \\
&=& P\big(|S_{10n}|\gs 4t_0\big)P\Big(\sup_{k\ls n}|S_k|\ls
3t_0\Big)\gs 2^{-121}.\nonumber
\end{eqnarray}
Hence by Lemma \ref{lem: lemat2} we obtain $t_0\gs c\sqrt{n}/4$
for large $n$.
\end{proof}
\begin{proof}[Proof of Theorem \ref{thm: rafal}]
By Corollary \ref{cor: wniosek1} for any $c^2<\Var(X)$ we have,
\begin{eqnarray}
P\Big(|S_{\tau_n}|\gs \frac{c}{20}\sqrt{an}\Big) & \gs &
P\bigg(\big|\frac{\tau_{n}}{n}-a\big| \ls \frac{a}{21},
\inf_{\frac{20}{21}an \ls k \ls \frac{22}{21}an}|S_k| \gs
\frac{c}{20}\sqrt{an}\bigg) \gs \nonumber
\\
& \gs & P\bigg(\inf_{\frac{20}{21}an\ls k\ls
\frac{22}{21}an}|S_k|\gs \frac{c}{4}\sqrt{\frac{2an}{21}}\bigg)
-P\Big(\big|\frac{\tau_n}{n}-a\big|>\frac{a}{21}\Big) \nonumber
\\ & \gs & 2^{-121}-P\Big(\big|\frac{\tau_n}{n}-a\big|>\frac{a}{21}\Big)
\gs 2^{-122} \nonumber
\end{eqnarray}
for sufficiently large $n$. Since $(n^{-1/2}S_{\tau_n})$ is
stochastically bounded, we immediately obtain $\Var(X_1)<\infty$.
If $E X_1\neq 0$ then
$$
\big|\frac{1}{\sqrt{n}}S_{\tau_n}\big|=\big|\frac{S_{\tau_n}}{\tau_n}\big|
\big|\frac{\tau_n}{n}\big|\sqrt{n}\rightarrow \infty \qquad
\textrm{in probability when} \quad n\rightarrow\infty.
$$
\end{proof}

\section{A Characterization of $\sqrt{n}$-CLTs}\label{section: CLT}

\noindent In this section we provide a generalization of Theorem
17.3.6 of \cite{MeynTweedie}. We obtain an if and only if
condition for the $\sqrt{n}$-CLT in terms of finiteness of the
second moment of a centered excursion from $\check{\alpha}.$
\begin{thm}\label{Thm1}
Suppose that $(X_n)_{n\gs 0}$ is ergodic and $\pi(g^2)<\infty$.
Let $\nu_m$ be the measure satisfying (\ref{eqn: small}), then the
$\sqrt{n}-$CLT holds  if and only if
\begin{equation} \label{cond: CTG 2}
\check{E}_{\nu_m^*}\bigg[\bigg(\sum_{n=0}^{\sigma_{\check{\alpha}}(0)}
Z_n(\bar{g})\bigg)^2\bigg] < \infty.
\end{equation}
Furthermore we have the following formula for variance
$$\sigma_g^2 = \frac{\varepsilon
\pi(C)}{m}\Bigg\{\check{E}_{\nu_m^*}\bigg[\bigg(
\sum_{n=0}^{\sigma_{\ptalfa}(0)}Z_n(\bar{g})\bigg)^2\bigg] + 2
\check{E}_{\nu_m^*}\bigg[\bigg(
\sum_{n=0}^{\sigma_{\ptalfa}(0)}Z_n(\bar{g})\bigg)
\bigg(\sum_{n=\sigma_{\ptalfa}(0)+1}^{\sigma_{\ptalfa}(1)}Z_n(\bar{g})\bigg)
\bigg]\Bigg\}.$$ 
\end{thm}
\begin{proof}
For $n\gs 0$ define
$$l_n:=\max\{k\gs 1:\;
m(\sigma_{\check{\alpha}}(k)+1)\ls  n\}$$ and for completeness
$l_n:=0$ if $m(\sigma_{\check{\alpha}}(0)+1)\gs n$. First we are
going to show that
\begin{equation}\label{eqn: nier0}
\bigg|\frac{1}{\sqrt{n}}\sum^{n-1}_{j=0}
\bar{g}(X_j)-\frac{1}{\sqrt{n}} \sum^{l_n-1}_{j=0} s_j\bigg|
\rightarrow 0 \qquad \textrm{in probability.}
\end{equation}
Thus we have to verify that the initial and final terms of the sum
do not matter. First observe that by the Harris recurrence
property of the chain $\sigma_{\check{\alpha}}(0)<\infty$,
$\check{P}_{\pi^{\ast}_0}$-a.s.  and hence
$\lim_{n\ra\infty}\check{P}_{\pi^{\ast}_0}(m\sigma_{\check{\alpha}}(0)\gs
n)=0$ and
$\check{P}_{\pi^{\ast}_0}(\sigma_{\check{\alpha}}(0)<\infty)=1.$
This yields
\begin{equation}\label{eqn: nier1}
\bigg|\frac{1}{\sqrt{n}}\sum^{n-1}_{j=0}
\bar{g}(X_j)-\frac{1}{\sqrt{n}}
\sum^{n-1}_{j=m(\sigma_{\check{\alpha}}(0)+1)} \bar{g}(X_j)\bigg|
\rightarrow 0,\quad \;\;\check{P}-\mbox{a.s.}
\end{equation}
The second point is to provide a similar argument for the tail
terms and to show that
\begin{equation}\label{eqn: nier2}
\bigg|\frac{1}{\sqrt{n}}\sum^{n-1}_{j=m(\sigma_{\check{\alpha}}(0)+1)}
\bar{g}(X_j)-\frac{1}{\sqrt{n}}\sum^{m\sigma_{\check{\alpha}}
(l_n)+m-1}_{j=m(\sigma_{\check{\alpha}}(0)+1)} \bar{g}(X_j)\bigg|
\rightarrow 0, \quad \;\;\mbox{in probability.}
\end{equation}
For $\va>0$ we have
\begin{eqnarray}
\check{P}_{\pi^{\ast}_0}\bigg(\Big|\frac{1}{\sqrt{n}}
\sum^{n-1}_{j=m(\sigma_{\check{\alpha}} (l_n)+1)}
\bar{g}(X_j)\Big| > \va \bigg) & \ls & \check{P}_{\pi^{\ast}_0}
\bigg( \frac{1}{\sqrt{n}}
\sum^{\sigma_{\check{\alpha}}(l_n+1)}_{j=\sigma_{\check{\alpha}}(l_n)+1}
Z_j(|\bar{g}|) > \va \bigg) \nonumber \\
& \ls & \sum^{\infty}_{k=0}\check{P}_{\check{\alpha}} \bigg(
\frac{1}{\sqrt{n}} \sum^{\tau_{\check{\alpha}}(1)}_{j=1}
Z_j(|\bar{g}|) > \va,\;\tau_{\check{\alpha}}(1)\gs
k\bigg).\nonumber
\end{eqnarray}
Now since
$\sum^{\infty}_{k=0}\check{P}_{\check{\alpha}}(\tau_{\check{\alpha}}(1)
\gs k)\ls
\check{E}_{\check{\alpha}}\tau_{\check{\alpha}}(1)<\infty$, where
we use that $\check{\alpha}$ is an atom for the split chain, we
deduce form the Lebesgue majorized convergence theorem that
(\ref{eqn: nier2}) holds. Obviously (\ref{eqn: nier1}) and
(\ref{eqn: nier2}) yield (\ref{eqn: nier0}).
\smallskip

\noindent We turn to prove that the condition (\ref{cond: CTG 2})
is sufficient for the CLT to hold. We will show that random
numbers $l_n$ can be replaced by their non-random equivalents.
Namely we apply the LLN (Theorem 17.3.2 in \cite{MeynTweedie})) to
ensure that \be\label{wb2} \lim_{n\ra\infty}
\frac{l_n}{n}=\lim_{n\ra\infty}\frac{1}{n}\sum^{[n/m]-1}_{k=1}
\mathbb{I}_{\{(X_{mk},Y_k)\in\check{\alpha}\}}=
\frac{\check{\pi}(\check{\alpha})}{m},\quad
\;\;\check{P}_{\pi_0^{\ast}}-\mbox{a.s.} \ee
Let
$$n^{\ast}:=\lfloor \check{\pi}(\check{\alpha})n
m^{-1}\rfloor,\qquad \underline{n}:=\lceil
(1-\va)\check{\pi}(\check{\alpha})n m^{-1}\rceil,\qquad
\overline{n}:=\lfloor (1+\va)\check{\pi}(\check{\alpha})n
m^{-1}\rfloor.$$ Due to the LLN we know that for any $\va>0$,
there exists $n_0$ such that for all $n\gs n_0$ we have
$\check{P}_{\pi^{\ast}_0}(\underline{n} \ls l_n\ls
\overline{n})\gs 1-\va$. Consequently
\begin{eqnarray}
\nonumber
\check{P}_{\pi^{\ast}_0}\bigg(\Big|\sum^{l_n-1}_{j=0}s_j-
\sum^{n^{\ast}}_{j=0}s_j \Big| > \sqrt{n}\beta \bigg) & \ls & \va
+\check{P}_{\pi^{\ast}_0}\bigg(\max_{\underline{n} \ls l \ls
n^{\ast}} \Big| \sum^{n^{\ast}}_{j=l}s_j \Big| > \beta
\sqrt{n}\bigg)+ \qquad \quad\\ && + \check{P}_{\pi^{\ast}_0}
\bigg( \max_{n^{\ast}+1\ls l \ls \overline{n}}
\Big|\sum^{l}_{j=n^{\ast}+1}s_j \Big| > \beta \sqrt{n}\bigg).
\label{eqn: nier4}
\end{eqnarray}
Since $(s_j)_{j\gs 0}$ are $1$-dependent, $M_k:=\sum^k_{j=0}s_j$
is not necessarily a martingale. Thus to apply the classical
Kolmogorov inequality we define $M^{0}_k=\sum_{j=0}^{\infty}
s_{2j} \mathbb{I}_{\{2j \leq k\}}$ and
$M^1_k=\sum^{\infty}_{j=0}s_{1+2j} \mathbb{I}_{\{1+2j \leq k\}}$,
which are clearly square-integrable martingales (due to
(\ref{cond: CTG 2})). Hence
\begin{eqnarray}
 \check{P}_{\pi^{\ast}_0}\big(\max_{\underline{n} \ls l \ls n^{\ast}}
|M_{n^{\ast}}-M_{l}|>\beta \sqrt{n}\big) & \ls &
\check{P}_{\pi^{\ast}_0}\Big(\max_{\underline{n} \ls l \ls
n^{\ast}} |M^0_{n^{\ast}}-M^0_{l}|>\frac{\beta \sqrt{n}}{2}\Big)+
\nonumber \\ && \qquad +
\check{P}_{\pi^{\ast}_0}\Big(\max_{\underline{n} \ls l \ls
n^{\ast}} |M^1_{n^{\ast}}-M^1_{l}|>\frac{\beta \sqrt{n}}{2}\Big)
\nonumber \\ & \ls & \frac{4}{n\beta^2}
\sum^{1}_{k=0}\big(\check{E}_{\pi^{\ast}_0}
|M^k_{n^{\ast}}-M^k_{\underline{n}}|^2 \big) \nonumber \\ &\ls &
C\va\beta^{-2}\check{E}_{\nu^{\ast}_m}(s_0^2),
\end{eqnarray}
where $C$ is a universal constant. In the same way we show that
$$\check{P}(\max_{n^{\ast}+1\ls l \ls \overline{n}}
|M_{l}-M_{n^{\ast}+1}|>\beta \sqrt{n})\ls C\va\beta^{-2}
\check{E}_{\nu^{\ast}_m}(s_0^2),$$ consequently, since $\va$ is
arbitrary, we obtain
\begin{equation}\label{eqn: main0}
\Big|\frac{1}{\sqrt{n}}\sum^{l_n-1}_{j=0} s_j-
\frac{1}{\sqrt{n}}\sum^{n^{\ast}}_{j=0}s_j \Big| \ra 0,\quad
\;\;\mbox{in probability.}
\end{equation}
The last step is to provide an argument for the CLT for
$1$-dependent, identically distributed random variables. Namely,
we have to prove that
\begin{equation}\label{eqn: main1}
\frac{1}{\sqrt{n}}\sum^{n}_{j=0}s_j \stackrel{d}{\ra} {\cal
N}(0,\bar{\sigma}^2),\quad \textrm{as} \quad n \to \infty,
\end{equation}
where
$$\bar{\sigma}^2:=\check{E}_{\nu^{\ast}_m}
(s_0(\bar{g}))^2+2\check{E}_{\nu^{\ast}_m}(s_0(\bar{g})s_1(\bar{g})).
$$
Observe that (\ref{eqn: nier1}), (\ref{eqn: nier2}), (\ref{eqn:
main0}) and (\ref{eqn: main1}) imply Theorem \ref{Thm1}. We fix
$k\gs 2$ and define
$\xi_j:=s_{kj+1}(\bar{g})+...+s_{kj+k-1}(\bar{g})$, consequently
$\xi_j$ are i.i.d. random variables and
\begin{equation}\label{eqn: nier5}
\frac{1}{\sqrt{n}}\sum^{n}_{j=0}s_j=
\frac{1}{\sqrt{n}}\sum^{\lfloor n/k \rfloor -1}_{j=0}\xi_j+
\frac{1}{\sqrt{n}}\sum^{\lfloor
n/k\rfloor}_{j=0}s_{kj}(\bar{g})+\frac{1}{\sqrt{n}}
\sum^{n}_{j=k[n/k]+1}s_j.
\end{equation}
Obviously the last term converges to $0$ in probability. Denoting
\begin{eqnarray}
\sigma_k^2 & := &
\check{E}_{\pi^{\ast}_0}(\xi_j)^2=(k-1)\check{E}_{\nu^{\ast}_m}
(s_0(\bar{g}))^2+2(k-2)\check{E}_{\nu^{\ast}_m}(s_0(\bar{g})s_1(\bar{g})),
\nonumber \\ \sigma_s^2 & := &\check{E}_{\nu^{\ast}_m}
(s_0(\bar{g}))^2.\nonumber
\end{eqnarray}
we use the classical CLT for i.i.d. random variables to see that
\begin{equation}\label{eqn: nier6}
\frac{1}{\sqrt{n}}\sum^{\lfloor n/k \rfloor-1}_{j=0}\xi_j
\stackrel{d}{\ra} {\cal N} (0,k^{-1}\sigma^2_k), \qquad
\mbox{and}\qquad \frac{1}{\sqrt{n}}\sum^{\lfloor n/k
\rfloor}_{j=0}s_{kj}(\bar{g})\stackrel{d}{\ra} {\cal
N}(0,k^{-1}\sigma^2_s).
\end{equation}
Moreover \be\label{eqn: nier7}
\lim_{n\ra\infty}\Big[\frac{1}{\sqrt{n}}\sum^{\lfloor n/k\rfloor
-1}_{j=0}\xi_j+ \frac{1}{\sqrt{n}}\sum^{\lfloor n/k \rfloor
}_{j=0}s_{kj}(\bar{g})\Big] \ee converges to ${\cal
N}(0,\sigma_g^2)$, with $k\ra\infty$. Since the weak convergence
is metrizable we deduce from (\ref{eqn: nier5}), (\ref{eqn:
nier6}) and (\ref{eqn: nier7}) that (\ref{eqn: main1}) holds.
\smallskip

\noindent The remaining part is to prove that (\ref{cond: CTG 2})
is also necessary for the CLT to hold. Note that if
$\sum^{n}_{k=0}\bar{g}(X_k)/\sqrt{n}$ verifies the CLT then
$\sum^{l_n-1}_{j=0}s_j$ is stochastically bounded by (\ref{eqn:
nier0}). We use the decomposition
$s_{i}=\underline{s}_i+\overline{s}_i$, $i\gs 0$ introduced in
Section \ref{section: tools}. By Lemma \ref{lemma: overline s} we
know that $\overline{s}_j$ is a sequence of $1$-dependent random
variables with the same distribution and finite second moment.
Thus from the first part of the proof we deduce that
$\sum^{l_{n}-1}_{j=0}\overline{s}_j/\sqrt{n}$ verifies a CLT and
thus is stochastically bounded. Consequently the remaining
sequence $\sum^{l_n-1}_{j=0}\underline{s}_j/\sqrt{n}$ also must be
stochastically bounded. Lemma \ref{lemma: underline s} states that
$(\underline{s}_j)_{j\gs 0}$ is a sequence of i.i.d. r.v.'s, hence
$\check{E}[\underline{s}_j^2]<\infty$ by Theorem \ref{thm: rafal}.
Also $l_n/n\ra \check{\pi}(\check{\alpha})m^{-1}$ by (\ref{wb2}).
Applying the inequality $(a+b)^2\ls 2(a^2+b^2)$ we obtain
$$
\check{E}_{\pi^{\ast}_0}[s_j]^2\ls 2
(\check{E}_{\pi^{\ast}_0}[\underline{s}_j^2]+
\check{E}_{\pi^{\ast}_0}[\overline{s}_j^2])<\infty
$$
which completes the proof.
\end{proof}
\begin{remark}
Note that in the case of $m=1$ we have $\bar{s}_i \equiv 0$ and
for Theorem \ref{Thm1} to hold, it is enough to assume $\pi|g|<
\infty$ instead of $\pi(g^2)<\infty.$ In the case of $m>1$,
assuming only $\pi |g|<\infty$ and (\ref{cond: CTG 2}) implies the
$\sqrt{n}$-CLT, but the proof of the converse statement fails, and
in fact the converse statement does not hold (one can easily
provide an appropriate counterexample).
\end{remark}

\section{Uniform Ergodicity} \label{section: CLT uniform}

\noindent In view of Theorem \ref{Thm1} providing a regeneration
proof of Theorem \ref{CLT uniform} amounts to establishing
conditions (\ref{cond: CTG 2}) and checking the formula for the
asymptotic variance. To this end we need some additional facts
about small sets for uniformly ergodic Markov chains.
\begin{thm}\label{thm: unif makes small} If $(X_n)_{n\gs0},$
a Markov chain on $(\stany,\borel)$ with stationary distribution
$\pi$ is uniformly ergodic, then $\stany$ is $\nu_m-$small for
some $\nu_m.$
\end{thm}
Hence for uniformly ergodic chains (\ref{eqn: small}) holds for
all $x \in {\cal X}.$ Theorem \ref{thm: unif makes small} is well
known in literature, in particular it results from Theorems 5.2.1
and 5.2.4 in \cite{MeynTweedie} with their $\psi=\pi.$
\smallskip

\noindent Theorem \ref{thm: unif makes small} implies that for
uniformly ergodic Markov chains (\ref{mieszanka}) can be rewritten
as
\begin{equation} \label{mieszanka 2}
P^m(x,\cdot)= \varepsilon \nu_m(\cdot) +
(1-\varepsilon)R(x,\cdot).
\end{equation}
The following mixture representation of $\pi$ will turn out very
useful.
\begin{lemma} \label{lemma: invariant}
If $(X_n)_{n\gs0}$ is an ergodic Markov chain with transition
kernel $P$ and (\ref{mieszanka 2}) holds, then
\begin{equation}\label{pi representation}
\pi = \varepsilon \mu := \varepsilon  \sum_{n=0}^{\infty} \nu_m
(1-\varepsilon)^n R^n.
\end{equation}
\end{lemma}
\begin{remark} This can be
easily extended to the more general setting than this of uniformly
ergodic chains, namely let
$P^m(x,\cdot)=s(x)\nu_m(\cdot)+(1-s(x))R(x,\cdot),$ $s:\stany \to
[0,1],$ $\pi s > 0.$ In this case $\pi = \pi s
\sum_{n=0}^{\infty}\nu_m R_\#^n,$ where
$R_\#(x,\cdot)=(1-s(x))R(x,\cdot).$ Related decompositions under
various assumptions can be found e.g. in \cite{Numm MC MCMCists},
\cite{Hob Rob} and \cite{Breyer Roberts} and are closely related
to perfect sampling algorithms, such as coupling form the past
(CFTP) introduced in \cite{Propp Wilson}.
\end{remark}
\begin{proof} First check that the measure in question is a probability measure.
$$\bigg(\varepsilon \sum_{n=0}^{\infty} \nu_m (1-\varepsilon)^n
R^n\bigg)(\stany)=\varepsilon \sum_{n=0}^{\infty}(1-\varepsilon)^n
\big(\nu_mR^n\big)(\stany)= 1.$$ It is also invariant for $P^m:$
\begin{eqnarray}
\bigg( \sum_{n=0}^{\infty} \nu_m (1-\varepsilon)^n R^n\bigg) P^m &
= & \bigg( \sum_{n=0}^{\infty}\nu_m (1-\varepsilon)^n
R^n\bigg)(\varepsilon \nu_m
+ (1-\varepsilon)R) \nonumber \\
& = & \varepsilon \mu \nu_m + \sum_{n=1}^{\infty} \nu_m
(1-\varepsilon)^{n} R^{n} = \sum_{n=0}^{\infty}\nu_m
(1-\varepsilon)^n R^n. \nonumber
\end{eqnarray}
Hence by ergodicity $\varepsilon \mu =\varepsilon \mu P^{nm} \to
\pi, \quad \textrm{as} \quad n\to \infty$. This completes the
proof.
\end{proof}
\begin{cor}\label{corr: for integrals}
 The decomposition in Lemma \ref{lemma: invariant}
implies that \begin{eqnarray}(i)&& \check{E}_{\nu_m^*}\big(
\sum_{n=0}^{\sigma(0)} \mathbb{I}_{\{X_{nm}\in A\}} \big)
=\check{E}_{\nu_m^*}\big( \sum_{n=0}^{\infty}
\mathbb{I}_{\{X_{nm}\in A\}}\mathbb{I}_{\{Y_0=0, \dots,
Y_{n-1}=0\}} \big) = \varepsilon^{-1}\pi(A),\nonumber \\
(ii)&& \check{E}_{\nu_m^*}\big( \sum_{n=0}^{\infty}
f(X_{nm},X_{nm+1},\dots;Y_n,Y_{n+1},\dots)\mathbb{I}_{\{Y_0=0,
\dots, Y_{n-1}=0\}} \big) = \nonumber \\ && \qquad \qquad \qquad
\qquad \qquad \qquad \qquad =
\varepsilon^{-1}\check{E}_{\pi^*}f(X_{0},X_{1},\dots;Y_0,Y_{1},\dots).\nonumber
\end{eqnarray}
\end{cor}
\begin{proof}
(i) is a direct consequence of (\ref{pi representation}). To see
(ii) note that $Y_n$ is a coin toss independent of $\{Y_0,\dots,
Y_{n-1}\}$ and $X_{nm},$ this allows for $\pi^*$ instead of $\pi$
on the RHS of (ii). Moreover the evolution of $\{X_{nm+1},
X_{nm+2},\dots ;Y_{n+1},Y_{n+2},\dots\}$ depends only (and
explicitly by (\ref{split 1}) and (\ref{split 0})) on $X_{nm}$ and
$Y_n.$ Now use (i).
\end{proof}
Our object of interest is
\begin{eqnarray}
I  &= &\check{E}_{\nu_m^*}\bigg[\bigg(
\sum_{n=0}^{\sigma(0)}Z_n(\bar{g})\bigg)^2\bigg] \; = \;
\check{E}_{\nu_m^*}\bigg[\bigg(
\sum_{n=0}^{\infty}Z_n(\bar{g})\mathbb{I}_{\{
\sigma_{\check{\alpha}}(0) \gs n \}}\bigg)^2\bigg]\nonumber\\&=&
\check{E}_{\nu_m^*}\bigg[
\sum_{n=0}^{\infty}Z_n(\bar{g})^2\mathbb{I}_{\{Y_0=0, \dots,
Y_{n-1}=0\}}\bigg] + \nonumber \\ && \qquad \qquad +
\;2\check{E}_{\nu_m^*}\bigg[\sum_{n=0}^{\infty}\sum_{k=n+1}^{\infty}
Z_n(\bar{g})\mathbb{I}_{\{ \sigma(0) \gs n \}} Z_k(\bar{g})
\mathbb{I}_{\{ \sigma_{\check{\alpha}}(0) \gs k \}} \bigg]
\nonumber\\&=&  A +B \label{I: terms A and B}
\end{eqnarray}
Next we use Corollary \ref{corr: for integrals} and then the
inequality $2ab \ls a^2+b^2$ to bound the term $A$ in (\ref{I:
terms A and B}).
\begin{eqnarray}
A & = & \varepsilon ^{-1}\check{E}_{\pi^*} Z_0(\bar{g})^2 =
\varepsilon^{-1} E_{\pi}\Big(\sum_{k=0}^{m-1}\bar{g}(X_k)\Big)^2
 \nonumber \\ & \ls & \varepsilon^{-1} m E_{\pi}\Big[\sum_{k=0}^{m-1} \bar{g}^2(X_k)
\Big] \ls \varepsilon^{-1} m^2 \pi \bar{g}^2 < \infty. \nonumber
\end{eqnarray}
We proceed similarly with the term $B$
\begin{eqnarray}
|B| & \ls & 2\check{E}_{\nu_m^*}\bigg[\sum_{n=0}^{\infty}
 |Z_n(\bar{g})| \mathbb{I}_{\{ \sigma_{\check{\alpha}}(0) \gs n \}}
\sum_{k=1}^{\infty}|Z_{n+k}(\bar{g})| \mathbb{I}_{\{
\sigma_{\check{\alpha}}(0) \gs n+k \}}\bigg] \nonumber \\ &=& 2
\varepsilon^{-1}\check{E}_{\pi^*} \bigg[ |Z_0(\bar{g})|
\sum_{k=1}^{\infty}|Z_{k}(\bar{g})| \mathbb{I}_{\{
\sigma_{\check{\alpha}}(0) \gs k \}}\bigg].\nonumber
\end{eqnarray}
By Cauchy-Schwarz,
\begin{eqnarray}
\check{E}_{\pi^*}\big[\mathbb{I}_{\{ \sigma_{\check{\alpha}}(0)
\gs k \}} |Z_0(\bar{g})| |Z_{k}(\bar{g})| \big] & \ls &
\sqrt{\check{E}_{\pi^*}\big[\mathbb{I}_{\{
\sigma_{\check{\alpha}}(0) \gs k \}} Z_0(\bar{g})^2\big]}
\sqrt{\check{E}_{\pi^*}Z_k(\bar{g})^2}  \nonumber \\
& = & \sqrt{\check{E}_{\pi^*}\big[\mathbb{I}_{\{ Y_0=0\}}
\mathbb{I}_{\{Y_1=0, \dots, Y_{k-1}=0 \}} Z_0(\bar{g})^2\big]}
\sqrt{\check{E}_{\pi^*}Z_0(\bar{g})^2}. \nonumber
\end{eqnarray}
Observe that $\{Y_1,\dots,Y_{k-1}\}$ and $\{X_0,\dots, X_{m-1}\}$
are independent. We drop $\mathbb{I}_{\{ Y_0=0\}}$ to obtain
$$
\check{E}_{\pi^*}\big[\mathbb{I}_{\{ \sigma_{\check{\alpha}}(0)
\gs k \}} |Z_0(\bar{g})| |Z_{k}(\bar{g})| \big]  \ls
(1-\varepsilon)^{\frac{k-1}{2}} \check{E}_{\pi^*}Z_0(\bar{g})^2
\ls (1-\varepsilon)^{\frac{k-1}{2}} m^2 \pi g^2.
$$
Hence $|B| < \infty$, and the proof of (\ref{cond: CTG 2}) is
complete. To get the variance formula note that the convergence we
have established implies
$$
I=\varepsilon ^{-1}\check{E}_{\pi^*}\bigg[
Z_0(\bar{g})\bigg]^2+2\varepsilon^{-1}\check{E}_{\pi^*} \bigg[
Z_0(\bar{g}) \sum_{k=1}^{\infty}Z_{k}(\bar{g}) \mathbb{I}_{\{
\sigma_{\check{\alpha}}(0) \gs k \}}\bigg].
$$
Similarly we obtain
\begin{eqnarray}
J & := & 2\check{E}_{\nu_m^*}\bigg[\big(
\sum_{n=0}^{\sigma_{\ptalfa}(0)}Z_n(\bar{g})\big)
\big(\sum_{n=\sigma_{\ptalfa}(0)+1}^{\sigma_{\ptalfa}(1)}
Z_n(\bar{g})\big)\bigg]\nonumber \\ &=&
2\varepsilon^{-1}\check{E}_{\pi^*} \bigg[ Z_0(\bar{g})
\sum_{k=\sigma_{\check{\alpha}(0)+1}}^{\infty}Z_{k}(\bar{g})
\mathbb{I}_{\{ \sigma_{\check{\alpha}}(1) \gs k \}}\bigg].
\nonumber \end{eqnarray}
Since $\pi(C)=1,$ we have $\sigma_g^2=\va m^{-1}(I+J).$ Next we
use Lemma \ref{lemma:conditional indep} and $\check{E}_{\pi^*}
Z_0(\bar{g}) = 0$ to drop indicators and since for $f: \stany \to
R,$ also $\check{E}_{\pi^*}f = E_{\pi}f,$ we have
\begin{eqnarray}
\va(I+J) & = & \check{E}_{\pi^*}\bigg[Z_0(\bar{g})
\bigg(Z_0(\bar{g}) + 2 \sum_{k=1}^{\infty}Z_k(\bar{g})\bigg)
\bigg] \nonumber \\ & = & E_{\pi}\bigg[Z_0(\bar{g})
\bigg(Z_0(\bar{g}) + 2 \sum_{k=1}^{\infty}Z_k(\bar{g})\bigg)
\bigg].\nonumber\end{eqnarray}
Now, since all the integrals are taken with respect to the
stationary measure, we can for a moment assume that the chain runs
in stationarity from $-\infty$ rather than starts at time $0$ with
$X_0 \sim \pi.$ Thus
\begin{eqnarray}
\sigma^2_g &= & m^{-1}E_{\pi} \bigg[Z_0(\bar{g}) \bigg(
\sum_{k=-\infty}^{\infty}Z_k(\bar{g})\bigg) \bigg] = m^{-1}
E_{\pi} \bigg[\sum_{l=0}^{m-1}\bar{g}(X_l) \bigg(
\sum_{k=-\infty}^{\infty}\bar{g}(X_k)\bigg) \bigg] \nonumber \\
& = & E_{\pi} \Big[\bar{g}(X_0)
\sum_{k=-\infty}^{\infty}\bar{g}(X_k)\Big] = \int_{\stany}
\bar{g}^2d\pi + 2\int_{\stany}\sum_{n=1}^{\infty}
\bar{g}(X_0)\bar{g}(X_n)d\pi. \nonumber \end{eqnarray}

\section{The difference between $m=1$ and $m \neq 1$} \label{section: problems}

Assume the small set condition (\ref{eqn: small}) holds and
consider the split chain defined by (\ref{split 1}) and
(\ref{split 0}). The following tours
$$\big\{\{X_{(\sigma(n)+1)m}, X_{(\sigma(n)+1)m +1}, \dots,
X_{(\sigma(n+1)+1)m-1}\}, n=0,1,\dots \big\}$$ that start whenever
$X_k \sim \nu_m$ are of crucial importance to the regeneration
theory and are eagerly analyzed by researchers. In virtually every
paper on the subject there is a claim these objects are
independent identically distributed random variables. This claim
is usually considered obvious and no proof is provided. However
this is not true if $m>1.$
\smallskip
In fact formulas (\ref{split 1}) and (\ref{split 0}) should be
convincing enough, as $X_{mn+1}, \dots, X_{(n+1)m}$ given $Y_n=1$
and $X_{nm}=x$ are linked in a way described by $P(x,dx_1) \cdots
P(x_{m-1}, dy).$ In particular consider a Markov chain on $\stany
= \{a,b,c,d,e\}$ with transition probabilities
\begin{eqnarray}
&&P(a,b)=P(a,c)=P(b,b)=P(b,d)=P(c,c)=P(c,e)=1/2,\qquad \qquad
\nonumber \\
&& P(d,a)=P(e,a)=1.\nonumber \end{eqnarray}
Let $\nu_4(d)=\nu_4(e)=1/2$ and $\varepsilon = 1/8.$ Clearly
$P^4(x,\cdot) \gs \varepsilon \nu_4(\cdot)$ for every $x \in
\stany,$ hence we established (\ref{eqn: small}) with $C=\stany.$
Note that for this simplistic example each tour can start with $d$
or $e.$ However if it starts with $d$ or $e$ the previous tour
must have ended with $b$ or $c$ respectively. This makes them
dependent. Similar examples with general state space $\stany$ and
$C\neq \stany$ can be easily provided. Hence Theorem \ref{Thm1} is
critical to providing regeneration proofs of CLTs and standard
arguments that involve i.i.d. random variables are not valid.

   \chapter{Fixed-Width Asymptotics} \label{Chapter fixed-width
asymptotics}

Determining the length of simulation for MCMC algorithms that
guarantees good quality of estimation is a fundamental problem.
One possible approach is to wait until width of an asymptotic
confidence interval based on the approximation by a normal
distribution becomes smaller then a user-specified value. This
requires estimating $\sigma^2_g$ the variance of the asymptotic
normal distribution. In this chapter we relax assumptions required
to obtain strongly consistent estimators of $\sigma^2_g$ in the
regenerative setting.

Results of this chapter (in particular the key Lemma
\ref{lemma_improved} and resulting from it Lemma \ref{lemma
improved brownian motion jasa} and Proposition \ref{proposition
inproved consistency jasa}) are based on the paper \cite{BeLa
JASA} and are joint work with Witold Bednorz.

The presentation of the fixed-width asymptotic approach is based
on \cite{JonesNaranCaffoNeath}. We provide only a quick sketch,
since the approach is well known in literature (see also
\cite{Geyer practical}, \cite{Mykland et al. regenerative
simulation}, \cite{HobertBiometrika}) and
\cite{JonesNaranCaffoNeath} is an excellent recent reference.

\section{Asymptotic Confidence Intervals}
\label{section: jasa
confidence interv}

Suppose that we are in the standard MCMC setting and our goal is
to estimate $I=E_{\pi}g=\int_{\stany}g(x)\pi(dx).$ Let $\lancuch$
be a time homogeneous, aperiodic and Harris recurrent Markov chain
with transition kernel $P$ and limiting invariant probability
distribution $\pi.$

Consider the estimator along one walk without burn-in, i.e.
\begin{equation}\label{eqn: jasa: estimator introduction}
\hat{I}_n = \frac{1}{n} \sum_{i=0}^{n-1}g(X_i) \end{equation}
of the unknown value $I.$ By Theorem \ref{thm introduction
convergence}, $\hat{I}_n \to I,$ as $n \to \infty,$ with
probability 1. Moreover, assume for a moment that a $\sqrt{n}-$CLT
holds and let $\sigma_g^2$ be the asymptotic variance, as defined
in (\ref{sqrt n CLTs 1}).

We will study the following sequential procedure. Let
$n^*=n^*(\varepsilon)$ be the first time that
\begin{equation}\label{eqn: jasa: stopping rule}
q_{\bullet} \frac{\hat{\sigma}_n}{\sqrt{n}} + p(n) \leq
\varepsilon,
\end{equation}
where $\hat{\sigma}_n^2$ is an estimate of $\sigma_g^2$ at time
$n,$ and $q_{\bullet}$ is an appropriate quantile, $p(n) > 0$ is a
strictly positive decreasing function on $Z_{+},$ and $\varepsilon
> 0$ is the desired half-width.

At time $n^*$ we build an interval
$I^*(\varepsilon):=[\hat{I}_{n^*}-\varepsilon,
\hat{I}_{n^*}+\varepsilon]$ of width $2\varepsilon.$ For
independent samples such procedures are known to work well and
belong to classical results of sequential statistics (c.f.
\cite{Chow Robbins}, \cite{Nadas sequential} and \cite{Liu W
sequential}). However in our context we have to apply the
following result form \cite{Glynn Whitt}.
\begin{thm}[Glynn \& Whitt 1992]\label{thm: jasa: Glynn Whitt}
If
\begin{itemize}
\item[(a)] A functional central limit theorem holds, i.e.
as $n \to \infty,$ the distribution of
$$Y_n(t) := \frac{1}{\sqrt{n}} \sum_{i=1}^{\lfloor nt \rfloor}
g(X_i)$$ converges to Brownian motion with variance $\sigma^2_g$
weakly in the Skorohod space on any finite interval,
\item[(b)] $\hat{\sigma}^2_n \to \sigma^2_g$ with probability $1$
as $n \to \infty,$
\item[(c)] The sequence $p(n)$ is strictly positive and
decreasing and $p(n) = o(n^{-1/2}),$
\end{itemize}
then
\begin{equation} \label{eqn: jasa: as validity}
P(I \in I^*(\varepsilon)) \to 1-\delta, \quad \textrm{as} \quad
\varepsilon \to 0.
\end{equation}
\end{thm}

Markov chains often enjoy a functional central limit theorem under
the same conditions that ensure the standard $\sqrt{n}-$CLT. In
particular the following results are well known:

\begin{thm} \label{thm: jasa: conditions for FCLT} Assume
$\lancuch$ is a Harris ergodic Markov chain. If one of the
following conditions holds, then a functional central limit
theorem also holds.
\begin{itemize}
\item[(a)] (due to \cite{Doukhan FCLT})
The chain is geometrically ergodic and\\
$E_{\pi}[g^2(x)(\log^+|g(x)|)] < \infty,$
\item[(b)] (due to \cite{RobRos hybrid})
The chain is geometrically ergodic,
reversible, and $E_{\pi} g^2(x) < \infty,$
\item[(c)] (due to \cite{Billingsley conv prob meas})
The chain is uniformly ergodic and
$E_{\pi}g^2(x)<\infty.$
\end{itemize}
\end{thm}

The goal of this chapter is to obtain additionally condition (b)
of Theorem \ref{thm: jasa: Glynn Whitt} for a suitable estimator
$\hat{\sigma}^2_n$ of $\sigma^2_g,$ under possibly weak
assumptions and consequently conclude (\ref{eqn: jasa: as
validity}). In particular we will need stronger assumptions then
those listed in Theorem \ref{thm: jasa: conditions for FCLT}, thus
condition (a) of Theorem \ref{thm: jasa: Glynn Whitt} will hold
automatically.

\section{Estimating Asymptotic Variance}

We will discuss two methods of estimating the asymptotic variance
described in \cite{JonesNaranCaffoNeath}, based on batch means and
regenerative simulation.

\subsection{Batch Means}
For the bath means estimator suppose that $n-1$ iterations of the
algorithm are performed and we partition the trajectory of length
$n$ into $a_n$ blocks of length $b_n$ i.e.
$$n \simeq a_n b_n$$
Define $\bar{Y}_1, \dots, \bar{Y}_{a_n}$ as
$$\bar{Y}_j := \frac{1}{b_n} \sum_{i=(j-1)b_n}^{jb_n-1}g(X_i).$$
Then the bath means estimate of $\sigma_g^2$ is
\begin{equation}\label{eqn: jasa: batch means estimator}
\hat{\sigma}^2_{BM} = \frac{b_n}{a_n-1} \sum_{j=1}^{a_n}
(\bar{Y}_j - \hat{I}_n)^2.\end{equation}

In the next section we provide an appropriate strategy for
choosing $a_n$ and $b_n$ for $\hat{\sigma}^2_{BM}$ to be a
consistent estimator.

\subsection{Regenerative Estimation} \label{section: jasa: reg
sim}

Assume that the following minorization condition with $m=1,$ as
introduced in Definition \ref{def: chains: minorization general}
holds.
\begin{equation}\label{jasa: minorization}
P(x, \cdot) \geq s(x) \nu(\cdot), \quad \textrm{for all}\quad x
\in \stany,
\end{equation}
and define the residual transition kernel $R(x, dy)$ as
$$R(x, dy) :=
\left\{\begin{array}{lll}(1-s(x))^{-1}(P(x,dy)-s(x)\nu(dy)) &
\textrm{if} & s(x)<1,\\ 0 & \textrm{if} & s(x) = 1.\end{array}
\right.$$
By straightforward modification of the split chain construction of
Section \ref{section: preliminaries} we obtain a bivariate process
$(X_n, Y_n)_{n\geq 0}$ that evolves according to the following
transition rule:
\begin{itemize}
\item given $X_n = x,$ draw $Y_n
\sim$ Bernoulli$(s(x))$
\item If $Y_n = 1,$ then draw $X_{n+1} \sim \nu(\cdot),$ otherwise
draw $X_{n+1} \sim R(x, \cdot).$
\end{itemize}

Moreover, the artificial atom $\check{\alpha}$ is now of the form
$\check{\alpha} = \stany \times \{1\}.$ Let us simplify the
notation of Section \ref{section: preliminaries} by setting
$\tau_n = \tau_{\check{\alpha}}(n),$ for $n=1,2,\dots$ Suppose
also that $X_0 \sim \nu$ and set $\tau_0=-1$ to keep notation
coherent with probabilistic behavior of the chain. Define also
$N_i = \tau_{i+1} - \tau_{i},$ for $i = 0,1,\dots,$ and recall
$s_i$ defined by (\ref{defi eqn si of g}). Since $m=1,$
$$s_i = \sum_{j=\tau_i + 1}^{\tau_{i+1}} g(X_j),$$
and observe that the $(N_i, s_i)$ pairs are iid random variables.

For regenerative estimation of the asymptotic variance we will
need $(Y_i)_{i\geq 0}$, thus we must simulate the split chain
$(X_i,Y_i)_{i\geq 0},$ not only the initial chain
$(X_i)_{i\geq0}.$ However the simulation from $R(x, \cdot)$ in
real life examples is often challenging. The following solution to
this problem is provided in \cite{Mykland et al. regenerative
simulation}.

Suppose that $P(x, \cdot)$ has a density $k(\cdot|x)$ and
$\nu(\cdot)$ has a density $v(\cdot)$ with respect to a reference
measure $dx.$ Given $X_i = x$ draw $X_{i+1} \sim k(\cdot|x)$ and
draw $Y_i$ from the distribution of $Y_i|X_i, X_{i+1},$ that is $$
Y_i \sim
\textrm{Bernoulli}\Big(\frac{s(X_i)v(X_{i+1})}{k(X_{i+1}|X_{i})}\Big).$$
The method is feasible in many settings of practical interest (cf.
\cite{Mykland et al. regenerative simulation},
\cite{JonesNaranCaffoNeath}).

Once we are able to simulate the split chain $(X_i,Y_i)_{i\geq
0},$ we can observe $\tau_0, \tau_1, \dots$ and compute the
following regenerative estimator of $I.$

\begin{equation}\label{jasa: eqn: regenerqtive est for I}
\hat{I}_{\tau_R} = \frac{1}{\tau_R+1} \sum_{j=0}^{\tau_R} g(X_j),
\end{equation}
where the fixed number $R$ is the total number of regenerations
observed. Note that $\hat{I}_{\tau_R}$ is a sum of fixed number of
iid. random variables. Thus if $E_{\nu}N_0^2 < \infty$ and
$E_{\nu}s_0^2 < \infty$ then
\begin{equation}\label{jasa: clt for reg sim}
\sqrt{R}(\hat{I}_{\tau_R} - I) \to N(0, \xi_g^2), \qquad
\textrm{as} \quad R \to \infty,
\end{equation}
where
$$\xi_g^2 = \frac{E_{\nu}(s_0
- N_0I)^2}{(E_{\nu}N_0)^2}.$$
Let $\bar{N} = R^{-1} (\tau_R + 1) =R^{-1} \sum^{R-1}_{i=0}N_i.$
As an approximation for $\xi_g^2$ one can take the following
regenerative estimator
\begin{equation} \label{jasa: estim for as var regenerative}
\hat{\xi}^2_{RS} := \frac{1}{R\bar{N}^2} \sum_{i=0}^{R-1}(s_i -
\hat{I}_{\tau_R}N_i)^2.
\end{equation}
Now observe that
\begin{eqnarray}
\hat{\xi}^2_{RS} - \xi_g^2 & = & \frac{1}{R\bar{N}^2}
\sum_{i=0}^{R-1}(s_i - \hat{I}_{\tau_R}N_i)^2 \pm
\frac{E_{\nu}(s_0 - N_0I)^2}{\bar{N}^2} - \frac{E_{\nu}(s_0 -
N_0I)^2}{(E_{\nu}N_0)^2} \nonumber \\ & = &
 \frac{1}{R\bar{N}^2}
\sum_{i=0}^{R-1}\left[ (s_i - \hat{I}_{\tau_R}N_i)^2 \pm
(s_i-N_iI)^2  - E_{\nu}(s_0 - N_0I)^2\right] + \nonumber \\ &&
\qquad \qquad \qquad\qquad \qquad \qquad + E_{\nu}(s_0 -
N_0I)^2\left[\frac{1}{\bar{N}^2} -
\frac{1}{(E_{\nu}N_0)^2}\right]. \nonumber
\end{eqnarray}
As noticed in \cite{JonesNaranCaffoNeath}, repeated application of
the strong law of large numbers (with $R \to \infty$) yields that
$\hat{\xi}^2_{RS}$ is a strongly consistent estimator of $\xi_g^2$
so it is enough to establish conditions $E_{\nu}N_0^2 < \infty$
and $E_{\nu}s_0^2 < \infty$ for the fixed width methodology to
work. This is deferred to the next section.

Clearly, in this modified regenerative setting an asymptotically
valid fixed-width result is obtained by terminating the simulation
the first time that
\begin{equation}\label{eqn: jasa: stopping rule regenerative}
q_{\bullet} \frac{\hat{\xi}_{RS}}{\sqrt{R}} + p(R) \leq
\varepsilon.
\end{equation}

\section{A Lemma and its Consequences}
\label{section: jasa: the lemma}

For geometrically ergodic Markov chains hitting times for sets of
positive stationary measure have geometrically decreasing tails.
In particular the following lemma is shown in
\cite{HobertBiometrika}.

\begin{lemma}[Lemma 2 of \cite{HobertBiometrika}]\label{jasa:
lemma geom tails} Let $\lancuch$ be a Harris ergodic chain and
assume that (\ref{jasa: minorization}) holds. If $\lancuch$ is
geometrically ergodic, then there exists a $\beta > 1,$ such that
$E_{\pi}\beta^{\tau_1} < \infty.$
\end{lemma}

Which immediately yields the following corollary.

\begin{cor} \label{jasa: corr geom tails}
Under the conditions of Lemma \ref{jasa: lemma geom tails}, for
any $a > 0,$
\begin{equation}\label{jasa: corr eqn geom tails}
\sum_{i=0}^{\infty} \Big(P_{\pi}(\tau_1 \geq i+1) \Big)^a \leq
\Big(E_{\pi} \beta^{\tau_1}\Big)^a \sum_{i=0}^{\infty}
\beta^{-a(i+1)} < \infty.
\end{equation}
\end{cor}

\begin{proof}
\begin{eqnarray}
\sum_{i=0}^{\infty} \Big(P_{\pi}(\tau_1 \geq i+1) \Big)^a & \leq &
\sum_{i=0}^{\infty} \Big(E_{\pi}(\mathbb{I}_{\{\tau_1 \geq i+1\}}
\beta^{\tau_1}\beta^{-(i+1)}) \Big)^a \nonumber \\ &=
&\sum_{i=0}^{\infty} \beta^{-a(i+1)}
\Big(E_{\pi}(\mathbb{I}_{\{\tau_1 \geq i+1\}}
\beta^{\tau_1}) \Big)^a \nonumber \\
& \leq &  \sum_{i=0}^{\infty} \beta^{-a(i+1)}\Big(E_{\pi}
\beta^{\tau_1}\Big)^a.\nonumber
\end{eqnarray}
\end{proof}

Observe also that we can integrate (\ref{jasa: minorization}) with
respect to $\pi$ and obtain $\pi(\cdot) \geq c\nu(\cdot),$ where
$c = E_{\pi}s.$ Thus for any function $h: \stany^{\infty} \to R,$
\begin{equation} \label{jasa: eqn: pi nu integral bound}
E_{\pi}|h(X_0, X_1, \dots)| \geq c E_{\nu}|h(X_0, X_1, \dots)|.
\end{equation}

Now we are in a position to prove our key result, namely the
following lemma.

\begin{lemma} \label{lemma_improved}
Let $\lancuch$ be a Harris ergodic Markov chain, assume the
minorization condition (\ref{jasa: minorization}) holds, and
$\lancuch$ is geometrically ergodic. Let $g: \stany \to R$ be a
real valued Borel function. Then, if
$$E_{\pi}|g|^{p+\delta} <
\infty \quad  \textrm{for some} \quad p
> 0 \quad  \textrm{and} \quad \delta>0,$$ then
$$E_{\nu}N^p_0<\infty \quad  \textrm{and} \quad E_{\nu}|s_0|^p <
\infty.$$
\end{lemma}

\begin{remark} Lemma \ref{lemma_improved} improves the two
following results:
\begin{itemize}
\item Theorem 2 of \cite{HobertBiometrika} that provides the
implication $$E_{\pi}|g|^{2+\delta}< \infty \Rightarrow
E_{\nu}N^2_0<\infty \;\;  \textrm{and} \;\; E_{\nu}|s_0|^2 <
\infty.$$
\item Lemma 1 of \cite{JonesNaranCaffoNeath} that for $p\geq 1$
provides implications
$$E_{\pi}|g|^{2^{(p-1)}+\delta} < \infty \Rightarrow
E_{\nu}N^p_0<\infty \;\;  \textrm{and} \;\; E_{\nu}|s_0|^p <
\infty.$$ and
$$E_{\pi}|g|^{2^{p}+\delta} < \infty \Rightarrow
E_{\nu}N^p_0<\infty \;\;  \textrm{and} \;\;
E_{\nu}|s_0|^{p+\delta} < \infty.$$
\end{itemize}
\end{remark}

\begin{remark}
Without additional restrictions $E_{\pi}|g|^{p} < \infty$ does not
imply $E_{\nu}|s_0|^p < \infty,$ so Lemma \ref{lemma_improved} can
not be improved. To see this note that Theorem \ref{Thm1} of
Chapter \ref{chapter: CLT} combined with the presumption that in
the setting of \mbox{Lemma \ref{lemma_improved}} $E_{\pi}|g|^{p} <
\infty$ implies $E_{\nu}|s_0|^p < \infty$ yields the Central Limit
Theorem for normalized sums of $g(X_i)$ for geometrically ergodic
Markov chains assuming only $E_{\pi}g^2 < \infty.$ This however is
not enough for the CLT, Bradley in \cite{Bradley} and also
H\"aggstr\"om in \cite{Haggstrom} provide counterexamples. Hence
to obtain the implication $E_{\pi}|g|^{p} < \infty \Rightarrow
E_{\nu}|s_0|^p < \infty,$ one needs stronger assumptions, e.g. if
$p=2$ then uniform ergodicity is enough, as proved in Chapter
\ref{chapter: CLT}.
\end{remark}
\begin{proof}[Proof of Lemma \ref{lemma_improved}]
First note that by (\ref{jasa: eqn: pi nu integral bound}) it is
enough to show that
$$E_{\pi}N^p_0<\infty \quad  \textrm{and}
\quad E_{\pi}|s_0|^p < \infty.$$ Moreover, since
$\max_{k}\big\{\frac{k^p}{\beta^k}\big\} < \infty$ for every $p>0$
and $\beta >1,$ by Lemma \ref{jasa: lemma geom tails} we obtain
immediately $E_{\pi}N^p_0<\infty.$ Thus we proceed to show that
$E_{\pi}|s_0|^p<\infty.$ To this end first note that
\begin{equation}\label{note} C:=\left(\left(E_{\pi}|
g(X_{i})|^{p+\delta}\right)^
{\frac{p}{p+\delta}}\right)^{1/p} < \infty.\end{equation}

 For $ p\geq 1$ we
use first the triangle inequality in $L^p$, then H\"{o}lder
inequality, then (\ref{note}) and finally Corollary \ref{jasa:
corr geom tails}.
    \begin{eqnarray} \left(E_{\pi}|s_0|^p\right)^{1/p} & \leq &
    \left[E_{\pi}\left(\sum_{i=0}^{\tau_1}|g(X_i)|
    \right)^p\right]^{1/p}
    \nonumber \\
    &=&
    \left[E_{\pi}\left(\sum_{i=0}^{\infty}\mathbf{1}(i \leq \tau_1
    )|g(X_i)|\right)^p\right]^{1/p}
    \nonumber \\
    & \leq & \sum_{i=0}^{\infty}\Big[E_{\pi}\mathbf{1}(i \leq \tau_1
    )|g(X_i)|^p\Big]^{1/p}
    \nonumber \\
    & \leq & \sum_{i=0}^{\infty}\left[\left(E_{\pi}\mathbf{1}(i \leq
    \tau_1)\right)^{\frac{\delta}{p+\delta}}
    \left(E_{\pi}|g(X_{i})|^{p+\delta}\right)^{\frac{p}{p+\delta}}
    \right]^{1/p}
    \nonumber \\ \label{p geq 1}
    & = & C\sum_{i=0}^{\infty}\left(P_{\pi}(\tau_1 \geq
    i)\right)^{\frac{\delta}{p(p+\delta)}} < \infty.
    \end{eqnarray}
For $0<p<1$ we use the fact $x^p$ is concave and then proceed
similarly as in (\ref{p geq 1}) to obtain
    \begin{eqnarray}
    E_{\pi}|s_0|^p & \leq & E_{\pi}\left(\sum_{i=0}^{\infty}
    \mathbf{1}(i \leq \tau_1
    )|g(X_i)|\right)^p \nonumber \\ &\leq&
     \sum_{i=0}^{\infty}E_{\pi}\mathbf{1}(i \leq \tau_1
    )|g(X_i)|^p
    \nonumber \\ &\leq& C^p \sum_{i=0}^{\infty}\left(
    P_{\pi}(\tau_1 \geq
    i)\right)^{\frac{\delta}{(p+\delta)}}<\infty. \nonumber
    \end{eqnarray}
\end{proof}

Lemma \ref{lemma_improved} allows us to restate results from
section 3.2 of \cite{JonesNaranCaffoNeath} with relaxed
assumptions. In particular in Lemma 2 and in Proposition 3 therein
it is enough to assume $E_{\pi}|g|^{2+\delta+\varepsilon}<\infty$
for some $\delta>0$ and some $\varepsilon > 0,$ instead of
$E_{\pi}|g|^{4+\delta}<\infty$ for some $\delta>0.$ Modifications
of the (rather long and complicated) proofs in
\cite{JonesNaranCaffoNeath} are straightforward. Hence we have
\begin{lemma}[Part b of Lemma 2 of \cite{JonesNaranCaffoNeath}]
\label{lemma improved brownian motion jasa} Let $\lancuch$ be a
Harris ergodic Markov chain with invariant distribution $\pi.$ If
$\lancuch$ is geometrically ergodic, (\ref{jasa: minorization})
holds and $E_{\pi}|g|^{2+\delta+\varepsilon}<\infty$ for some
$\delta>0$ and some $\varepsilon > 0,$ then there exists a
constant $0 < \sigma_g < \infty,$ and a sufficiently large
probability space such that
$$ \left|\sum_{i=1}^n g(X_i) - n E_{\pi}g - \sigma_g B(n) \right|
= O(\gamma (n))$$
with probability $1$ as $n \to \infty,$ where
$\gamma(n)=n^{\alpha}\log n,$ $\alpha=1/(2+\delta),$ and
$B=\{B(t), t\geq 0 \}$ denotes a standard Brownian motion.
\end{lemma}
\begin{prop}[Proposition 3 of \cite{JonesNaranCaffoNeath}]
\label{proposition inproved consistency jasa} Let $\lancuch$ be a
Harris ergodic Markov chain with invariant distribution $\pi.$
Further, suppose $\lancuch$ is geometrically ergodic, (\ref{jasa:
minorization}) holds and
$E_{\pi}|g|^{2+\delta+\varepsilon}<\infty$ for some $\delta>0$ and
some $\varepsilon > 0.$ If
  \begin{enumerate}
    \item $a_n \to \infty,$ as $n \to \infty,$
    \item $b_n \to \infty$ and $b_n/n \to 0$ as $n \to \infty,$
    \item $b_n^{-1}n^{2\alpha}[\log n]^3 \to 0$ as $n \to \infty,$
    where $\alpha=1/(2+\delta),$
    \item there exists a constant $c\geq 1,$ such that
    $\sum_{n=1}^{\infty}(b_n/n)^c < \infty,$
  \end{enumerate}
Then $\hat{\sigma}^2_{BM} \to \sigma^2_g$ w.p.1 as $n\to \infty.$
\end{prop}

\begin{concluding remark}
Compare the foregoing result with Section \ref{section: jasa: reg
sim} or with Proposition 1 of \cite{JonesNaranCaffoNeath} to see
that both methods described here, i.e. regenerative simulation
(RS) and batch means (CBM), provide strongly consistent estimators
of $\sigma^2_g$ under the same assumption for the target
\mbox{function $g.$}
\end{concluding remark}

   \chapter{Fixed-Width Nonasymptotic Results under Drift Condition}
\label{CHAPTER Fixed-Width Nonasymptotic Results under Drift
Condition}

In this Chapter we establish nonasymptotic fixed width estimation.
We assume a drift condition towards a small set and bound the mean
square error of estimators obtained by taking averages along a
single trajectory of a Markov chain Monte Carlo algorithm. We use
these bounds to determine the length of the trajectory and the
burn-in time that ensures $(\varepsilon-\alpha)-$approximation,
i.e. desired precision of estimation with given probability. Let
$I$ be the value of interest and $\hat{I}$ its MCMC estimate.
Precisely, our lower bounds for the length of the trajectory and
burn-in time ensure that
\begin{equation} \nonumber P(|\hat{I}-I|\leq \varepsilon)\geq
1-\alpha \end{equation}
and depend only and explicitly on drift parameters, $\varepsilon$
and $\alpha.$ Next we introduce an MCMC estimator based on the
median of multiple shorter runs. It turns out that this estimation
scheme allows for sharper bounds for the total simulation cost
required for the $(\varepsilon-\alpha)-$approximation. For both
estimation schemes numerical examples are provided that include
practically relevant Gibbs samplers for a hierarchical random
effects model.
\section{Introduction}
Recall the estimation strategies introduced in Section
\ref{Section algorithms} and described by (\ref{est along one
walk}-\ref{est multiple run}). \textit{Estimation Along one Walk}
uses average along a single trajectory of the underlying Markov
chain and discards the initial part to reduce bias. The estimate
of the unknown value $I=\int_{\stany}f(x) \pi(dx)$ is of the form
\begin{equation}\label{est along one walk 2}
\hat{I}_{t,n}=\frac{1}{n}\sum_{i=t}^{t+n-1}f(X_i)\end{equation}
and $t$ is called the burn-in time.

The strategy is believed to be more efficient then
\textit{estimation along one walk with spacing} and
\textit{multiple run} described in Section \ref{Section
algorithms} and is usually the practitioners choice. Some precise
results are available for reversible Markov chains. Geyer in
\cite{Geyer practical} shows that using spacing as in (\ref{est
along one walk with spacing}) is ineffective (in terms of
asymptotic variance) and Chan and Yue in \cite{chan yue - jrssb -
asymptotic} prove that (\ref{est along one walk 2}) is
asymptotically efficient in a class of linear estimators (in terms
of mean square error).

The goal of this chapter is to derive lower bounds for $n$ and $t$
in (\ref{est along one walk 2}), that minimize the total
computation cost $n+t,$ and that ensure the following condition of
$(\varepsilon, \alpha)-$approximation:
 \begin{equation} \label{eps-alpha} P(|\hat{I}_{t,n}-I|\leq \varepsilon)\geq
1-\alpha,\end{equation} where $\varepsilon$ is the precision of
estimation and $1-\alpha,$ the confidence level. Due to results in
\cite{Geyer practical} and \cite{chan yue - jrssb - asymptotic} no
other linear modifications of the estimation scheme in (\ref{est
along one walk 2}) are analyzed. To decrease the total simulation
cost for (\ref{eps-alpha}) we introduce instead a nonlinear
estimator based on the median of multiple shorter runs.

Results of this or related type have been obtained for discrete
state space $\stany$ and bounded target function $f$ by Aldous in
\cite{Aldous}, Gillman in \cite{Gillman} and recently by Le\'on
and Perron in \cite{Perron}. Niemiro and Pokarowski in
\cite{NiePo} give results for relative precision estimation. For
uniformly ergodic chains on continuous state space $\stany$ and
bounded function $f,$ Hoeffding type inequalities are available
(due to Glynn and Ormonait in \cite{Glynn}, and an improved bound
due to Meyn et al. in \cite{MeynKonto}) and can easily lead to the
desired
$(\varepsilon-\alpha)-$approximation. 
To our best knowledge there are no explicit bounds for $n$ and $t$
in more general settings, especially when $f$ is not bounded and
the chain is not uniformly ergodic. A remarkable presentation of
the state of the art approach to dealing with this problem is
provided by Jones at al. in the recent paper
\cite{JonesNaranCaffoNeath}. They suggest two procedures for
constructing consistent estimators for the variance of the
asymptotic normal distribution for geometrically ergodic split
chains and thus under the additional assumption of
$E_{\pi}|f|^{2+\delta}< \infty$ for some $\delta > 0$ (see Chapter
\ref{Chapter fixed-width asymptotics} here for this weakened
assumption and details of the procedure).

Our approach is to assume a version of the well known drift
condition towards a small set (Assumption \ref{ass:drift}) and
give explicit lower bounds on $n$ and $t$ in terms of drift
parameters defined in Assumption \ref{ass:drift} and approximation
parameters defined in (\ref{eps-alpha}).

\medskip

The rest of the Chapter is organized as follows. In Section
\ref{sec:drift} we introduce the drift condition assumption and
preliminary results. In Section \ref{sec:MainResult} we obtain an
explicit bound for the mean square error of the estimator defined
in (\ref{est along one walk}). In Section \ref{sec:eps-alpha} we
construct two different $(\varepsilon-\alpha)-$approximation
procedures, one based on the sample mean of one long trajectory
and the other based on the median of multiple shorter runs. We
close with examples in Sections \ref{sec: Toy Example} and
\ref{sec:example HREM}, in particular we show how to obtain
explicit lower bounds for $t$ and $n$ that guarantee the
$\varepsilon-\alpha-$approximarion for a hierarchical random
effects model of practical relevance.

\section{A Drift Condition and Preliminary Lemmas}
\label{sec:drift}
Since in what follows we deal with integrals of unbounded
functions $f$ with respect to probability measures, the very
common total variation distance defined by (\ref{total var dist})
is inappropriate for measuring distances between probability
measures and we need to use the $V-$norm and $V-$norm distance
introduced in Section \ref{sec: chap chains: stationarity and
ergodicity}.

\medskip
We analyze the MCMC estimation along a single trajectory under the
following assumption of a drift condition towards a small set.
\begin{assu} \label{ass:drift}
\begin{itemize}
\item[(A.1)]{Small set.} There exist $C\in \borel,$ $\tilde{\beta}>0$ and a
probability measure $\nu$ on $(\stany, \borel)$ such that for all
$x\in C$ and $A \in \borel$ $$P(x,A)\geq \tilde{\beta} \nu (A).$$
\item[(A.2)]{Drift.} There exist a function $V:\stany \to
[1,\infty)$ and constants $\lambda<1$ and $K<\infty$ satisfying
$$PV(x) \leq \left\{\begin{array}{lcc}
\lambda V(x), & \text{if} & x\notin C, \\ K, & \text{if} & x \in
C.
\end{array}
\right.$$
\item[(A.3)]{Aperiodicity.} There exists $\beta>0$ such that
$\tilde{\beta} \nu(C) \geq \beta.$
\end{itemize}
\end{assu}

In the sequel we refer to $\tilde{\beta}, V(x), \lambda, K, \beta$
as drift parameters.

\begin{remark} Establishing a drift condition for real life examples
is usually not an easy task. As indicated in \cite{MeynTweedie}
polynomials are often suitable candidates for a drift function $V$
and also functions proportional to $\pi^{1/2}$ may turn out to be
a lucky choice. Computable toy and real life examples of
\cite{Bax} and \cite{Jones Hobert Gibbs for rand eff mod} confirm
this observations.
\end{remark}
\begin{remark} There is a strong probabilistic intuition behind Assumption
\ref{ass:drift}. Every time the chain visits the small set $C,$ it
regenerates with probability $\tilde{\beta}.$ The role of the
drift condition (A.2) is to guarantee that the chain visits the
small set $C$ frequently enough. Typically $C$ is in the
,,center'' of the state space $\stany$ and the drift function $V$
takes small values on $C$ and increases as it goes away from $C.$
Assume first that $X_n = x \notin C.$ The condition $PV(x) \leq
\lambda V(x)$ means that $X_{n+1} \sim P(x, \cdot)$ is on average
getting closer to $C$ (closer in terms of $V$). Whereas $PV(x)
\leq K$ for $X_n = x \in C$ means that $X_{n+1}$ will perhaps jump
out of $C,$ but not too far away, i.e. the integral of $V$ with
respect to the distribution of $X_{n+1}$ is bounded (by the same
value) for all $x \in C.$ Assumption (A.3) together with (A.1)
imply aperiodicity.
\end{remark}

Assumption \ref{ass:drift} is often used and widely discussed in
Markov chains literature. Substantial effort has been devoted to
establishing convergence rates for Markov chains under the drift
condition (A.1-3) or related assumptions. For discussion of
various drift conditions and their relation see Meyn and Tweedie
\cite{MeynTweedie}. For quantitative bounds on convergence rates
of Markov chains see the survey paper by Roberts and Rosenthal
\cite{RobRos survey} and references therein. In the sequel we make
use of the recent convergence bounds obtained by Baxendale in
\cite{Bax}.
\begin{thm}[Baxendale \cite{Bax}] \label{thm:bax}
Under Assumption \ref{ass:drift} $(X)_{n\geq 0}$ has a unique
stationary distribution $\pi$ and $\pi V < \infty.$ Moreover,
there exists $\rho < 1$ depending only and explicitly on
$\tilde{\beta}, \beta, \lambda$ and $K$ such that whenever $\rho <
\gamma < 1$ there exists $M < \infty$ depending only and
explicitly on $\gamma, \tilde{\beta}, \beta, \lambda$ and $K$ such
that for all $n\geq 0$
 \begin{equation}\label{eqn: bax}
 |||P^n-\pi |||_{V} \leq M\gamma^n.\end{equation}
\end{thm}

When we refer in the sequel to $V-$uniform ergodicity, we mean the
convergence determined by (\ref{eqn: bax}). There are different
formulas for $\rho$ and $M$ for general operators, self adjoint
operators and self adjoint positive operators in both atomic and
nonatomic case. We give them in Section \ref{sec:Bax form} for the
sake of completeness. To our knowledge the above-mentioned theorem
gives the best available explicit constants.

\begin{cor} \label{cor: burn-in} Under
Assumption \ref{ass:drift}
\[\|\pi_0 P^n-\pi \|_{V} \leq
 \min\{\pi_0 V, \|\pi_0 - \pi \|_{V}\} M\gamma^n,\]
where $M$ and $\gamma$ are such as in Theorem \ref{thm:bax}.
\end{cor}

\begin{proof} From Theorem \ref{thm:bax} we have
$\|P^n(x,\cdot)-\pi(\cdot)\|_{V} \leq M\gamma^n V(x),$ which
yields \begin{eqnarray} \pi_0 V M \gamma^n & \geq & \int_{\stany}
\|P^n(x,\cdot)-\pi(\cdot)\|_{V} \pi_0(dx) \geq \sup_{|g|\leq V}
\int_{\stany}|P^n(x,\cdot)g-\pi g|\pi_0(dx) \nonumber \\ & \geq &
\sup_{|g|\leq V} |\pi_0 P^ng-\pi g|= \|\pi_0 P^n-\pi
\|_{V}.\nonumber \end{eqnarray}
 Now let $b_V=\inf_{x\in \stany}V(x).$
 Since $|||\cdot|||_{V}$ is an operator norm and
$\pi$ is invariant for $P$, we have
\begin{eqnarray} \|\pi_0 P^n-\pi \|_{V} & = & b_V|||\pi_0 P^n-\pi
|||_{V} = b_V|||(\pi_0 - \pi)(P^n-\pi) |||_{V} \nonumber \\ & \leq
& b_V|||\pi_0 - \pi|||_{V}|||P^n-\pi |||_{V} = \|\pi_0 - \pi
\|_{V}|||P^n-\pi |||_{V}. \nonumber \\ & \leq & \|\pi_0-\pi\|_V
M\gamma^n. \nonumber \end{eqnarray}
\end{proof}

Now we focus on the following simple but useful observation.
\begin{lemma} \label{lem:hoelder} If for a Markov chain $(X_n)_{n \geq 0}$ on
$\stany$ with transition kernel $P$ Assumption \ref{ass:drift}
holds with parameters $\tilde{\beta}, V(x), \lambda, K, \beta,$ it
holds also with $\tilde{\beta}_r:=\tilde{\beta},$
$V_r(x):=V(x)^{1/r},$ $\lambda_r:=\lambda^{1/r},$ $K_r:=K^{1/r},$
$\beta_r:=\beta$ for every $r>1.$
\end{lemma}
\begin{proof}
It is enough to check (A.2). For $x \notin C$ by Jensen inequality
we have
\[ \lambda V(x)\geq 
\int_{\stany} V(y)P(x,dy) \geq \left(\int_{\stany}
V(y)^{1/r}P(x,dy)\right)^{r}
\]
and hence $PV(x)^{1/r}\leq \lambda^{1/r}V(x)^{1/r},$ as claimed.
Similarly for $x\in C$ we obtain $PV(x)^{1/r}\leq K^{1/r}.$
\end{proof}

Lemma \ref{lem:hoelder} together with Theorem \ref{thm:bax} yield
the following corollary.
\begin{cor} \label{cor:hoelder} Under Assumption \ref{ass:drift}
we have
\[|||P^n-\pi |||_{V^{1/r}} \leq M_r \gamma_r^n,\]
where $M_r$ and $\gamma_r$ are constants defined as in Theorem
\ref{thm:bax} resulting from drift parameters defined in Lemma
\ref{lem:hoelder}.
\end{cor}

Integrating the drift condition with respect to $\pi$ yields the
following bound on $\pi V.$
\begin{lemma} \label{lemma: pi V bound}
Under Assumption  \ref{ass:drift}
\[ \pi V \leq \pi(C) \frac{K-\lambda}{1-\lambda} \leq \frac{K-\lambda}{1-\lambda}. \]
\end{lemma}

Let $f_c = f - \pi f.$ The next lemma provides a bound on
$||f_c|^p|_V$ in terms of $||f|^p|_V$ without additional effort.
\begin{lemma} \label{lemma: f_c^p V-norm}
Under Assumption  \ref{ass:drift}
\[
||f_c|^p|_V^{2/p} \leq \Big(C_{f^p_V}^{1/p} + \frac{\pi
(C)}{b_V^{1/p}} K_{p, \lambda} \Big)^2
 \leq \big(C_{f^p_V}^{1/p} + K_{p, \lambda} \big)^2,
\]
 where $b_V = \inf_{x \in \stany}
V(x),$ $C_{f^p_V} = ||f|^p|_V$ and $K_{p,\lambda} =
\frac{K^{1/p}-\lambda^{1/p}}{1-\lambda^{1/p}}.$
\end{lemma}

\begin{proof} Note that $\pi
V^{1/p} \leq \pi (C) K_{p,\lambda} \leq K_{p, \lambda}$ by Lemma
\ref{lemma: pi V bound} and proceed:
\begin{eqnarray}
||f_c|^p|_V & = & \sup_{x \in \stany} \frac{|f(x) - \pi
f|^p}{V(x)} \leq \sup_{x \in \stany} \frac{\Big(C_{f^p_V}^{1/p}
V^{1/p}(x) + \pi |f|\Big)^p}{V(x)} \nonumber \\ \nonumber
 & \leq & \sup_{x \in \stany} \frac{\Big(C_{f^p_V}^{1/p}
V^{1/p}(x) + \pi (C) K_{p,\lambda}\Big)^p}{V(x)}
 \leq C_{f^p_V} \bigg(1+
 \frac{\pi (C) K_{p, \lambda}}{b_V^{1/p} C^{1/p}_{f^p_V}
 }\bigg)^p.
\end{eqnarray}
\end{proof}

\section{MSE Bounds} \label{sec:MainResult}

By $MSE(\hat{I}_{0,n})$ we denote the mean square error of
$\hat{I}_{0,n},$ i.e.
$$MSE(\hat{I}_{0,n})=E_{\pi_0}[\hat{I}_{0,n}-I]^2.$$ Bonds on
$MSE(\hat{I}_{0,n})$ are essential to establish
$(\varepsilon-\alpha)-$approximation of type (\ref{eps-alpha}) and
are also of independent interest.

\begin{thm}\label{MSE bound general}
Assume the Drift Condition \ref{ass:drift} holds and $X_0\sim
\pi_0.$ Then for every measurable function $f:\mathcal{X}\to R,$
every $p\geq 2$ and every $r\in [\frac{p}{p-1},p]$
\begin{equation}\label{MSE bound general bound}
MSE(\hat{I}_{0,n}) \leq
\frac{||f_c|^p|_V^{2/p}}{n}\left(1+\frac{2M_r
\gamma_r}{1-\gamma_r}\right)\left(\pi V + \frac{M\min\{\pi_0
V,\|\pi_0-\pi\|_{V}\}}{n(1-\gamma)}\right),\end{equation}
 where $f_c=f-\pi f$ and constants $M, \gamma, M_r, \gamma_r$ depend
 only and explicitly on
 $\tilde{\beta}, \beta, \lambda$ and $K$ from Assumption \ref{ass:drift} as
 in Theorem \ref{thm:bax} and Corollary \ref{lem:hoelder}.
\end{thm}
The formulation of the foregoing Theorem \ref{MSE bound general}
is motivated by a trade-off between small $V$ and small $\lambda$
in Assumption \ref{ass:drift}. It should be intuitively clear that
establishing the drift condition for a quickly increasing $V$
should result in smaller $\lambda$ at the cost of bigger $\pi V.$
So it may be reasonable to look for a valid drift condition with
$V\geq C||f_c|^p|$ for some $p>2$ instead of the natural choice of
$p=2.$ Lemma \ref{lem:hoelder} should strengthen this intuition.
The most important special case for $p=r=2$ is emphasized below as
a corollary.%

The unknown value $\pi_0 V$ in (\ref{MSE bound general bound})
depends on $\pi_0$ which is users choice and usually a
deterministic point. Also, in many cases a fairly small bound for
$\pi V$ should be possible to obtain by direct calculations, since
in the typical setting $\pi$ is exponentially concentrated whereas
$V$ is a polynomial of degree 2. These calculations should
probably borrow from those used to obtain the minorization and
drift conditions. However, in absence of a better bound for $\pi
V$ Lemma \ref{lemma: pi V bound} is at hand. Similarly Lemma
\ref{lemma: f_c^p V-norm} bounds the unknown value
$||f_c|^p|_V^{2/p}$ in terms of $||f|^p|_V.$ Note that in
applications both $f$ and $V$ have explicit formulas known to the
user and $||f|^p|_V$ can be evaluated directly or easily bounded.

\begin{proof}
Note that $|f|_{V^{1/r}}^r=||f|^r|_V.$ Without loss of generality
consider $f_c$ instead of $f$ and assume $||f_c|^p|_V=1.$ In this
setting
 $|f_c^2|_V\leq 1,$ $Var_{\pi}f_c=\pi f_c^2 \leq \pi V,$
$MSE(\hat{I}_{0,n})=E_{\pi_0}(\hat{I}_{0,n})^2,$ and also for
every $r\in [\frac{p}{p-1},p],$
$$|f_c|_{V^{1/r}}\leq ||f_c|^{p/r}|_{V^{1/r}}=1\quad
\textrm{and}\quad |f_c|_{V^{1-1/r}}\leq
||f_c|^{p-p/r}|_{V^{1-1/r}}=1.$$
 Obviously
\begin{eqnarray} \label{eqn:MSE deco} nMSE(\hat{I}_{0,n})&=&
\frac{1}{n}\sum_{i=0}^{n-1}E_{\pi_0}f_c(X_i)^2 +
\frac{2}{n}\sum_{i=0}^{n-2}\sum_{j=i+1}^{n-1}
E_{\pi_0}f_c(X_i)f_c(X_j).
\end{eqnarray}
We start with a bound for the first term of the right hand side of
(\ref{eqn:MSE deco}). Since $f_c^2(x)\leq V(x),$ we use Corollary
\ref{cor: burn-in} for $f_c^2.$ Let $C=\min\{\pi_0
V,\|\pi_0-\pi\|_{V}\}$ and proceed
\begin{equation} \label{eqn:MSE first}
\frac{1}{n}\sum_{i=0}^{n-1}E_{\pi_0}f_c(X_i)^2
=\frac{1}{n}\sum_{i=0}^{n-1}\pi_0 P^i f_c^2 \leq \pi
f_c^2+\frac{1}{n}\sum_{i=0}^{n-1}CM\gamma^{i}
 \leq  \pi V + \frac{CM}{n(1-\gamma)}.
\end{equation}
To bound the second term of the right hand side of (\ref{eqn:MSE
deco}) note that $|f_c|\leq V^{1/r}$ and use Corollary
\ref{cor:hoelder}.
\begin{eqnarray}
\frac{2}{n}\sum_{i=0}^{n-2}\sum_{j=i+1}^{n-1}
E_{\pi_0}f_c(X_i)f_c(X_j) & = &
\frac{2}{n}\sum_{i=0}^{n-2}\sum_{j=i+1}^{n-1} \pi_0 \left(P^i
\left(f_cP^{j-i}f_c\right)\right) \nonumber \\
& \leq & \frac{2}{n}\sum_{i=0}^{n-2}\sum_{j=i+1}^{n-1} \pi_0
\left(P^i
\left(|f_c||P^{j-i}f_c|\right)\right) \nonumber \\
&\leq& \frac{2M_r}{n}\sum_{i=0}^{n-2}\sum_{j=i+1}^{\infty}
\gamma_r^{j-i} \pi_0 \left(P^i \left(|f_c|V^{1/r}\right)\right)
\nonumber \\
&\leq& \frac{2M_r \gamma_r}{n(1-\gamma_r)}\sum_{i=0}^{n-2} \pi_0
\left(P^i \left(|f_c|V^{1/r}\right)\right)=\spadesuit \nonumber
\end{eqnarray}
Since $|f_c|\leq V^{1/r}$ and $|f_c|\leq V^{1-1/r},$ also
$|f_cV^{1/r}|\leq V$ and we use Corollary \ref{cor: burn-in} for
$|f_c|V^{1/r}.$
\begin{eqnarray} \label{eqn:MSE second}
\spadesuit &\leq& \frac{2M_r
\gamma_r}{n(1-\gamma_r)}\sum_{i=0}^{n-2} \left(\pi
\left(|f_c|V^{1/r}\right)+CM\gamma^i \right) \leq \frac{2M_r
\gamma_r}{1-\gamma_r}\left(\pi V + \frac{CM
}{n(1-\gamma)}\right).\qquad
\end{eqnarray}

Combine (\ref{eqn:MSE first}) and (\ref{eqn:MSE second}) to obtain
\begin{eqnarray} MSE(\hat{I}_{0,n}) & \leq &
\frac{||f_c|^p|_V^{2/p}}{n}\left(1+\frac{2M_r
\gamma_r}{1-\gamma_r}\right)\left(\pi V + \frac{CM
}{n(1-\gamma)}\right). \nonumber
\end{eqnarray}
\end{proof}

\begin{cor}\label{MSE bound typical}
In the setting of Theorem \ref{MSE bound general}, we have in
particular
\begin{equation} \label{MSE bound typical bound} MSE(\hat{I}_{0,n})\leq
\frac{|f_c^2|_V}{n}\left(1+\frac{2M_2
\gamma_2}{1-\gamma_2}\right)\left(\pi V + \frac{M\min\{\pi_0
V,\|\pi_0-\pi\|_{V}\}}{n(1-\gamma)}\right).\end{equation}
\end{cor}

The foregoing bound is easy to interpret: $\pi V|f_c^2|_V$ should
be close to $Var_{\pi}f$ for an appropriate choice of $V,$
moreover $2M_2\gamma_2/(1-\gamma_2)$ corresponds to the
autocorrelation of the chain and the last term $M\min\{\pi_0
V,\|\pi_0-\pi\|_{V}\}/n(1-\gamma)$ is the price for
nonstationarity of the initial distribution. See also Theorem
\ref{thm:as var} for further interpretation.

Theorem \ref{MSE bound general} is explicitly stated for
$\hat{I}_{0,n},$ but the structure of the bound is flexible enough
to cover most typical settings as indicated below.

\begin{cor} \label{cor:MSE bounds various}
In the setting of Theorem \ref{MSE bound general},
\begin{equation}\label{MSE bound perfect}
MSE(\hat{I}_{0,n})  \leq  \frac{\pi
V||f_c|^p|_V^{2/p}}{n}\left(1+\frac{2M_r
\gamma_r}{1-\gamma_r}\right), \quad \textrm{if }
\pi_0=\pi,\end{equation}
\begin{equation}\label{MSE bound deterministic start} MSE(\hat{I}_{0,n}) \leq
\frac{||f_c|^p|_V^{2/p}}{n}\left(1+\frac{2M_r
\gamma_r}{1-\gamma_r}\right)\left(\pi V +
\frac{MV(x)}{n(1-\gamma)}\right), \quad \textrm{if }
\pi_0=\delta_x, \end{equation}
\begin{equation}\label{MSE bound burn-in} MSE(\hat{I}_{t,n})  \leq
\frac{||f_c|^p|_V^{2/p}}{n}\left(1+\frac{2M_r
\gamma_r}{1-\gamma_r}\right)\left(\pi V + \frac{M^2\gamma^t
V(x)}{n(1-\gamma)}\right), \quad \textrm{if } \pi_0=\delta_x.
 \end{equation}
\end{cor}

\begin{proof}
Only (\ref{MSE bound burn-in}) needs a proof. Note that $X_t \sim
\delta_x P^t.$ Now use Theorem \ref{thm:bax} to see that
$\|\delta_x P^t - \pi \|_{V} \leq M\gamma^tV(x),$ and apply
Theorem \ref{MSE bound general} with $\pi_0 = \delta_x P^t.$
\end{proof}
Bound (\ref{MSE bound perfect}) corresponds to the situation when
a perfect sampler is available. For deterministic start without
burn-in and with burn-in (\ref{MSE bound deterministic start}) and
(\ref{MSE bound burn-in}) should be applied respectively.

Next we derive computable bounds for the asymptotic variance
$\sigma^2_f$ 
in central limit theorems for Markov chains under the assumption
of the Drift Condition \ref{ass:drift}.

\begin{thm}\label{thm:as var}
Under the Drift Condition \ref{ass:drift} the Markov chain
$(X_n)_{n\geq 0}$ and a function $f,$ such that $|f^2_c|_V <
\infty$ (or equivalently $|f^2|_V<\infty$), admit a central limit
theorem, i.e:
 \begin{equation} \label{eqn: CLT} \sqrt{n}(\hat{I}_{0,n}-I)
\stackrel{d}{\to} N(0,\sigma^2_f) \quad \textrm{as} \quad n \to
\infty,\end{equation} moreover
\begin{equation} \label{eqn:as var}
\sigma^2_f=\lim_{n\to \infty}nE_{\pi}[\hat{I}_{0,n}-I]^2 \leq \pi
V||f_c|^p|_V^{2/p}\left(1+\frac{2M_r\gamma_r}{1-\gamma_r}\right).\end{equation}
\end{thm}

\begin{proof}
The CLT (i.e. (\ref{eqn: CLT}) and the equation in (\ref{eqn:as
var})) is a well known fact and results from $V-$uniform
ergodicity implied by Theorem \ref{thm:bax} combined with Theorems
17.5.4 and 17.5.3 of \cite{MeynTweedie}. Theorem \ref{MSE bound
general} with $\pi_0 = \pi$ yields the bound for $\sigma^2_f$ in
(\ref{eqn:as var}).
\end{proof}

\begin{remark} For reversible Markov chains significantly sharper
bounds for $\sigma^2_f$ can be obtained via functional analytic
approach. For a reversible Markov chain its transition kernel $P$
is a self-adjoint operator on $L^2_{\pi}.$ Let $f \in L^2_{\pi}$
and $\pi f =0.$ If we denote by $E_f$ the positive measure on
$(-1,1)$ associated with $f$ in the spectral decomposition of $P,$
we obtain (cf. \cite{KipnisVaradhan}, \cite{Geyer practical})
\begin{equation} \label{eqn: as var spectral}
\sigma^2_f=\int_{(-1,1)}\frac{1+\lambda}{1-\lambda}E_f(d\lambda)
\leq \frac{1+\rho}{1-\rho} Var_{\pi}f \leq \frac{1+\rho}{1-\rho}
\pi V |f_c^2|_V.
\end{equation}
Where the first inequality in (\ref{eqn: as var spectral}) holds
if we are able to bound the spectral radius of $P$ acting on
$L^2_{\pi}$ by some $\rho < 1$ (cf. \cite{Geyer practical},
\cite{RobRos hybrid}). Corollary 6.1 of \cite{Bax} yields the
required bound with $\rho$ defined as in Theorem \ref{thm:bax}.
\end{remark}

\section{$(\varepsilon-\alpha)-$Approximation}\label{sec:eps-alpha}

$(\varepsilon-\alpha)-$approximation is an easy corollary of $MSE$
bounds by the Chebyshev inequality.

\begin{thm}[$(\varepsilon-\alpha)-$approximation]\label{thm:eps-alpha}
Let
 \begin{eqnarray} \label{def: b}
 b &=& \frac{\pi
V||f_c|^p|_V^{2/p}}{\varepsilon^2 \alpha}\left(1+\frac{2M_r
\gamma_r}{1-\gamma_r}\right),
 \\  \label{def: c} c &=& \frac{M\min\{\pi_0
V,\|\pi_0-\pi\|_{V}\}||f_c|^p|_V^{2/p}}{\varepsilon^2
\alpha(1-\gamma)}\left(1+\frac{2M_r\gamma_r}{1-\gamma_r}\right),\\
 \label{def: n(t)}
 n(t)&=&\frac{b+\sqrt{b^2+4c(t)}}{2}, \\
 \label{def: c(t)}
c(t) &=& \frac{M^2\gamma^t V(x) ||f_c|^p|_V^{2/p}}{\varepsilon^2
\alpha(1-\gamma)}\left(1+\frac{2M_r\gamma_r}{1-\gamma_r}\right),\\
\label{def: wave c} \tilde{c} & = & \frac{M^2V(x)
  ||f_c|^p|_V^{2/p}}{\varepsilon^2
\alpha(1-\gamma)}\left(1+\frac{2M_r\gamma_r}{1-\gamma_r}\right).
 \end{eqnarray}
 Then under Assumption \ref{ass:drift},
\begin{eqnarray}\label{eps-alpha no burn-in}
 P(|\hat{I}_{0,n}-I| \leq \varepsilon) \geq  1-\alpha,&
\textrm{if} &
  X_0 \sim \pi_0, \quad n \geq \frac{b+\sqrt{b^2+4c}}{2}.\\
\label{eps-alpha with burn-in}
  P(|\hat{I}_{t,n}-I| \leq  \varepsilon) \geq 1-\alpha, & \textrm{if}
&  \left\{ \begin{array}{l}  X_0 \sim \delta_x, \\
 t \geq \max\left\{0, 
 \log_{\gamma}\left(
 \frac{2+\sqrt{4+b^2\ln^2\gamma}}{\tilde{c}\ln^2\gamma}\right)\right\},\quad\\
  n \geq n(t).
  \end{array} \right.
 \end{eqnarray}
 And the above bounds in (\ref{eps-alpha with burn-in}) give the
 minimal length of the trajectory $(t+n)$ resulting from (\ref{MSE bound burn-in}).
\end{thm}

\begin{proof}
From the Chebyshev's inequality we get
 \begin{eqnarray} \label{Czybyszew} P(|\hat{I}_{t,n}-I|\leq
\varepsilon) &=& 1- P(|\hat{I}_{t,n}-I|\geq \varepsilon) \nonumber \\
&\geq &1-\frac{MSE(\hat{I}_{t,n})}{\varepsilon^2} \geq  1-\alpha
\quad \textrm{if}\quad MSE(\hat{I}_{t,n})\leq \varepsilon^2
\alpha. \qquad
\end{eqnarray}
To prove (\ref{eps-alpha no burn-in}) set $C=\min\{\pi_0
V,\|\pi_0-\pi\|_{V}\},$ and combine (\ref{Czybyszew}) with
(\ref{MSE bound general bound}) to get $$n^2-n \frac{\pi
V||f_c|^p|_V^{2/p}}{\varepsilon^2
\alpha}\left(1+\frac{2M_r\gamma_r}{1-\gamma_r}\right) -\frac{M
C||f_c|^p|_V^{2/p}}{\varepsilon^2
\alpha(1-\gamma)}\left(1+\frac{2M_r\gamma_r}{1-\gamma_r}\right)\geq
0,$$ and hence $n  \geq \frac{b+\sqrt{b^2+4c}}{2},$ where $b$ and
$c$ are defined by (\ref{def: b}) and (\ref{def: c}) respectively.
The only difference in (\ref{eps-alpha with burn-in}) is that now
we have $c(t)$ defined by (\ref{def: c(t)}) instead of $c.$ It is
easy to check that the best bound on $t$ and $n$ (i.e. that
minimizes $t+n$) is such that
$$n\geq n(t) \qquad \textrm{and} \qquad t \geq \max\left\{0,
\min\{t\in N: n'(t)\geq -1\}\right\},$$
 where $n(t)$ is defined by (\ref{def: n(t)}).
 Standard calculations show that
 $$\min\{t\in N: n'(t)\geq -1\} =
 \min\{t\in N: (\gamma^t)^2\tilde{c}^2\ln^2\gamma -
 \gamma^t 4 \tilde{c} - b^2 \leq 0\},$$
 where $\tilde{c}$ is defined by (\ref{def: wave c}). Hence we obtain
 $$t \geq \max\left\{0,(\ln\gamma)^{-1}\ln\left(
 \frac{2+\sqrt{4+b^2\ln^2\gamma}}{\tilde{c}\ln^2\gamma}\right)\right\}
 \qquad \textrm{and} \qquad n\geq n(t).$$
 This completes the proof.
\end{proof}

\begin{remark} \label{rem: burn-in} The formulation of Theorem
\ref{thm:eps-alpha} and the above proof indicate how the issue of
a sufficient burn-in should be understood. The common description
of $t$ as \textit{time to stationarity} and the often encountered
approach that $t^*=t(x,\tilde{\varepsilon})$ should be such that
$\rho (\pi, \delta_xP^{t^*})\leq \tilde{\varepsilon}$ (where
$\rho(\cdot, \cdot)$ is a distance function for probability
measures, e.g. total variation distance, or $V-$norm distance)
seems not appropriate for such a natural goal as
$(\varepsilon-\alpha)-$approximation. The optimal burn-in time can
be much smaller then $t^*$ and in particular cases it can be $0.$
Also we would like to emphasize that in the typical drift
condition setting, i.e. if $\mathcal{X}$ is not compact and the
target function $f$ is not bounded, the $V-$norm should be used as
a measure of convergence, since $||\pi_t-\pi||_{tv} \to 0$ does
not even imply $\pi_tf \to \pi f.$
\end{remark}

Next we suggest an alternative estimation scheme that allows for
sharper bounds for the total simulation cost needed to obtain
$(\varepsilon-\alpha)-$approximation for small $\alpha.$ We will
make use of the following simple lemma taken from the more
complicated setting of \cite{NiePo}.

\begin{lemma} \label{lem: median}
Let $m \in N$ be an odd number and let $\hat{I}_1,\dots,
\hat{I}_m$ be independent random variables, such that
$P(|\hat{I}_k - I| \leq \varepsilon) \geq 1-a > 1/2,$ for
$k=1,\dots, m.$ Define $\hat{I}:=
\textup{med}\{\hat{I}_1,\dots,\hat{I}_m\}.$ Then
\begin{equation} \label{eqn: median}
P(|\hat{I}-I| \leq \varepsilon) \geq 1 - \alpha, \quad \textrm{if}
\quad m \geq \frac{2\ln (2\alpha)}{\ln [4a(1-a)]}.
\end{equation} \end{lemma}

\begin{proof} Since $P(|\hat{I}_k-I|>\varepsilon) \leq a < 1/2,$
by elementary arguments we obtain
 \begin{eqnarray}P(|\hat{I}-I| > \varepsilon) &\leq &\sum_{k=(m+1)/2}^{m}
 \binom{m}{k}a^k (1-a)^{n-k} \nonumber \\ &\leq &
 2^{m-1} a^{m/2} (1-a)^{m/2} \nonumber \\
 & = & \frac{1}{2}\exp \left\{\frac{m}{2}\ln (4a(1-a))\right\}.
 \nonumber \end{eqnarray}
 The last term does not exceed $\alpha$ if $m \geq 2\ln (2\alpha) / \ln
 [4a(1-a)],$ as claimed.
\end{proof}

Hence $(\varepsilon-\alpha)-$approximation can be obtained by the
following Algorithm \ref{Alg: median}, where Theorem
\ref{thm:eps-alpha} should be used to find $t$ and $n$ that
guarantee $(\varepsilon-a)-$approximation and $m$ results from
Lemma \ref{lem: median}.

\begin{alg} \label{Alg: median} ~
 \begin{enumerate}
\item Simulate $m$ independent runs of length $t+n$ of the
underlying Markov chain, $$X_{0}^{(k)}, \dots, X_{t+n-1}^{(k)},
\quad k=1,\dots, m.$$
\item Calculate $m$ estimates of $I,$ each based on a single run,
$$\hat{I}_{k}=\hat{I}_{t,n}^{(k)}=\frac{1}{n}\sum_{i=t}^{t+n-1}
f(X_{i}^{(k)}), \quad k=1,\dots,m.$$
\item For the final estimate take
$$\hat{I}=\textup{med}\{\hat{I}_1,\dots,\hat{I}_m\}.$$
\end{enumerate}
\end{alg}

The total cost of Algorithm \ref{Alg: median} amounts to
\begin{equation} \label{eqn: cost} C=C(a)=m(t+n) \end{equation}
 and depends on $a$ (in addition to previous
parameters). The optimal $a$ can be found numerically, however it
is worth mentioning $a=0,11969$ is an acceptable arbitrary choice
(cf. \cite{NiePo}). A closer look at equation (\ref{eqn: cost})
reveals that the leading term is
$$mb=\frac{1}{a\ln \{[4a(1-a)]^{-1}\}}\left\{\frac{2\ln
\{(2\alpha)^{-1}\} \pi
V||f_c|^p|_V^{2/p}}{\varepsilon^2}\left(1+\frac{2M_r
\gamma_r}{1-\gamma_r}\right)\right\},$$ where $b$ is defined by
(\ref{def: b}).
Function $a\ln \{[4a(1-a)]^{-1}\}$ has one maximum on $(0,1/2)$ at
$a \approx 0,11969.$

\section{A Toy Example - Contracting Normals} \label{sec: Toy Example}

To illustrate the results of previous sections we analyze the
\textit{contracting normals} example studied by Baxendale in
\cite{Bax} (see also \cite{Roberts Tweedie}, \cite{Roberts
Rosenthal shift} and \cite{Rosenthal gibbs}), where Markov chains
with transition probabilities $P(x,\cdot) = N(\theta x, 1-
\theta^2)$ for some parameter $\theta \in (-1,1)$ are considered.

Similarly as in \cite{Bax} we take a drift function $V(x) = 1+
x^2$ and a small set $C = [-c,c]$ with $c > 1,$ which allows for
$\lambda = \theta^2 + \frac{2(1-\theta^2)}{1+c^2} < 1$ and $K = 2
+ \theta^2(c^2-1).$ We also use the same minorization condition
with $\nu$ concentrated on $C,$ such that $\tilde{\beta} \nu (dy)
= \min_{x \in C} (2\pi (1-\theta^2))^{-1/2} \exp (-\frac{(\theta x
- y)^2}{2(1-\theta^2)})dy.$ This yields $\tilde{\beta} =
2[\Phi(\frac{(1+|\theta|)c}{\sqrt{1-\theta^2}}) -
\Phi(\frac{|\theta|c}{\sqrt{1-\theta^2}})],$ where $\Phi$ denotes
the standard normal cumulative distribution function.

 Baxendale in \cite{Bax} indicated that the chain is
reversible with respect to its invariant distribution $\pi =
N(0,1)$ for all $\theta \in (-1, 1)$ and it is reversible and
positive for $\theta > 0.$

Moreover, in Lemma \ref{lemma: contracting normals convergence} we
observe a relationship between marginal distributions of the chain
with positive and negative values of $\theta.$ By $\mathcal{L}(X_n
| X_0, \theta)$ denote the distribution of $X_n$ given the
starting point $X_0$ and the parameter value $\theta.$
\begin{lemma} \label{lemma: contracting normals convergence}
\begin{equation}
\mathcal{L}(X_n|X_0, \theta) = \mathcal{L}(X_n|(-1)^n X_0, -
\theta).
\end{equation} \end{lemma}
\begin{proof}
Let $Z_1, Z_2, \dots $ be an iid $N(0,1)$ sequence, then
\begin{eqnarray}
\mathcal{L}(X_n|X_0, \theta) & = & \mathcal{L}\Big(\theta^n X_0 +
\sum_{k=1}^n \theta^{n-k} \sqrt{1-\theta^2} Z_k \Big)
\nonumber \\
& = & \mathcal{L}\Big((-\theta)^n (-1)^n X_0 + \sum_{k=1}^n
(-\theta)^{n-k} \sqrt{1-\theta^2} Z_k \Big)
\nonumber \\
 &=& \mathcal{L}(X_n|(-1)^n
X_0, - \theta), \nonumber
\end{eqnarray}
and we used the fact that $Z_k$ and $-Z_k$ have the same
distribution.
\end{proof}
For $\theta < 0$ using Lemma \ref{lemma: contracting normals
convergence} and the fact that $V(x) = 1+x^2$ is symmetric we
obtain
 \begin{eqnarray}
||\mathcal{L}(X_n|X_0, \theta) - \pi||_V & = & ||\mathcal{L}(X_n|
(-1)^n X_0, - \theta) - \pi||_V \leq M \gamma^n V((-1)^n
X_0)\nonumber \\ \nonumber &=& M \gamma^n V(X_0) = M \gamma^n
(1+X_0^2).\end{eqnarray} Thus for all $\theta \in (-1,1)$ we can
bound the $V-$norm distance between $\pi$ and the distribution of
$X_n$ via Theorem \ref{thm:bax} with $\rho$ and $M=M(\gamma),$
where $\gamma \in (\rho, 1),$ computed for reversible and positive
Markov chains (see Appendix \ref{sec: Bax form rev pos} for
formulas).
\begin{table}[!htb] \label{table
contracting normals - Bax} \begin{center}Table \ref{table
contracting normals - Bax} - bounds based on Baxendale's
$V-$uniform ergodicity constants.
\begin{tabular}{c@{ }@{ }c@{ }@{ }c@{ }@{ }c@{ }@{ }c@{ }@{ }c@{ }@{ }
c@{ }@{ }c@{ }@{ }c@{ }@{ }c@{ }@{ }c@{ }@{ }c@{ }@{ }c}
$|\theta|$ & $\varepsilon$  & $\alpha$ &  $\rho$ & $\rho_2$ &
$\gamma$ & $\gamma_2$ & $M$ & $M_2$ & $m$ & $t$ & $n$ & total cost
\\ \hline
$.5$ & $.1$ & $.1$      & .895 & .899 & .915 & .971 & 36436 & 748 & 1 & 218 & 6.46e+09& 6.46e+09\\
$.5$ & $.1$ & $10^{-5}$ & .895 & .899 & .915 & .971 & 36436 & 748 & 1 & 218 & 6.46e+13& 6.46e+13\\
$.5$ & $.1$ & $10^{-5}$ & .895 & .899 & .915 & .971 & 36436 & 748 & 27& 218 & 5.39e+09& 1.46e+11\\
\end{tabular}\end{center}
\end{table}
The choice of $V(x) = 1+ x^2$ allows for
$(\varepsilon-\alpha)-$approximation of $\int_{\stany}f(x)
\pi(dx)$ if $|f^2|_V < \infty$ for the possibly unbounded function
$f.$ In particular the MCMC works for all linear functions on
$\stany.$ We take $f(x) = x$ where $|f^2|_V = 1$ as an example. We
have to provide parameters and constants required for Theorem
\ref{thm:eps-alpha}. In this case the optimal starting point is
$X_0=0$ since it minimizes $V(x) = 1+x^2.$ To bound $\pi V$ we use
Lemma \ref{lemma: pi V bound} and Lemma \ref{lemma: f_c^p V-norm}
yields a bound on $||f_c|^2|_V^{2/p} = |f_c^2|_V.$

Examples of bounds for $t$ and $n$ for the one walk estimator, or
$t, $ $n$ and $m$ for the median of multiple runs estimator are
given in Table \ref{table contracting normals - Bax}. The bounds
are computed for $c=1.6226$ which minimizes $\rho_2$ (rather than
$\rho$) for $\theta = 0.5.$ Then a grid search is performed to
find optimal values of $\gamma$ and $\gamma_2$ that minimize the
total simulation cost. Note that in Baxendale's constant $M$
depends on $\gamma$ and $M$ goes relatively quickly to $\infty$ as
$\gamma \to \rho.$ This is the reason why optimal $\gamma$ and
$\gamma_2$ are far from $\rho$ and $\rho_2$ and this turns out to
be the main weakness of Baxendale's bounds. Also for small $\alpha
= 10^{-5}$ we observe a clear computational advantage of the
median of multiple runs estimation. The $m=27$ shorter runs have
significantly lower total cost then the single long run.
%
%
%
R functions for computing this example and also the general bounds
resulting from Theorem \ref{thm:eps-alpha} are available at
http://akson.sgh.waw.pl/\~{}klatus/

\section{The Example - a Hierarchical Random Effects Model} \label{sec:example HREM}

In this section we describe a hierarchical random effects model
which is a widely applicable example and provides a typical target
density $\pi$ that arises in Bayesian statistics. Versions of this
model and the efficiency of MCMC sampling have been analyzed e.g.
by Gelfand and Smith in \cite{Gelfand Smith}, Rosenthal in
\cite{Rosenthal gibbs}, \cite{Rosenthal drift} and many other
authors. In particular Hobert and Geyer in \cite{Hobert Geyer}
analyzed a Gibbs sampler and a block Gibbs sampler for this model
and showed the underlying Markov chains are in both cases
geometrically ergodic (we describe these samplers in the sequel).
Jones and Hobert in \cite{Jones Hobert Gibbs for rand eff mod}
derived computable bounds for the geometric ergodicity parameters
and consequently computable bounds for the total variation
distance $\|P^t(x,\cdot)-\pi \|_{tv}$ to stationarity in both
cases. They used these bounds to determine the burn-in time. Their
work was a breakthrough in analyzing the hierarchical random
effects model, however, mere bounds on burn-in time do not give a
clue on the total amount of simulation needed. Also, bounding the
total variation distance seems inappropriate when estimating
integrals of unbounded functions, as indicated in Remark \ref{rem:
burn-in}. In this section we establish the
$(\varepsilon-\alpha)-$approximation for the hierarchical random
effects model. This consists of choosing a suitable sampler,
establishing the Drift Condition \ref{ass:drift} with explicit
constants, computing $V-$uniform ergodicity parameters, and
optimizing lower bounds for $t$ and $n$ in case of estimation
along one walk or for $t,$ $n$ and $m$ in (\ref{eqn: cost}) for
the median of shorter runs. This may turn out to be a confusing
procedure, hence we outline it here in detail, discuss
computational issues and provide necessary R functions.

\subsection{The Model} \label{subsection: the Model}

Since we will make use of the drift conditions established by
Jones and Hobert in \cite{Jones Hobert Gibbs for rand eff mod} we
also try to follow their notation in the model description. Let
$\mu$ and $\lambda_{\theta}$ be independent and distributed as
$$\mu \sim N(m_0, s_0^{-1}) \quad \textrm{and} \quad
\lambda_{\theta} \sim \textrm{Gamma}(a_1,b_1),$$
where $m_0 \in \mathbb{R}, s_0>0, a_1>0,$ and $b_1>0$ are known
constants.

At the second stage, conditional on $\mu$ and $\lambda_{\theta},$
random variables $\theta_1,\dots \theta_K$ and $\lambda_e$ are
independent and distributed as
$$\theta_i | \mu, \lambda_{\theta} \sim N(\mu,
\lambda_{\theta}^{-1}) \quad \textrm{and} \quad \lambda_e \sim
\textrm{Gamma}(a_2,b_2),$$
where $a_2>0, b_2>0$ are known constants.

Finally in the third stage, conditional on $\theta =
(\theta_1,\dots, \theta_K)$ and $\lambda_e,$ the observed data
$y=\{Y_{ij}\}$ are independent with
$$Y_{ij}|\theta, \lambda_e \sim N(\theta_i,\lambda_e^{-1}),$$
where $i=1,\dots, K$ and $j=1,\dots, m_i.$

The Bayesian approach involves conditioning on the values of the
observed data $\{Y_{ij}\}$ and considering the joint distribution
of all $K+3$ parameters given this data. Thus we are interested in
the posterior distribution, that is, the following distribution
defined on the space $\stany = (0,\infty)^2\times
\mathbb{R}^{K+1},$
\begin{eqnarray}\label{Vrianace Components Posterior 1}
\mathcal{L}(\theta_1,\dots, \theta_K, \mu, \lambda_{\theta},
\lambda_e | \{Y_{ij}\}) &=& \pi (\theta, \mu, \lambda | y)
\\ &\varpropto & \nonumber d(y|\theta, \lambda_e)
d(\theta|\mu, \lambda_{\theta})
d(\lambda_e)d(\lambda_{\theta})d(\mu)=\clubsuit,
\end{eqnarray}
where $d$ denotes a generic density and hence the final formula
for the unnormalised density takes the form of
\begin{eqnarray}\label{Vrianace Components Posterior 2}
\clubsuit &=& e^{-b_1\lambda_{\theta}}
\lambda_{\theta}^{a_1-1}e^{-b_2\lambda_e} \lambda_e^{a_2-1}
e^{-\frac{1}{2}s_0(\mu - m_0)^2} \nonumber \\ && \times
\prod_{i=1}^K
\left[e^{-\frac{1}{2}\lambda_{\theta}(\theta_i-\mu)^2}\lambda_{\theta}^{1/2}\right]
\times \prod_{i=1}^K \prod_{j=1}^{m_i}
\left[e^{-\frac{1}{2}\lambda_{e}(y_{ij}-\theta_i)^2}\lambda_{e}^{1/2}\right],
\end{eqnarray}
and we have to deal with a density that is high-dimensional,
irregular, strictly positive in $\stany$ and concentrated in the
,,center'' of $\stany,$ which is very typical for MCMC situations
\cite{RobRos survey}. Computing expectations with respect to $\pi
(\theta, \mu, \lambda | y)$ is crucial for bayesian inference
(e.g. to obtain bayesian estimators) and requires MCMC techniques.

\subsection{Gibbs Samplers for the Model} \label{subsection:
samplers for HREM}

Full conditional distributions required for a Gibbs sampler can be
computed without difficulty. Let
$$\bar{y}_i:=\frac{1}{m_i}\sum_{j=1}^{m_i}y_{ij}, \qquad M:=\sum_i
m_i, \qquad \bar{\theta}=\frac{1}{K}\sum_i \theta_i,$$
$$\theta_{-i}:=(\theta_1,\dots,
\theta_{i-1},\theta_{i+1},\dots,\theta_K), \qquad \nu_1(\theta,
\mu):= \sum_{i=1}^K(\theta_i-\mu)^2,$$ $$
\nu_2(\theta):=\sum_{i=1}^K(\theta_i-\bar{y}_i)^2, \qquad
SSE:=(y_{ij}-\bar{y}_i)^2.$$
Now the conditionals are
\begin{eqnarray} \label{variance comp cond lambda_theta}
\lambda_{\theta} | \theta, \mu, \lambda_e, y & \sim &
\textrm{Gamma}\left(\frac{K}{2}+a_1,
\frac{\nu_1(\theta,\mu)}{2}+b_1\right), \\ \label{variance comp
cond lambda_e} \lambda_{e} | \theta, \mu, \lambda_{\theta}, y &
\sim & \textrm{Gamma}\left(\frac{M}{2}+a_2,
\frac{\nu_2(\theta)+SSE}{2}+b_2\right),
\\\label{variance comp cond theta_i}
\theta_i | \theta_{-i}, \mu, \lambda_{\theta}, \lambda_e, y & \sim
& N\left(\frac{\lambda_{\theta}\mu+m_i \lambda_e
\bar{y}_i}{\lambda_{\theta}+m_i \lambda_e},
\frac{1}{\lambda_{\theta}+m_i \lambda_e}\right), \\
\label{variance comp cond mu} \mu | \theta, \lambda_{\theta},
\lambda_e, y & \sim &
N\left(\frac{s_0m_0+K\lambda_{\theta}\bar{\theta}}{s_0+K\lambda_{\theta}},
\frac{1}{s_0+K\lambda_{\theta}}\right).
\end{eqnarray}
Gibbs samplers for the variance components model and its versions
have been used and studied by many authors. We consider the two
Gibbs samplers analyzed by Jones and Hobert in \cite{Jones Hobert
Gibbs for rand eff mod}.

\begin{itemize}
\item The \textbf{fixed-scan Gibbs sampler} that updates $\mu,$
then $\theta = (\theta_1, \dots \theta_K),$ then
$\lambda=(\lambda_{\theta},\lambda_e).$ Note that $\theta_i$'s are
conditionally independent given $(\mu, \lambda)$ and so are
$\lambda_{\theta}$ and $\lambda_e$ given $(\theta, \mu).$ Thus the
one step Markov transition density $(\mu', \theta', \lambda') \to
(\mu, \theta, \lambda)$ of this Gibbs sampler is
\begin{eqnarray}
\label{eqn: fixed scan Gibbs sampler transition density} p(\mu,
\theta, \lambda|\mu', \theta', \lambda') & = & d(\mu|\theta',
\lambda', y) \left[\prod_{i=1}^K d(\theta_i|\mu, \lambda',
y)\right]\qquad \\ && \qquad \times \textrm{ } d(\lambda_{\theta}
| \theta, \mu, y) d(\lambda_e| \theta, \mu, y).\nonumber
\end{eqnarray}
Where $d$ denotes a generic density and $y=\{Y_{ij}\},$
$i=1,\dots,K;$ $j=1, \dots m_i,$ is the observed data.
\item Hobert and Geyer in \cite{Hobert Geyer} introduced a more
efficient \textbf{block Gibbs sampler} (also analyzed by Jones and
Hobert in \cite{Jones Hobert Gibbs for rand eff mod}), in which
all the components of $$\xi = (\theta_1, \dots \theta_K, \mu)$$
are updated simultaneously. It turns out that $$\xi|\lambda, y
\sim N(\xi^*, \Sigma)\quad \textrm{where}\quad\xi^*=\xi^*(\lambda,
y) \quad \textrm{and} \quad \Sigma=\Sigma(\lambda, y).$$ Thus the
one step Markov transition density $(\lambda', \xi') \to (\lambda,
\xi)$ of the block Gibbs sampler is
\begin{equation} \label{eqn: block Gibbs sampler trans density}
p(\lambda, \xi|\lambda', \xi') = d(\lambda_{\theta}|\xi', y)
d(\lambda_{e}|\xi', y)d(\xi|\lambda, y).
\end{equation}
We give now the formulas for $\xi^*$ and $\Sigma$ derived in
\cite{Hobert Geyer}. Let
$$ \tau = \sum_{i=1}^{K} \frac{m_i
\lambda_{\theta}\lambda_{e}}{\lambda_{\theta+m_i\lambda_{e}}},$$
then
 \begin{eqnarray}
E(\mu|\lambda) & = & \frac{1}{s_0 +\tau}\bigg[\sum_{i=1}^{K}
\frac{m_i \lambda_{\theta}\lambda_{e}
\bar{y}_i}{\lambda_{\theta+m_i\lambda_{e}}} + m_0 s_0 \bigg],
 \nonumber \\ \nonumber
E(\theta_i|\lambda) & = & \frac{\lambda_{\theta} E(\mu|\lambda)
}{\lambda_{\theta}+m_i \lambda_{e}} + \frac{m_i \lambda_{e}
\bar{y}_i}{\lambda_{\theta+m_i\lambda_{e}}}.
 \end{eqnarray}
and
\begin{eqnarray} \nonumber
Var(\theta_i|\lambda) & = &  \frac{1}{\lambda_{\theta}+m_i
\lambda_{e}}\bigg[1+ \frac{\lambda_{\theta}^2}{(\lambda_{\theta} +
m_i \lambda_{e})(s_0+\tau)}\bigg],
\\ \nonumber
 Cov(\theta_i,
\theta_j|\lambda) & = & \frac{\lambda_{\theta}^2}{\lambda_{\theta}
+ m_i \lambda_{e})(\lambda_{\theta} + m_j \lambda_{e})(s_0+\tau)},
\\ \nonumber
 Cov(\theta_i, \theta_j|\lambda) & = &
\frac{\lambda_{\theta}^2}{\lambda_{\theta} + m_i
\lambda_{e})(\lambda_{\theta} + m_j \lambda_{e})(s_0+\tau)},
\\ \nonumber
 Var(\mu|\lambda) & = & \frac{1}{s_0 +\tau}. \end{eqnarray}
\end{itemize}
\subsection{Relations between Drift Conditions} \label{subsection:
drifts - lemmas}

A crucial step for $(\varepsilon-\alpha)-$approximation is
establishing the drift condition \ref{ass:drift} which in the
sequel will be referred to as the Baxendale-type drift condition.
To this end we use the Rosenthal-type (cf. \cite{Rosenthal drift})
and Roberts-and-Tweedie-type (cf. \cite{Roberts Tweedie}) drift
conditions established by Jones and Hobert in \cite{Jones Hobert
Gibbs for rand eff mod} combined with their type of a small set
condition.

In the following definitions and lemmas $P$ denotes the transition
kernel of the Markov chain $\lancuch$ and the subscripts of drift
condition parameters indicate the type of drift condition they
refer to.

\begin{assu}[The Rosenthal-type drift condition] ~
\begin{itemize}
\item[(R.1)]
There exists a function $V_R: \mathcal{X} \to [0,\infty)$ and
constants $0<\lambda_R < 1$ and $K_R<\infty$ satisfying
\begin{equation} \label{eqn: Rosenthals drift}
PV_{R}(x)\leq \lambda_R V_R(x) + K_R. \end{equation}
\item[(R.2)]
Let $C_R=\{x \in \mathcal{X}: V_R(x) \leq d_R\},$ where $d_R >
2K_R/(1-\lambda_R).$ There exists a probability measure $\nu_R$ on
$\mathcal{X}$ and $\tilde{\beta}_R>0,$ such that for all $x \in
C_R$ and $A\in \borel,$
\begin{equation}\label{eqn: Rosenthals small set}
P(x,A)\geq \tilde{\beta}_R \nu_R(A).\end{equation}
\end{itemize}
\end{assu}
\begin{assu}[The Roberts-and-Tweedie-type drift condition] ~
\begin{itemize}
\item[(RT.1)]
There exists a function $V_{RT}: \mathcal{X} \to [1,\infty)$ and
constants $0<\lambda_{RT} < 1$ and $K_{RT}<\infty$ satisfying
\begin{equation} \label{eqn: Roberts-Tweedie drift}
PV_{RT}(x)\leq \lambda_{RT} V_{RT}(x) +
K_{RT}\mathbb{I}_{C_{RT}}(x),
\end{equation}
where $C_{RT}=\{x \in \mathcal{X}: V_{RT}(x) \leq d_{RT}\},$ and
$d_{RT} \geq \frac{K_{RT}}{1-\lambda_{RT}}-1.$
\item[(RT.2)]
There exists a probability measure $\nu_{RT}$ on $\mathcal{X}$ and
$\tilde{\beta}_{RT}>0,$ such that for all $x \in C_{RT}$ and $A\in
\borel,$
\begin{equation}\label{eqn: Roberts-Tweedie small set}
P(x,A)\geq \tilde{\beta}_{RT} \nu_{RT}(A).\end{equation}
\end{itemize}
\end{assu}
The following lemma relates the two drift conditions.
\begin{lemma}[Lemma 3.1 of \cite{Jones Hobert Gibbs for rand eff
mod}] Assume that the Rosenthal-type drift condition holds. Then
for any $d>0$ the Roberts-and-Tweedie-type drift condition holds
with parameters
\[
V_{RT}=V_{R}+1, \quad \lambda_{RT}= \lambda_{RT}(d) =
\frac{d+\lambda_{R}}{d+1}, \quad K_{RT}=K_{R}+1-\lambda_R, \quad
\tilde{\beta}_{RT}=\tilde{\beta}_{R},\]
\[C_{RT}=C_{RT}(d)=\bigg\{x \in \mathcal{X}:
V_{RT}(x) \leq \frac{(d+1)K_{RT}}{d(1-\lambda_{RT})}\bigg\} \quad
 \textrm{and} \quad \nu_{RT}=\nu_{R}.
\]
\end{lemma}
The Baxendale-type drift condition we work with results from each
of the above conditions and the following lemma is easy to verify
by simple algebra.
\begin{lemma} \label{lemma: drift Rosenth to Bax}
If the Rosenthal-type or the Roberts-and-Tweedie-type drift
condition holds, then the Baxendale-type drift condition (A.1-2)
verifies with
\[
V=V_{RT}=V_R+1, \qquad \lambda= \lambda(d)=
\lambda_{RT}=\frac{d+\lambda_{R}}{d+1},\]
\[ \nu
=\nu_{RT} = \nu_{R}, \qquad C=C(d)=C_{RT},  \qquad
\tilde{\beta}=\tilde{\beta}_{RT}=\tilde{\beta}_{R}, \]
\[\quad K=K(d)=K_{RT}+
\lambda_{RT}d_{RT} = (K_R +1 - \lambda_R)\frac{d^2 + 2d +
\lambda_R}{d(1-\lambda_R)}.
\]
\end{lemma}

Observe next that integrating each of the drift conditions yields
a bound on $\pi V$ similar to the one obtained in Lemma
\ref{lemma: pi V bound} and the best available bound should be
used in Theorem \ref{thm:as var} and Theorem \ref{thm:eps-alpha}.
In particular, if the Baxendale-type drift condition is obtained
from the Roberts-and-Tweedie-type drift condition via Lemma
\ref{lemma: drift Rosenth to Bax}, integrating the latter always
leads to a better bound on $\pi V.$ Also, if one starts with
establishing the Rosenthal-type drift condition, the value of $d$
used for bounding $\pi V$ does not have to be the same as the one
used for establishing the Baxendale-type drift and minorization
condition and it should be optimized. Moreover
$\frac{K_R}{1-\lambda_R} + 1 < \frac{K_{RT}}{1- \lambda_{RT}} <
\frac{K-\lambda}{1-\lambda}$ for every $d > 0.$ This leads to the
following lemma which can be checked by straightforward
calculations.

\begin{lemma} \label{lemma: pi V bound min}
Provided the drift functions are as in Lemma \ref{lemma: drift
Rosenth to Bax}, the bound on $\pi V$ can be optimized as follows
\begin{equation} \label{eqn: pi V bound min}
\pi V  \leq  \min \left\{\inf_d \Big\{\pi (C_{RT}(d))
\frac{K_{RT}}{1- \lambda_{RT}(d)}\Big\}, \frac{K_R}{1-\lambda_R} +
1 \right\} \leq  \frac{K_R}{1-\lambda_R} + 1.
\end{equation}
\end{lemma}

\subsection{Drift and Minorization Conditions for the Samplers}
\label{subsection: drifts for samplers}

For the fixed-scan Gibbs sampler and the block Gibbs sampler of
Section \ref{subsection: samplers for HREM} Jones and Hobert in
\cite{Jones Hobert Gibbs for rand eff mod} (Section 4 and 5
therein) obtained the following drift and minorization conditions.
See their paper for derivation and more elaborative commentary of
these results.

\subsubsection{Drift and Minorization for the block Gibbs Sampler}

Assume $m'= \min \{m_1, \dots, m_K\} \geq 2$ and $K \geq 3.$
Moreover define
\[ \delta_1 = \frac{1}{2a_1+K-2}, \quad
\delta_2=\frac{1}{2a_2+M-2}, \quad \delta_3=(K+1)\delta_2, \quad
\delta_4 = \delta_2 \sum_{i=1}^K m_i^{-1}, \]
\[ \delta = \max \{\delta_1, \delta_3\}, \quad
c_1=\frac{2b_1}{2a_1+K-2}, \quad c_2=\frac{2b_2+SSE}{2a_2+M-2}.
\]
Observe that $0<\delta_i<1$ for $i=1,2,3,4.$ Also let $\triangle$
denote the length of the convex hull of the set $\{\bar{y}_1,
\dots,  \bar{y}_K, m_0\}.$
\begin{prop}[Drift for unbalanced case] Fix $\lambda_R \in (\delta,
1)$ and let $\phi_1$ and $\phi_2$ be positive numbers such that
$\frac{\phi_1 \delta_4}{\phi_2}+\delta < \lambda_R.$ Define the
drift function as
\begin{equation} \label{eqn: drifr block unbalanced}
V_1(\theta, \mu) = \phi_1 \nu_1(\theta, \mu) + \phi_2
\nu_2(\theta),\end{equation}
 where $\nu_1(\theta, \mu)$ and $\nu_2(\theta)$ are defined in
 Section \ref{subsection: samplers for HREM}. With this drift
 function the block Gibbs sampler satisfies the Rosenthal-type drift
 condition with
 \begin{equation} \label{eqn: drift block unbalanced K}
 K_R = \phi_1\big[c_1 + c_2 \frac{\delta_4}{\delta_2} + K
 \triangle^2 \big] +\phi_2 \big[c_2(K+1) + M \triangle^2 \big].
\end{equation}
\end{prop}
A better drift condition can be obtained in the balanced case,
when $m_i=m \geq 2$ for $i=1, \dots, K.$ Let $\delta_5=K
\delta_2.$
\begin{prop}[Drift for balanced case] Fix $\lambda_R \in (\delta,
1)$ and let $\phi$ be a positive number such that $\phi
\delta_5+\delta < \lambda_R.$ Define the drift function as
\begin{equation} \label{eqn: drifr block balanced}
V_2(\theta, \mu) = \phi \nu_1(\theta, \mu) + m^{-1}
\nu_2(\theta).\end{equation}
 With this drift
 function the block Gibbs sampler satisfies the Rosenthal-type drift
 condition with
 \begin{equation} \label{eqn: drift block balanced K}
 K_R = \phi c_1 + (\phi K +K +1) \frac{c_2}{m} + \max\{\phi,1\}
 \sum_{i=1}^K \max \big\{(\bar{y}-\bar{y}_i)^2,
 (m_0-\bar{y}_i)^2\big\},
\end{equation}
where $\bar{y}:=K^{-1}\sum_{i=1}^K \bar{y}_i.$
\end{prop}
Proposition \ref{prop:minorization for block gibbs} (Proposition
4.1 of \cite{Jones Hobert Gibbs for rand eff mod}) provides a
minorization condition for the Rosenthal-type drift-minorization
condition for the block Gibbs sampler for both, the balanced and
unbalanced case. Note that the balanced case drift function $V_2$
is a special case of the unbalanced drift function $V_1,$ hence we
focus on $V_1.$

Now consider the candidate $C_R=\{(\theta, \mu): V_1(\theta, \mu)
\leq d_R\}$ for a small set. Note that $C_R$ is contained in
$S_B=S_{B_1} \cap S_{B_2},$ where $S_{B_1}=\{(\theta, \mu) :
\nu_1(\theta, \mu) < d_R/\phi_1\}$ and $S_{B_2}=\{(\theta, \mu) :
\nu_2(\theta) < d_R/\phi_2\}.$ Hence it is enough to establish a
minorization condition that holds for $S_B.$

Let $\Gamma(\alpha, \beta; x)$ denote the value of the
Gamma$(\alpha,\beta)$ density at $x$ and define functions
$h_1(\lambda_{\theta})$ and $h_2(\lambda_{e})$ as follows:
\[
h_1(\lambda_{\theta}) = \left\{\begin{array}{ll}
 \Gamma\big(\frac{K}{2}+a_1, b_1 ; \lambda_{\theta}\big), &
 \lambda_{\theta} < \lambda_{\theta}^*, \\
 \Gamma\big(\frac{K}{2}+a_1, \frac{d_R}{2\phi_1} + b_1;
 \lambda_{\theta} \big), & \lambda_{\theta} \geq
 \lambda_{\theta}^*, \end{array} \right.
\]
where
\[
\lambda_{\theta}^*= \frac{\phi_1(K+2 a_1)}{d_R}
\log\big(1+\frac{d_R}{2b_1\phi_1}\big)
\]
and
\[
h_2(\lambda_{e}) = \left\{\begin{array}{ll}
 \Gamma\big(\frac{M}{2}+a_2,\frac{SSE}{2} + b_2 ; \lambda_{e}\big), &
 \lambda_{e} < \lambda_{e}^*, \\
 \Gamma\big(\frac{M}{2}+a_2, \frac{d_R+\phi_2 SSE}{2\phi_2} + b_2;
 \lambda_{e} \big), & \lambda_{e} \geq
 \lambda_{e}^*, \end{array} \right.
\]
where
\[
\lambda_{e}^*= \frac{\phi_2(M+2 a_2)}{d_R}
\log\big(1+\frac{d_R}{\phi_2(2b_2+SSE)}\big).
\]
Now define a density $q(\lambda, \theta, \mu)$ on $R_{+}^2\times
R^K \times R$ by
\[
q(\lambda, \theta, \mu)=
\Big(\frac{h_1(\lambda_{\theta})}{\int_{R_{+}}h_1(\lambda_{\theta})
d\lambda_{\theta}}\Big)
\Big(\frac{h_2(\lambda_{e})}{\int_{R_{+}}h_2(\lambda_{e})
d\lambda_{e}}\Big) d(\xi|\lambda, y),
\]
where $d(\xi|\lambda, y)$ is the normal density in (\ref{eqn:
block Gibbs sampler trans density}) resulting from the block Gibbs
sampler construction.
Next define
\[
\tilde{\beta}_R= \bigg(\int_{R_{+}}h_1(\lambda_{\theta})
d\lambda_{\theta}\bigg) \bigg(\int_{R_{+}}h_2(\lambda_{e})
d\lambda_{e}\bigg).
\]
Also recall $p(\lambda, \xi | \lambda', \xi') = p(\lambda, \theta,
\mu| \lambda', \theta', \mu'),$ the Markov transition density of
the block Gibbs sampler as specified in (\ref{eqn: block Gibbs
sampler trans density}).

We are in a position to state the minorization condition.
\begin{prop}[Minorization Condition] \label{prop:minorization for block gibbs}
The Markov transition density for the block Gibbs sampler
satisfies the following minorization condition:
\begin{equation} \label{eqn:minorization for block gibbs}
p(\lambda, \theta, \mu| \lambda', \theta', \mu') \geq
\tilde{\beta}_R q(\lambda, \theta, \mu) \quad \textrm{for every}
\quad (\theta', \mu') \in S_B.
\end{equation}
\end{prop}
\subsubsection{Drift and Minorization for the
fixed-scan  Gibbs sampler}
As before assume that $K \geq 3$ and
\[
2 \leq m' = \min \{m_1, \dots, m_K\} \leq \max \{m_1, \dots,
m_K\}= m''.
\]
Define
\[
\delta_6 = \frac{K^2+2K a_1}{2s_0+K^2+2Ka_1} \quad \textrm{and}
\quad \delta_7=\frac{1}{2(a_1-1)}.
\]
Clearly $\delta_6 \in (0,1)$ and if $a_1 > 3/2$ then also
$\delta_7 \in (0,1).$ Moreover if $a_1 > 3/2,$ then since $2s_0
b_1 > 0,$ there exists $\rho_1 \in (0,1)$ such that
\begin{equation}\label{eqn: tech for drift fixed-scan}
\Big(K+\frac{\delta_6}{\delta_7}\Big) \delta_1 < \rho_1.
\end{equation}
Define also
\[
\nu_3(\theta, \lambda) =
\frac{K\lambda_{\theta}}{s_0+K\lambda_{\theta}}(\bar{\theta}-\bar{y})^2
\quad \textrm{and} \quad s^2 = \sum_{i=1}^{K}(\bar{y}_i -
\bar{y})^2.
\]
\begin{prop}[Drift Condition]
Assume that $a_1 > 3/2, $ $5m'>m''$ and let $\rho_1 \in (0,1)$
satisfy (\ref{eqn: tech for drift fixed-scan}). Fix
\[
c_3 \in (0, \min\{b_1, b_2\}) \quad \textrm{and} \quad \lambda_R
\in (\max\{\rho_1, \delta_6, \delta_7\}, 1).
\]
Define the drift function as
\begin{equation} \label{eqn: drift fixed-scan}
V_3(\theta, \lambda) = e^{c_3 \lambda_{\theta}} + e^{c_3
\lambda_{e}} + \frac{\delta_7}{K\delta_1 \lambda_{\theta}} +
\nu_3(\theta, \lambda).
\end{equation}
With this drift function the fixed-scan Gibbs sampler satisfies
the Rosenthal-type drift condition with
\begin{equation} \label{eqn: drift K constant fixed scan}
K_R= \big(\frac{b_1}{b_1-c_3}\big)^{a_1+\frac{K}{2}} +
\big(\frac{b_2}{b_2-c_3}\big)^{a_2+\frac{M?}{2}} + (\delta_6 +
\delta_7) \big[\frac{1}{s_0}+(m_0-\bar{y})^2+ \frac{s^2}{K}\big]+
\frac{2b_1 \delta_7}{K}.
\end{equation}
\end{prop}
We now turn to the minorization condition for the fixed-scan Gibbs
sampler provided in Section 5.2 of \cite{Jones Hobert Gibbs for
rand eff mod}.
Similarly as before, consider the candidate $C_R = \{(\theta,
\lambda): V_3 \leq d_R\}$ for a small set and let
\[
c_4 = \frac{\delta_7}{K\delta_1 d_R}, \quad c_l = \bar{y}-
\sqrt{(m_0-\bar{y})^2 +d_R} \quad \textrm{and} \quad c_u = \bar{y}
+ \sqrt{(m_0-\bar{y})^2 +d_R}.
\]
The minorization condition will be given on a set $S_G$ such that
\[C_R \subseteq S_G = S_{G_1} \cap S_{G_2} \cap S_{G_3},\] where
\begin{eqnarray}
\nonumber
 S_{G_1} & = & \Big\{(\theta, \lambda) : c_4 \leq
\lambda_{\theta} \leq \frac{\log d_R}{c_3} \Big\},
\\ \nonumber
 S_{G_2} & = & \Big\{(\theta, \lambda) : 0 <
\lambda_{e} \leq \frac{\log d_R}{c_3} \Big\},
\\ \nonumber
 S_{G_3} & = & \Big\{(\theta, \lambda) : c_l \leq
\frac{s_0m_0+ K \lambda_{\theta} \bar{\theta}}{s_0 +
K\lambda_{\theta}} \leq c_u \Big\}.
\end{eqnarray}
Moreover to assure that $S_{G_1} \cap S_{G_2}$ is nonempty, choose
$d_R$ such that
\[d_R \log d_R >
\frac{c_3\delta_7}{K\delta_1}.\]
Let $N(\zeta, \sigma^2; x)$ denote the value of the $N(\zeta,
\sigma^2)$ density at $x$ and define functions $g_1(\mu, \theta)$
and $g_2(\mu)$ as follows:
\[
g_1(\nu, \theta) = \big(\frac{c_4}{2\pi}\big)^{\frac{K}{2}}
\exp\bigg\{ - \frac{\log d_R}{2 c_3} \sum_{i=1}^K \big[
(\theta_i-\mu)^2 +m_i (\theta_i -\bar{y}_i)^2
 \big]\bigg\}
\]
and
\[
g_2(\mu) = \left\{ \begin{array}{ll}
 N \big( c_u, \big[ s_0 +\frac{K \log(d_R)}{c_3} \big]^{-1}; \mu \big),
 &  \mu \leq \bar{y}, \\
N \big( c_l, \big[ s_0 +\frac{K \log(d_R)}{c_3} \big]^{-1}; \mu
\big),
 &  \mu > \bar{y}. \end{array} \right.
\]
Now define a density on $R \times R^K \times R_{+}^2$ by
\[
q(\mu, \theta, \lambda) = \Big(\frac{g_1(\mu, \theta)g_2(\mu)}{
\int_R \int_{R^K} g_1(\mu, \theta)g_2(\mu) d\theta d\mu } \Big)
d(\lambda| \mu, \theta, y),
\]
where $d(\lambda| \mu, \theta, y)$ is the joint Gamma distribution
of $\lambda_{\theta}$ and $\lambda_{e}$ in (\ref{eqn: fixed scan
Gibbs sampler transition density}) resulting from the fixed-scan
Gibbs sampler construction. Next define
\[
\tilde{\beta}_R = \Big(\frac{s_0 + K c_4}{s_0 + \frac{K \log
d_R}{c_3}}\Big)^{1/2} \Big(\int_R \int_{R^K} g_1(\mu,
\theta)g_2(\mu) d\theta d\mu  \Big).
\]
Also recall $p(\mu, \theta, \lambda| \mu', \theta', \lambda'),$
the Markov transition density of the fixed-scan Gibbs sampler as
specified in (\ref{eqn: fixed scan Gibbs sampler transition
density}).
We are in a position to state the minorization condition.
\begin{prop}
The Markov transition density for the fixed-scan Gibbs sampler
satisfies the following minorization condition
\begin{equation}\label{eqn: minorization fixed-scan}
p(\mu, \theta, \lambda | \mu', \theta', \lambda') \geq
\tilde{\beta}_R q(\mu, \theta, \lambda) \quad \textrm{for every}
\quad (\theta', \lambda') \in S_G.
\end{equation}
\end{prop}
Moreover Jones and Hobert in \cite{Jones Hobert Gibbs for rand eff
mod} obtained a closed form expression for $\tilde{\beta}_R$ in
(\ref{eqn: minorization fixed-scan}) involving the standard normal
cumulative distribution function $\Phi.$ Let
\begin{eqnarray}
\nonumber
 \nu & = & \bigg[ s_0 + \frac{\log d_R }{c_3} \Big( K + \sum_{i=1}^{K}
 \frac{m_i}{1+m_i} \Big)\bigg]^{-1},
\\ \nonumber
 m_l & = & \nu \bigg[ c_l s_0 + \frac{\log d_R }{c_3} \Big( K c_l +
 \sum_{i=1}^{K}
 \frac{\bar{y}_i m_i}{1+m_i} \Big)\bigg],
\\ \nonumber
 m_u & = & \nu \bigg[ c_u s_0 + \frac{\log d_R }{c_3} \Big( K c_u +
 \sum_{i=1}^{K}
 \frac{\bar{y}_i m_i}{1+m_i} \Big)\bigg].
\end{eqnarray}
Then
\begin{eqnarray}
\nonumber
 \tilde{\beta}_R & = & \Big(\frac{c_4 c_3}{\log d_R}
 \Big)^{\frac{K}{2}}
 \sqrt{\nu(s_0 + K c_4)} \sqrt{\prod_{i=1}^{K}  \frac{1}{1+m_i}}
 \exp\bigg\{-\frac{\log d_R}{2c_3} \sum_{i=1}^{K}\frac{\bar{y}_i^2
 m_i}{1+ m_i}\bigg\}
 \\ \nonumber
 && \times \Bigg[ \exp\bigg\{-\frac{c_u^2 s_0}{2}- \frac{Kc_u^2 \log d_R}{2c_3}
 + \frac{m_u^2}{2\nu}\bigg\} \Phi \Big(\frac{\bar{y} - m_u}{\sqrt{\nu}}\Big)
 \\ \nonumber
 && \qquad \qquad + \exp\bigg\{-\frac{c_l^2 s_0}{2}- \frac{Kc_l^2 \log d_R}{2c_3}
 + \frac{m_l^2}{2\nu}\bigg\}
 \bigg(1- \Phi \Big(\frac{\bar{y} - m_l}{\sqrt{\nu}}\Big)\bigg)
 \Bigg].
\end{eqnarray}
\subsection{Obtaining the Bounds} \label{subsec: obtaining the
bounds}

We focus on obtaining the bounds for
$(\varepsilon-\alpha)-$approximation for bayesian estimators of
parameters $\mu, \lambda_{\theta}, \lambda_e$ and $\theta_i.$ This
involves integrating one dimensional projections of the identity
function on parameter space. The drift function $V$ has to be at
least of order $f^2$ since $|f^2|_{V}$ has to be finite. Note that
for the two described samplers different drift conditions has been
established and neither of them majorizes quadratic functions in
all the parameters. Thus specifying a parameter, say $\lambda_e$
implies the choice of the fixed-scan Gibbs sampler with the drift
function $V_3,$ whereas for $\mu$ the block-scan Gibbs sampler
with drift function $V_1$ or $V_2$ is the only option.

Once the sampler and the type of the drift condition is chosen,
the user must provide his choice of $\lambda_R, \phi$ and $d_R$
for the Rosenthal-type drift-minorization condition. The next step
is the right choice of $d$ in Lemma \ref{lemma: drift Rosenth to
Bax} which yields the parameters of the Baxendale-type drift
condition. Provided the Baxendale-type drift condition is
established with computable parameters, there are still four
parameters left to the user, namely the mutually dependent
$\gamma$ and $M$ in Baxendale's Theorem \ref{thm:bax} and their
counterparts $\gamma_2$ and $M_2$ from Corollary
\ref{cor:hoelder}. Unfortunately the bounds on $t$ and $n$ or $t,$
$n$ and $m$ are very complicated functions of these parameters
subject to users choice and finding optimal values analytically
seems impossible. Also, in our experience, small changes in these
quantities usually result in dramatically different bounds.

Similarly as burn-in bounds in \cite{Jones Hobert Gibbs for rand
eff mod}, final bounds for $(\varepsilon-\alpha)-$approximation
also strongly depend on the hyperparameter setting and the
observed data.

Thus we provide appropriate R functions for approximating optimal
bonds on the simulation parameters. This functions are available
on http://akson.sgh.waw.pl/\~{}klatus/
\section{Concluding Remarks}

To our best knowledge, in the above setting of an unbounded target
function $f$ and without assuming uniform ergodicity of the
underlying Markov chain (which in practice means the state space
$\stany$ is not compact) we derived first explicit bounds for the
total simulation cost required for
$(\varepsilon-\alpha)-$approximation. These bounds are sometimes
feasible and sometimes infeasible on a PC, and probably always
exceed the true values by many orders of magnitude. Although
$10^9$ iterations in our Toy Example takes about 1 minute on a
standard PC, sampling more realistic chains will take more time
and the bound will be even more conservative.

However, the message of the Chapter is a very positive one: the
current theoretical knowledge of Markov chains has reached the
stage when for many MCMC algorithms of practical relevance applied
to difficult problems, i.e. estimating expectations of unbounded
functions, we are able to provide a rigorous, nonasymptotic, a
priori analysis of of the quality of estimation. This is much more
then the often used in practice visual assessment of convergence ,
more sophisticated a posteriori convergence diagnostics, bounding
only burn in time or even using asymptotic confidence intervals.

We also notice the following:
\begin{itemize}
\item The leading term in the bound for $n$ is
$b = \frac{\pi V |f_c^2|_V}{\varepsilon^2 \alpha}(1+ \frac{2M_2
\gamma_2}{1-\gamma_2})$ (where we took $p=r=2$ for simplicity).
$\pi V |f_c^2|_V$ should be of the order of $Var_{\pi}f,$ thus
this term is inevitable. $\varepsilon^{-2}$ results from
Chebyshev's inequality, since we proceed by bounding the mean
square error. $\alpha^{-1}$ can be reduced to $\log(\alpha^{-1})$
for small $\alpha$ by Lemma \ref{lem: median} and Algorithm
\ref{Alg: median} which in fact results in an exponential
inequality. The last term $1+ \frac{2M_2 \gamma_2}{1-\gamma_2}$ is
of the same order as a general bound for the ratio of the
asymptotic variance and the stationary variance, under drift
condition and without reversibility as indicated by Theorem
\ref{thm:as var}. Thus it also seems to be inevitable. However we
acknowledge this bound seems to be very poor due to the present
form of $V-$uniform ergodicity constants.
\item The term $1+ \frac{2M_2 \gamma_2}{1-\gamma_2}$ is the
bottleneck of the approach. Here good bounds on $\gamma$ and the
somewhat disregarded $M(\gamma)$ are equally important.
Improvements in Baxendale-type convergence bounds may lead to
dramatic improvement of the bounds on the total simulation cost
(e.g. by applying the preliminary results of \cite{Bednorz
bounds}).
\item Improvements of drift parameters (i.e. establishing better drift
functions and minorization conditions) imply significant
improvement of the convergence bounds in Baxendale's Theorem.
\item The drift conditions we used as well as the Baxendale's
theorem are far from optimal and subject to improvement.
\item We applied the theoretical results to the toy example of Section
\ref{sec: Toy Example} where the drift and minorization conditions
are available without much effort and to the Hierarchical Random
Effects Model with drift and minorization conditions established
in \cite{Jones Hobert Gibbs for rand eff mod}. Even more general
models are feasible in this setting, in particular in the recent
paper \cite{Johnson Jones} Johnson and Jones established drift and
minorization conditions for a bayesian hierarchical version of a
general linear mixed model.
\item Establishing drift conditions might be difficult. A good first try
may be $V(x)$ proportional to $\pi(x)^{-1/2}$ or to some suitable
quadratic function.
\end{itemize}


\section{Appendix - Formulas for $\rho$ and M} \label{sec:Bax form}


In the sequel the term \textit{atomic case} and \textit{nonatomic
case} refers to $\tilde{\beta}=1$ and $\tilde{\beta} < 1$
respectively. If $\tilde{\beta} < 1,$ define
\[
\alpha_1 = 1+ \frac{\log\frac{K-\tilde{\beta}}{1-\beta}}{\log
\lambda^{-1}}, \quad \alpha_2 = \left\{\begin{array}{ll} 1, &
\textrm{if } \nu(C)=1, \\
1 + \frac{\log\tilde{K}}{\log \lambda^{-1}}, & \textrm{if }
\nu(C)+\int_{C^c} V d \nu \leq \tilde{K},
\\
1 + \big(\log\frac{K}{\tilde{\beta}}\big) \big/
(\log\lambda^{-1}), & \textrm{otherwise.}
\end{array} \right.
\]
Then let
\[
R_0 = \min\{\lambda^{-1}, (1-\tilde{\beta})^{-1/\alpha_1}\},
\qquad L(R) = \left\{ \begin{array}{lll}
\frac{\tilde{\beta}R^{\alpha_2}}{1-(1-\tilde{\beta})R^{\alpha_1}},
& \textrm{if} & 1 < R < R_0,\\
\infty & \textrm{if} & R=R_0. \end{array} \right.
\]
\subsection{Formulas for general operators} \label{sec: Bax form
general}
 For $\beta > 0,$ $ R> 1$ and $L>1,$ let $R_1=R_1(\beta,
R, L)$ be the unique solution $r \in (1, R)$ of the equation
\[
\frac{r-1}{r(\log(R/r))^2}=\frac{e^2\beta (R-1)}{8(L-1)}
\]
and for $1 < r < R_1,$ define
\[
K_1(r,\beta, R, L) = \frac{2 \beta + 2 (\log N)(\log (R/r))^{-1} -
8 N e^{-2} (r-1)r^{-1}(\log (R/r))^{-2}}{ (r-1) [\beta - 8N e^{-2}
(r-1) r^{-1} (\log (R/r))^{-2}]},
\]
where $N = (L-1)/(R-1).$

For the \textit{atomic case} we have $\rho = 1/R_1(\beta,
\lambda^{-1},\lambda^{-1} K)$ and for $\rho < \gamma < 1,$
\begin{eqnarray}
M & = & \frac{\max(\lambda, K-\lambda/\gamma)}{\gamma-\lambda} +
\frac{K(K-\lambda/\gamma)}{\gamma (\gamma - \lambda)}
K_1(\gamma^{-1}, \beta, \lambda^{-1}, \lambda^{-1}K) \nonumber \\
\label{eqn: M general atomic} && + \frac{(K-\lambda / \gamma)
\max(\lambda, K-\lambda)}{(\gamma - \lambda)(1-\lambda)} +
\frac{\lambda(K-1)}{(\gamma - \lambda)(1-\lambda)}.
\end{eqnarray}
For the \textit{nonatomic case} let $\tilde{R}=
\textrm{arg}\max_{1<R<R_0} R_1(\beta, R, L(R)).$ Then we have
$\rho = 1/R_1(\beta, \tilde{R},L(\tilde{R}))$ and for $\rho <
\gamma < 1,$
\begin{eqnarray}
M & = & \frac{\gamma^{-\alpha_2 -1}(K\gamma-\lambda)}{(\gamma -
\lambda)[1-(1-\tilde{\beta})\gamma^{-\alpha_1}]^2} \times \left(
\frac{\tilde{\beta} \max(\lambda, K-\lambda)}{1-\lambda} +
\frac{(1-\tilde{\beta})(\gamma^{-\alpha_1}-1)}{\gamma^{-1} - 1}
\right)
 \nonumber \\
\nonumber
 && +
\frac{\max(\lambda, K-\lambda/\gamma)}{\gamma-\lambda} +
\frac{\tilde{\beta}\gamma^{-\alpha_2
-2}K(K\gamma-\lambda)}{(\gamma -
\lambda)[1-(1-\tilde{\beta})\gamma^{-\alpha_1}]^2}
K_1(\gamma^{-1}, \beta, \tilde{R}, L(\tilde{R}))
\\ \nonumber
&&+ \frac{\gamma^{-\alpha_2}\lambda
(K-1)}{(1-\lambda)(\gamma-\lambda)[1-(1-\tilde{\beta})\gamma^{-\alpha_1}]}
+\frac{K[K\gamma-\lambda-\tilde{\beta}(\gamma-\lambda)]}{\gamma^2
(\gamma - \lambda)[1-(1-\tilde{\beta})\gamma^{-\alpha_1}]}
 \\ \label{eqn: M general nonatomic}
  && + \frac{K-\lambda
-\tilde{\beta}(1-\lambda)}{(1-\lambda)(1-\gamma)}
 \left((\gamma^{-\alpha_2}-1) +
 (1-\tilde{\beta})(\gamma^{-\alpha_1}-1)/\tilde{\beta}
 \right).
\end{eqnarray}

\subsection{Formulas for self-adjoint operators} \label{sec: Bax form
reversible} A Markov chain is said to be reversible with respect
to $\pi$ if
$\int_{\stany} Pf(x) g(x) \pi(dx) = \int_{\stany} f(x) Pg(x) \pi
(dx) $ for all $f, g \in L^2(\pi).$ For reversible Markov chains
the following tighter bounds are available.

For the \textit{atomic case} define
\[
R_2 = \left\{\begin{array}{lll} \min \left\{ \lambda^{-1}, r_s
\right\},  & \textrm{if} & K
> \lambda + 2\beta, \\ \lambda^{-1}, & \textrm{if} & K \leq \lambda +
2\beta,
\end{array} \right.
\]
where $r_s$ is the unique solution of $1 + 2 \beta r = r^{1+(\log
K)(\log \lambda^{-1})}.$
Then $\rho = R_2^{-1}$ and for $\rho < \gamma < 1$ take $M$ as in
(\ref{eqn: M general atomic}) with $K_1(\gamma^{-1}, \beta,
\lambda^{-1}, \lambda^{-1}K)$ replaced by $K_2 = 1+
1/(\gamma-\rho).$

For the \textit{nonatomic case} let
\[
R_2 = \left\{\begin{array}{lll} r_s, & \textrm{if} & L(R_0) > 1 +
2\beta R_0,
\\R_0, & \textrm{if} & L(R_0) \leq 1 + 2\beta R_0,
\end{array} \right.
\]
where $r_s$ is the unique solution of $1 + 2 \beta r = L(r).$
Then $\rho = R_2^{-1}$ and for $\rho < \gamma < 1$ take $M$ as in
(\ref{eqn: M general nonatomic}) with $K_1(\gamma^{-1}, \beta,
\tilde{R}, L(\tilde{R}))$ replaced by $K_2 = 1 +
\sqrt{\tilde{\beta}}/(\gamma-\rho).$

\subsection{Formulas for self-adjoint positive operators}
\label{sec: Bax form rev pos} A Markov chain is said to be
positive if $\int_{\stany}Pf(x)f(x)\pi(dx) \geq 0$ for every $f
\in L^2(\pi).$ For reversible and positive markov chains take
$M$'s as in Section \ref{sec: Bax form reversible} with $\rho =
\lambda$ in the \textit{atomic case} and $\rho = R_0^{-1}$ in the
\textit{nonatomic case.}

   \chapter{Convergence Results for Adaptive Monte Carlo} \label{CHAPTER Convergence Results for Adaptive Monte
Carlo}

Ergodicity results for adaptive Monte Carlo algorithms usually
assume \emph{time-stability} of transition kernels. On the other
hand, a large class of time-inhomogeneous Markov Chains is
ergodic. This suggests existence of adaptive MC algorithms which
fail to satisfy the \emph{time-stability}
condition but are still ergodic. We present a modification of Atchad%
\'{e}-Rosenthal ergodicity Theorems (3.1 and 3.2 in \cite{aro})
that does not assume \emph{time-stability} of transition kernels.
We use a weaker \emph{path-stability} condition instead, that
results from \emph{time-stability} condition by the triangle
inequality.
%
%
%
\section{Introduction}
As before, we deal with computation of analytically intractable
integral
\[ I=\int_{\mathcal{X}} f(x) \pi (x)dx. \]
For computational efficiency of the Markov chain Monte Carlo
approach, the simulated Markov chain should converge to its
stationary distribution reasonably quickly. This can sometimes be
achieved by careful design of the transition kernel $P$ of the
chain, on the basis of a detailed preliminary analysis of $\pi$.
Intuitively, the more features of $\pi$ are known, the better $P$
can be designed. So a non-Markovian approach might be to allow the
transition kernel of the simulated stochastic process $(X_{n})_{n
\geq 0}$ to adapt whenever new features of $\pi$ are encountered
during the process run. Simulations show that this approach can
indeed outperform algorithms based on classical ideas. For
numerous examples and an insight of how to tune the transition
kernel ''on the fly'' see \cite{RobRos Adapt Examples} and
references therein. However, since in this case $(X_{n})_{n \geq
0}$ is not a Markov chain any more, it may fail to converge to the
expected asymptotic distribution even if each participating
transition kernel is ergodic and has the same stationary
distribution. A simple but nonintuitive example is given in
Section~\ref{sec:ex}. Difficulty to obtain general ergodicity
results appears to be the main problem in adaptive Monte Carlo.

For versions of adaptive MC and related work we refer to e.g.
\cite{Fishman}, \cite{Evans}, \cite{GelSa}. In more recent papers
 \cite{Gilks} showed adaptation of the transition
kernel can be performed (without damaging the ergodicity of the
algorithm) on regeneration times. The idea of adaptive MC through
regeneration was then investigated in \cite{Kadane} and
\cite{Sahu}. Convergence results in fairly general setting have
been derived in \cite{Haario} which was followed by refined
theorems in \cite{aro} and a discrete state space version of those
results presented in \cite{Nott}.

In each of the above mentioned papers ergodicity results either on
regeneration times, or fit within the so called diminishing
adaptation framework and assume the \emph{time-stability}
condition for transition kernels. Yet the existence of ergodic
inhomogeneous Markov chains suggests the \emph{time-stability} of
transition kernels is not necessary for ergodicity of adaptive MC
algorithms. After introductory examples in Section~\ref{sec:ex},
in Section~\ref{sec:thm} we give ergodicity theorems that use a
weaker \emph{path-stability} condition, which results from the
\emph{time-stability} condition by triangle inequality. However we
have to pay the price for it and formulate the \emph{uniform
ergodicity} condition in the time inhomogeneous setting, which
makes it more complicated then in the original Atchad\'{e} and
Rosenthal's theorems. In Section~\ref{sec:proof} we prove the main
result of this Chapter.
%
%
%
%

%
%
%
\section{One Intuitive and One Not-so-Intuitive Example}
\label{sec:ex}
We begin with a simple example where we briefly analyze two
stochastic processes using the same two transition matrices.

Consider the state space $\mathcal{X}=\{0,1\}$ and $\pi $, the
uniform distribution on $\mathcal{X} $. Let
\[ P_{1}=\left[
\begin{array}{cc}
1/2 & 1/2 \\ 1/2 & 1/2
\end{array}
\right] \quad \textrm{and} \quad P_{2}=(1-\varepsilon )\left[
\begin{array}{cc}
1 & 0 \\ 0 & 1
\end{array}
\right] +\varepsilon P_{1} \quad \textrm{for some} \quad
\varepsilon
>0. \]
Note that $\pi $ is the stationary distribution for both, $P_{1}$ and $P_{2}$. Let $\wp $ be some probability distribution on $\{P_{1},P_{2}\}$. Let $%
P^{(0)},P^{(1)},P^{(2)},...$ be an iid sample from $\wp $. In the
sequel we will use the convention $\min\varnothing=\infty$ and
$\max\varnothing=-\infty$.
\begin{ex} Let $(X_{n})_{n\geq 0}$ be a stochastic process
with an initial distribution $p_{0}$, evolving in step $k$
according to the transition matrix$\ P^{(k)}$. $(X_{n})_{n \geq
0}$ is clearly an in-homogeneous Markov Chain and $p_{n}$ (the
distribution of $X_{n}$) converges to the stationary distribution
$\pi $: let $U_{n}:=\{k: k \leq n, P^{(k)}=P_{1}\}$ and
$u_{n}=\max U_{n}$. The distribution of $X_{n}$, given $u_{n}\neq
-\infty$ is $\pi$, so we have the following bound on the total
variation distance between $p_{n}$ and $\pi$: \[\|p_{n}-\pi
\|_{tv} \leq P(u_{n}=-\infty) \stackrel{n \to
\infty}{\longrightarrow}0 \quad a.s.
\]
\end{ex}
\begin{ex}(due to W. Niemiro). Now consider $(Y_{n})_{n\geq 0}$
with an initial distribution $q_{0}$ and an initial transition matrix $Q_{0}$%
, evolving for $n\geq 1$ according to the following adaptive rule:
\[Q_{k}=\left\{
\begin{array}{ccc}
P_{1} & \text{if} & Y_{k-1}=0 \\ P_{2} & \text{if} & Y_{k-1}=1
\end{array}
\right. \]

Note that after two consecutive 1 (and this occurs with
probability at least $\frac{1}{4}$ for any $k,k+1$) $Y_{n}$ is
trapped in 1 and can escape only with probability $\varepsilon $.
Let $\bar{q}_{1}=\lim_{n\to \infty }P(Y_{n}=1)$ and
$\bar{q}_{0}=\lim_{n\to \infty}P(Y_{n}=0).$ Now it is clear, that for small $\varepsilon $ we will have $%
\bar{q}_{1} \gg \bar{q}_{0}$ and the procedure fails to give the
expected asymptotic distribution.
\end{ex}
Both processes $(X_{n})_{n \geq 0}$ and $(Y_{n})_{n \geq 0}$ are
allowed to use essentially different transition matrices in two
consecutive steps. But one of them converges to the desired
distribution $\pi $ and the other one fails to converge. In our
opinion it is not the ''time stability'' condition, that is
crucial for convergence of an adaptive Monte Carlo algorithm. It
is the ''path-stability'' condition, that reads ''if the path is
similar, the transition kernel should be similar as well''.
Obviously $(X_{n})_{n \geq 0}$ satisfies this condition and
$(Y_{n})_{n \geq 0}$ does not.

In the following section we will try to formalize this intuition.
\section{Convergence Results}\label{sec:thm}
We will similarly as in \cite{aro} analyze a stochastic process
$(X_{n})_{n \geq 0}$ on a general state space $\mathcal{X}$,
generated by the following algorithm:
\begin{alg}\label{al:alg}
Assuming we have an initial transition kernel $P_{x_{0}}$ and an
initial point $x_{0} \in \mathcal{X}$, the algorithm proceeds as
follows:
\begin{enumerate}
\item If for time $n \geq 0$ we have $X_{n}=x$ and a transition
kernel $P_{n, \widetilde{X}_{n}}$, which is allowed to depend on
the path $\widetilde{X}_{n} = (X_{0},\dots,X_{n}) \in
\mathcal{X}^{n+1}$; then sample from
$P_{n,\widetilde{X}_{n}}(x,\cdot).$
\item Use $\widetilde{X}_{n+1} = (X_{0},\dots,X_{n+1})$ to build a new transition kernel $P_{n, \widetilde{X}_{n+1}}$ to
be used at time $n+1$.
\end{enumerate}
\end{alg}
For $(X_{n})_{n \geq 0}$ generated by Algorithm \ref{al:alg} we
shall write $P_{\mu}$ to denote its distribution on
$(\mathcal{X}^{\infty},\mathcal{F}^{\infty})$ when $X_{0} \sim
\mu$, and $E_{\mu}$ to denote the expectation with respect to
$P_{\mu}$. If $\mu=\delta_{x}$, we usually write $E_{x}$ and
$P_{x}$ instead of $E_{\mu}$ and $P_{\mu}$. By $P_{\mu,n}$ we will
denote the marginal distribution of $X_{n}$ induced by $P_{\mu}$,
thus $P_{\mu,n}$ is a probability measure on $\mathcal{X}$.
To denote two trajectories of length $n+k+1$, that have a common
initial part of length $n+1$ and then split, we will write
$(\tilde{x}_{n},\tilde{y}_{k})$ and
$(\tilde{x}_{n},\tilde{y}_{k}')$.

We will prove ergodicity theorems similar to Theorem 3.1 and 3.2
in \cite{aro}, but under modified assumptions.

\begin{assu} \label{amain} There exist a measurable function
$V:\mathcal{X} \rightarrow [1,\infty)$ and real number sequences
$(\tau_{n}), (a_{n}), (R_{n})$, such that $(\tau_{n}),(R_{n})
\rightarrow 0$ as $n \to \infty$ and:
\begin{itemize}
\item[A.1] (uniform ergodicity) For all $j \geq 1, n \geq 0, x \in \mathcal{X}$ and $\tilde{x}_{n} \in
\mathcal{X}^{n+1}$, there exists
$\tilde{y}_{j}'=(y_{1}',\dots,y_{j}')$ and $0 \leq l \leq j-1$
such that
\begin{equation}\label{a1}
\Big\|\prod_{i=0}^{j-1}
P_{n+i,(\tilde{x}_{n},y_{1}',\dots,y_{i}')}(x,\cdot) -
\pi_{n+l,(\tilde{x}_{n},y_{1}',\dots,y_{l}')}(\cdot) \Big\|_{V}
\leq R_{j}V(x).
\end{equation}
\item[A.2] (path-stability) For all $x \in \mathcal{X}, \tilde{x}_{n} \in
\mathcal{X}^{n+1}$, there exists $\tilde{y}_{k}' \in
\mathcal{X}^{k}$, such that $\tilde{x}_{n}$ and $\tilde{y}_{k}'$
satisfy (\ref{a1}) with $j=k$ and for all $\tilde{y}_{k} \in
\mathcal{X}^{k}$,
\begin{equation} \label{a2}
\big\| P_{n+k,(\tilde{x}_{n},\tilde{y}_{k})}(x,\cdot) -
P_{n+k,(\tilde{x}_{n},\tilde{y}_{k}')}(x,\cdot) \big\|_{V} \leq
K_{1}\tau_{n} a_{k}V(x).
\end{equation}
\item[A.3] For all $x \in \mathcal{X}, \tilde{x}_{n} \in
\mathcal{X}^{n+1}, \tilde{y}_{k} \in \mathcal{X}^{k}$,
\begin{equation}\label{a3}
\big\| \pi_{n+k,(\tilde{x}_{n}, \tilde{y}_{k})} -
\pi_{n,\tilde{x}_{n}} \big\|_{V} \leq K_{2}\tau_{n} a_{k}.
\end{equation}
\item[A.4] For all $n \geq 1$,
\begin{eqnarray}\label{a4}
\lefteqn{\int V^{2}(x_{n}) P_{\mu,n}(dx_{n})={}}\\ &=\int \dots
\int V^{2}(x_{n}) P_{n-1,\tilde{x}_{n-1}}({x}_{n-1},dx_{n}) \dots
P_{0,\tilde{x}_{0}}({x}_{0},dx_{1}) \leq
K_{3}V^{2}(x_{0})\nonumber
\end{eqnarray}

and
\begin{equation} \label{a4'}
\sup_{n,\tilde{x}_{n}} \pi_{n,\tilde{x}_{n}}(V) < \infty.
\end{equation}
\item[A.5] For any finite constants $c_{1}, c_{2}$, define \[B(c_{1},
c_{2}, n) := \min_{1 \leq k \leq n}
(c_{1}\phi_{k}\tau_{n-k}+c_{2}R_{k}),\] where
$\phi_{n}=\sum_{k=1}^{n}a_{k}$. Assume that $B(c_{1}, c_{2},
n)=\mathcal{O}(\frac{1}{n^{\varepsilon}})$ for some $\varepsilon >
0$.
\end{itemize}
\end{assu}
\medskip
Under these assumptions we will prove two ergodicity theorems:
\medskip
\begin{thm}\label{thm:conv} Let $(X_{n})_{n \geq 0}$ be the stochastic process
generated by Algorithm \ref{al:alg} with $X_{0}=x_{0}$. Under
A.1-A.4 there exist constants $k_{1},k_{2} < \infty$ such that for
any measurable function $f:\mathcal{X} \to R$ with $|f| \leq V$,
\begin{equation}
\big| E_{x_{0}}(f(X_{n})-\pi_{n,\tilde{X}_{n}}(f)) \big| \leq
B(k_{1},k_{2},n)V(x_{0}).
\end{equation}
\end{thm}
\begin{thm}\label{thm:lln} Under A.1-A.5, for any measurable
function $f:\mathcal{X} \to R$ and $|f| \leq V$, for any starting
point $x_{0} \in \mathcal{X}$,
\begin{equation}
\frac{1}{n} \sum_{i=0}^{n-1}
\big(f(X_{i})-\pi_{i,\tilde{x}_{i}}(f)\big) \to 0, \quad
\textrm{as } n \to \infty, \quad P_{x_{0}} \textrm{- a.s.}
\end{equation}
\end{thm}
\begin{remark}
\begin{enumerate}
\item If $\pi_{n,\tilde{x}_{n}} \equiv \pi$, as it usually occurs in
Monte Carlo setting ($\pi$ is the invariant target distribution),
then Theorem \ref{thm:conv} gives a bound on the rate of
convergence of the distribution of $X_{n}$ to $\pi$ and Theorem
\ref{thm:lln} provides a law of large numbers type result.
\item In
this typical case ($\pi_{n,\tilde{x}_{n}} \equiv \pi$) the theory
of inhomogeneous Markov chains can be applied to check Assumption
A.1 (compare \cite{DMR}).
\item In
particular this theorems can be applied in case when
$\pi_{n,\tilde{x}_{n}} \equiv \pi$ and $P_{n,\tilde{x}_{n}} \geq
\varepsilon \pi$, for some $\varepsilon
> 0$, as considered in \cite{Nott}.
\item Assumptions used here differ from those in \cite{aro}, where
A.1 and A.2 are as follows:
\begin{itemize}
\item[A.1'] \emph{(uniform ergodicity)} For all $j>0$, $n \geq 0$, $x \in \mathcal{X}$ and
$\tilde{x}_{n} \in \mathcal{X}^{n+1}$, \[
\big\|P^{j}_{n,\tilde{x}_{n}}(x,\cdot) -
\pi_{n,\tilde{x}_{n}}(\cdot) \big\|_{V} \leq R_{j}V(x). \]
\item[A.2'] \emph{(time-stability)} For all $x \in \mathcal{X}$, $\tilde{x}_{n} \in
\mathcal{X}^{n+1}$, $\tilde{y}_{k} \in \mathcal{X}^{k}$, \[ \big\|
P_{n+k,(\tilde{x}_{n},\tilde{y}_{k})}(x,\cdot) -
P_{n,\tilde{x}_{n}}(x,\cdot) \big\|_{V} \leq K_{1}\tau_{n}
a_{k}V(x). \]
\end{itemize}
The \emph{path-stability} condition results from the
\emph{time-stability} condition by the triangle inequality, so
assumption A.2 presented here is weaker. Assumptions A.1 here and
A.1' in \cite{aro} are incomparable.
$\|P^{j}_{n,\tilde{x}_{n}}(x,\cdot) - \pi_{n,\tilde{x}_{n}}(\cdot)
\|_{V} $ does not have to converge even if A.1-5 hold. It involves
some computation, similar to this in the proof of Lemma
\ref{lem:main}, to show A.1', A.2' together with A.3-4 imply
$$\Big\|\prod_{i=0}^{j-1}
P_{n+i,(\tilde{x}_{n},y_{1}',\dots,y_{i}')}(x,\cdot) -
\pi_{n+l,(\tilde{x}_{n},y_{1}',\dots,y_{l}')}(\cdot) \Big\|_{V}
\leq B(k_{1},k_{2},j)V(x),$$
 so if additionally A.5 holds,
$$\Big\|\prod_{i=0}^{j-1}
P_{n+i,(\tilde{x}_{n},y_{1}',\dots,y_{i}')}(x,\cdot) -
\pi_{n+l,(\tilde{x}_{n},y_{1}',\dots,y_{l}')}(\cdot)
\Big\|_{V}=\mathcal{O}\big(\frac{1}{n^{\varepsilon}}\big).$$

 However our version is more
complicated and might turn out to be difficult to check even if
$\pi_{n,\tilde{x}_{n}} \equiv \pi$.
\item \emph{Path-stability} instead of \emph{time-stability}
condition enables to apply this ergodicity theorems to Monte Carlo
algorithms that are inhomogeneous in their nature, like simulated
annealing. In other words we can adapt Monte Carlo methods based
on inhomogeneous Markov chains as well.
\item Finally, the theorem handles our introductory toy examples
i.e. $(X_{n})_{n \geq 0}$ that converges to the desired
distribution satisfies A.1-A.4 (but does not satisfy A.2' in
\cite{aro}). $(Y_{n})_{n \geq 0} $ that fails to converge, fails
to satisfy assumption A.2 as well.
\end{enumerate}
\end{remark}
\section{Proofs}\label{sec:proof}
We now proceed to prove theorems from Section \ref{sec:thm}. The
proof follows closely Atchad\'{e} and Rosenthal \cite{aro}.
Crucial point of the proof is Lemma \ref{lem:main}. Once Lemma
\ref{lem:main} is shown under our modified assumptions, we derive
Theorems \ref{thm:conv} and \ref{thm:lln} in essentially identical
manner as in \cite{aro}. This part of the proof is purely
expository and presented here for the sake of completeness.

Let $(\mathcal{F}_{n})_{n=-\infty}^{\infty}$ be a filtration
defined by: \begin{equation} \label{filtr} \mathcal{F}_{n}:=
\left\{
\begin{array}{ll} \{\O,\Omega\} & \textrm{if $n < 0$}\\ \sigma
(X_{0}, \dots, X_{n}) & \textrm{if $n \geq 0$} \end{array} \right.
\end{equation} and $g_{k,\tilde{X}_{k}}(x):=f(x)-\pi_{k,\tilde{X}_{k}}(f).$
\begin{lemma} \label{lem:main} Assume A.1-A.4 hold. Then there
are some constants $0<k_{1},k_{2}< \infty$ such that for any $n
\geq 0$, $j \geq 1$ and any measurable function $f$ with $|f| \leq
V$, we have:
\begin{equation}
\Big\| E_{x_{0}} \big( g_{n+j,\tilde{X}_{n+j}}(X_{n+j}) |
\mathcal{F}_{n} \big) \Big\|_{2} \leq B(k_{1},k_{2},j)V(x_{0}).
\end{equation}
\end{lemma}
\medskip
The proof of Lemma \ref{lem:main} is given later in this section.
We start with Theorems \ref{thm:conv} and \ref{thm:lln}.
\medskip
\begin{proof}[Proof of Theorem \ref{thm:conv}] Let $n=0$ in Lemma
\ref{lem:main}. We obtain the following: \[\Big\| E_{x_{0}} \big(
g_{j,\tilde{X}_{j}}(X_{j}) | \mathcal{F}_{0} \big) \Big\|_{2} =
\big|E_{x_{0}} \big( f(X_{j})-\pi_{j,\tilde{X}_{j}}(f) \big) \big|
\leq  B(k_{1},k_{2},j)V(x_{0}),\] for all $|f| \leq V$, which is
Theorem \ref{thm:conv}.
\end{proof}
\medskip

\begin{proof} [Proof of Theorem \ref{thm:lln}] To prove Theorem
\ref{thm:lln} we will use the theory of mixingales. Theorem
\ref{thm:mix} used here is presented in Appendix. Let
\begin{equation}
Y_{n}:=f(X_{n})-\pi_{n,\tilde{X}_{n}}(f)-E_{x_{0}}\big(f(X_{n})-\pi_{n,\tilde{X}_{n}}(f)\big).
\end{equation}
The proof will proceed according to the following plan:
\begin{enumerate}
\item Show that
\begin{equation} \label{st1} E_{x_{0}}\big(f(X_{n})-\pi_{n,\tilde{X}_{n}}(f)\big) \to
0\textrm{ as } n \to \infty. \end{equation}
\item Show that $(Y_{n})_{n\geq 0}$ is a mixingale of size
$-\frac{\varepsilon}{2}$ and use Theorem \ref{thm:mix} to conclude
that \begin{equation}\label{st2} \frac{1}{n} \sum_{i=0}^{n-1}
Y_{i} \to 0 \textrm{ as } n \to \infty \quad P_{x_{0}} \textrm{
a.s.}\end{equation}
\item The foregoing results in \begin{equation} \label{st3} \frac{1}{n} \sum_{i=0}^{n-1} \big( f(X_{i})-\pi_{i,\tilde{X}_{i}}(f) \big) \to 0 \textrm{ as } n \to
\infty \quad P_{x_{0}} \textrm{ a.s.}\end{equation} This states
Theorem \ref{thm:lln}.
\end{enumerate}

To see that (\ref{st1}) holds, it is enough to recall Theorem
\ref{thm:conv} and Assumption A.5.

To prove (\ref{st2}) consider first condition (\ref{mix:2}). Since
the filtration is defined by (\ref{filtr}), we have
$E(Y_{n}|\mathcal{F}_{n+j})=Y_{n}$ and (\ref{mix:2}) is satisfied
for any positive number sequences $(c_{n})$ and $(\psi_{n})$.

Condition (\ref{mix:1}) is obviously satisfied for $j \geq n$, for
any positive number sequences $(c_{n})$ and $(\psi_{n})$ as well,
since $EY_{n}=0$. For the case $j < n$ we will use Lemma
\ref{lem:main}:
{\setlength\arraycolsep{2pt}
\begin{eqnarray} \big\| E_{x_{0}}(Y_{n} | \mathcal{F}_{n-j})
\big\|_{2} & = & \Big\| E_{x_{0}} \Big( g_{n,\tilde{X}_{n}}(X_{n})
-  \big( E_{x_{0}}(g_{n,\tilde{X}_{n}}(X_{n}))\big) \big|
\mathcal{F}_{n-j} \Big) \Big\|_{2} \nonumber\\
 & \leq &  \Big\|
E_{x_{0}} \Big( g_{n,\tilde{X}_{n}}(X_{n}) | \mathcal{F}_{n-j}
\Big) \Big\|_{2} + {} \nonumber\\&&
 {}+ \Big\|E_{x_{0}} \Big(
E_{x_{0}} \big( g_{n,\tilde{X}_{n}}(X_{n})\big) \big|
\mathcal{F}_{n-j} \Big) \Big\|_{2} \nonumber\\
 & = & \Big\| E_{x_{0}} \Big( g_{n,\tilde{X}_{n}}(X_{n}) | \mathcal{F}_{n-j}
\Big) \Big\|_{2} + \Big\|E_{x_{0}} \Big(
g_{n,\tilde{X}_{n}}(X_{n}) \big| \mathcal{F}_{0} \Big) \Big\|_{2}
\nonumber\\
 & \leq & B(k_{1},k_{2},j)V(x_{0}) + B(k_{1},k_{2},n)V(x_{0}) \nonumber\\
  & = & \mathcal{O}(j^{-\varepsilon}) +
  \mathcal{O}(n^{-\varepsilon}) =
  \mathcal{O}(j^{-\varepsilon}) \nonumber
\end{eqnarray}}

Now we set in (\ref{mix:1}) $c_{n}\equiv 1$ and take appropriate
$\psi_{j}$, such that $\psi_{j}=\mathcal{O}(j^{-\varepsilon})$.
Hence $(Y_{n})_{n\geq 0}$ is a mixingale of size
$-\frac{\varepsilon}{2}$. Since
$\frac{c_{n}}{n}=\mathcal{O}(n^{-1})$ and $-1 <
\min\{-\frac{1}{2}, \frac{\varepsilon}{2} - 1\}$, we can apply
Theorem \ref{thm:mix} and conclude that $\frac{1}{n}
\sum_{i=0}^{n-1} Y_{i} \to 0 \textrm{ as } n \to \infty \quad
P_{x_{0}} \textrm{ a.s.}$

Combining (\ref{st1}) and (\ref{st2}), we get (\ref{st3}) by an
elementary argument.
\end{proof}
\medskip
Now we proceed to prove Lemma \ref{lem:main}.
\medskip
\begin{proof} [Proof of Lemma \ref{lem:main}]
Note that \begin{equation} \label{gcenter}
\pi_{n,\tilde{X}_{n}}(g_{n,\tilde{X}_{n}})=
\pi_{n,\tilde{X}_{n}}(f-\pi_{n,\tilde{X}_{n}}(f))=0 \quad
P_{x_{0}} \textrm{ a.s.}
\end{equation}
The idea of the proof is to split the quantity $\big\| E_{x_{0}}
\big( g_{n+j,\tilde{X}_{n+j}}(X_{n+j}) | \mathcal{F}_{n} \big)
\big\|_{2}$ into two terms, say $A$ and $B$ and bound them using
Assumptions A.1, A.2 and A.3.

Denote by
$(\tilde{x}_{n},\tilde{y}_{j})=(\tilde{x}_{n},y_{1},\dots,y_{j})$
a trajectory of length $n+j$. According to this notation we will
usually write $y_{i}$ for $x_{n+i}$. Given
$(X_{0},\dots,X_{n})=\tilde{x}_{n}$ we have
{\setlength\arraycolsep{2pt}
\begin{eqnarray} &&E_{x_{0}} \Big(g_{n,\tilde{X}_{n}}(X_{n+j}) |
\tilde{X}_{n}=\tilde{x}_{n} \Big) = \nonumber \\&&
 \qquad = \int
g_{n,\tilde{x}_{n}}(y_{j})P_{n+j-1,(\tilde{x}_{n},y_{1},\dots,y_{j-1})}
 (y_{j-1},dy_{j}) \dots P_{n,\tilde{x}_{n}}(x_{n},dy_{1}) \nonumber\\&&
 \qquad = \eta_{j-1}(\tilde{x}_{n}) + \nonumber \\&&
 \qquad +\int
 g_{n,\tilde{x}_{n}}(y_{j})P_{n+j-1,(\tilde{x}_{n},\tilde{y}_{j-1}')}
 (y_{j-1},dy_{j}) P_{n+j-2,(\tilde{x}_{n},\tilde{y}_{j-2})}
 (y_{j-2},dy_{j-1}) \dots \nonumber\\&&
 \qquad \dots  P_{n,\tilde{x}_{n}}(x_{n},dy_{1}), \nonumber
\end{eqnarray}}
where $\tilde{y}_{j}'=(y_{1}', \dots, y_{j}')$ is as in Assumption
A.2, and {\setlength\arraycolsep{2pt}
\begin{eqnarray} && \eta_{j-1}(\tilde{x}_{n})= \nonumber\\&&
 \qquad = \int
g_{n,\tilde{x}_{n}}(y_{j})\Big(P_{n+j-1,(\tilde{x}_{n},\tilde{y}_{j-1})}
 (y_{j-1},dy_{j})-P_{n+j-1,(\tilde{x}_{n},\tilde{y}_{j-1}')}
 (y_{j-1},dy_{j})\Big) \nonumber\\&&
 \qquad P_{n+j-2,(\tilde{x}_{n},\tilde{y}_{j-2})}
 (y_{j-2},dy_{j-1}) \dots
 \dots  P_{n,\tilde{x}_{n}}(x_{n},dy_{1}). \nonumber
\end{eqnarray}}
By exchanging transition kernels for all coordinates, we get:
\begin{equation} \label{split} E_{x_{0}} \Big(g_{n,\tilde{X}_{n}}(X_{n+j}) |
\tilde{X}_{n}=\tilde{x}_{n} \Big) =
 \sum_{k=1}^{j-1} \eta_{k}(\tilde{x}_{n}) + \Big( \prod_{i=0}^{j-1}
 P_{n+i,(\tilde{x}_{n},\tilde{y}_{i}')} \Big)g_{n,\tilde{x}_{n}}(x_{n}),
\end{equation}
where {\setlength\arraycolsep{2pt}
\begin{eqnarray} \label{eta} \eta_{k}(\tilde{x}_{n})&=& \int \Big( \prod_{i=k+1}^{j-1}
 P_{n+i,(\tilde{x}_{n},\tilde{y}_{i}')} \Big)g_{n,\tilde{x}_{n}}(y_{k+1})
 \nonumber\\&&
 \Big(P_{n+k,(\tilde{x}_{n},\tilde{y}_{k})}
 (y_{k},dy_{k+1})-P_{n+k,(\tilde{x}_{n},\tilde{y}_{k}')}
 (y_{k},dy_{k+1})\Big) \nonumber\\&&
 P_{n+k-1,(\tilde{x}_{n},\tilde{y}_{k-1})}
 (y_{k-1},dy_{k}) \dots
 \dots  P_{n,\tilde{x}_{n}}(x_{n},dy_{1}).
\end{eqnarray}}
Consider the second term of the right hand side of (\ref{split}):
{\setlength\arraycolsep{2pt}
\begin{eqnarray} \label{right} &&\Bigg| \Big( \prod_{i=0}^{j-1}
 P_{n+i,(\tilde{x}_{n},\tilde{y}_{i}')}
 \Big)g_{n,\tilde{x}_{n}}(x_{n}) \Bigg| = \nonumber\\
 &&\qquad= \Bigg| \Big( \prod_{i=0}^{j-1}
 P_{n+i,(\tilde{x}_{n},\tilde{y}_{i}')}
 \Big) \big( f - \pi_{n,\tilde{x}_{n}}(f) \big) \Bigg| \nonumber\\
 && \qquad= \Bigg| \Big( \prod_{i=0}^{j-1}
 P_{n+i,(\tilde{x}_{n},\tilde{y}_{i}')}
 \Big) f(x_{n}) - \pi_{n,\tilde{x}_{n}}(f)\Bigg|
 \nonumber\\&&
 \qquad = \Bigg| \Big( \prod_{i=0}^{j-1}
 P_{n+i,(\tilde{x}_{n},\tilde{y}_{i}')}
 \Big) f(x_{n}) -\pi_{n+l,(\tilde{x}_{n},\tilde{y}_{l}')}(f) \Bigg|
 +\Big| \pi_{n+l,(\tilde{x}_{n},\tilde{y}_{l}')}(f) - \pi_{n,\tilde{x}_{n}}(f)\Big|
 \nonumber\\&&
 \qquad \leq \Big\|\prod_{i=0}^{j-1}
 P_{n+i,(\tilde{x}_{n},\tilde{y}_{i}')}(x_{n},\cdot) -
 \pi_{n+l,(\tilde{x}_{n},\tilde{y}_{l}')}(\cdot) \Big\|_{V}
 + \big\| \pi_{n+l,(\tilde{x}_{n}, \tilde{y}_{l}')} -
 \pi_{n,\tilde{x}_{n}} \big\|_{V} \nonumber\\&&
 \qquad \leq R_{j}V(x_{n})+K_{2}\tau_{n} a_{j},
\end{eqnarray}}
where the inequalities result from Assumptions A.1 and A.3.

We will now bound the first term of the right hand side of
(\ref{split}). Note that since
$g_{n,\tilde{x}_{n}}(y_{k+1})=f(y_{k+1})-\pi_{n,\tilde{x}_{n}}(f)$
and $\pi_{n,\tilde{x}_{n}}(f)$ given $\tilde{x}_{n}$ is some real
number, we obtain:
 {\setlength\arraycolsep{2pt}
\begin{eqnarray} \label{left:inside} &&
\Big( \prod_{i=k+1}^{j-1}
 P_{n+i,(\tilde{x}_{n},\tilde{y}_{i}')}
 \Big)g_{n,\tilde{x}_{n}}(y_{k+1})= \nonumber\\&&
 \qquad \qquad \qquad \qquad
 = \Big( \prod_{i=k+1}^{j-1}
 P_{n+i,(\tilde{x}_{n},\tilde{y}_{i}')}
 \Big)f(y_{k+1}) - \pi_{n,\tilde{x}_{n}}(f)
\end{eqnarray}}
and
{\setlength\arraycolsep{2pt}
\begin{eqnarray} \label{eta:inside} \int \pi_{n,\tilde{x}_{n}}(f)
 \Big(P_{n+k,(\tilde{x}_{n},\tilde{y}_{k})}
 (y_{k},dy_{k+1})-P_{n+k,(\tilde{x}_{n},\tilde{y}_{k}')}
 (y_{k},dy_{k+1})\Big) && \nonumber\\
 P_{n+k-1,(\tilde{x}_{n},\tilde{y}_{k-1})}
 (y_{k-1},dy_{k}) \dots
 \dots  P_{n,\tilde{x}_{n}}(x_{n},dy_{1}) &=& 0.
\end{eqnarray}}
Hence using (\ref{left:inside}) and (\ref{eta:inside}) we get
 {\setlength\arraycolsep{2pt}
\begin{eqnarray} \label{eta:f} \eta_{k}(\tilde{x}_{n})&=&
 \int \Big( \prod_{i=k+1}^{j-1}
 P_{n+i,(\tilde{x}_{n},\tilde{y}_{i}')} \Big)f(y_{k+1})
 \nonumber\\&&
 \Big(P_{n+k,(\tilde{x}_{n},\tilde{y}_{k})}
 (y_{k},dy_{k+1})-P_{n+k,(\tilde{x}_{n},\tilde{y}_{k}')}
 (y_{k},dy_{k+1})\Big) \nonumber\\&&
 P_{n+k-1,(\tilde{x}_{n},\tilde{y}_{k-1})}
 (y_{k-1},dy_{k}) \dots
 \dots  P_{n,\tilde{x}_{n}}(x_{n},dy_{1}).
\end{eqnarray}}
Since for each $0 \leq l \leq j-k-2$ and $\tilde{y}_{l}''$ we have
  {\setlength\arraycolsep{2pt}
\begin{eqnarray} && \Big|\Big( \prod_{i=k+1}^{j-1}
 P_{n+i,(\tilde{x}_{n},\tilde{y}_{i}')} \Big) f(y_{k+1}) \Big| =
 \Big| \Big( \prod_{i=k+1}^{j-1}
 P_{n+i,(\tilde{x}_{n},\tilde{y}_{i}')} \Big) f(y_{k+1})
 \nonumber\\&& \qquad \qquad \qquad \phantom{\leq}
 -\pi_{n+k+1+l,(\tilde{x}_{n},\tilde{y}_{k+1}',\tilde{y}_{l}'')}(f)
 +\pi_{n+k+1+l,(\tilde{x}_{n},\tilde{y}_{k+1}',\tilde{y}_{l}'')}(f)\Big|
 \nonumber\\&& \qquad \qquad \qquad
 \leq \Big\|\prod_{i=k+1}^{j-1}
 P_{n+i,(\tilde{x}_{n},\tilde{y}_{i}}(y_{k+1},\cdot) -
 \pi_{n+k+1+l,(\tilde{x}_{n},\tilde{y}_{k+1}',\tilde{y}_{l}'')}(\cdot)
 \Big\|_{V} \nonumber\\ && \qquad \qquad \qquad \phantom{\leq}
 +
 |\pi_{n+k+1+l,(\tilde{x}_{n},\tilde{y}_{k+1}',\tilde{y}_{l}'')}(f)|,
\end{eqnarray}}
 we can apply A.1 and write an analogous equality to
 (\ref{eta:inside}) for
 $\pi_{n+k+1+l,(\tilde{x}_{n},\tilde{y}_{k+1}',\tilde{y}_{l}'')}(f)$
 resulting from A.1 to get:
  {\setlength\arraycolsep{2pt}
\begin{eqnarray} \label{eta:V} |\eta_{k}(\tilde{x}_{n})| & \leq &
 \sup_{I:\mathcal{X} \to \{-1,1\}} \int R_{j-1-k}V(y_{k+1})
 I(y_{k+1}) \nonumber\\&&
 \qquad \Big(P_{n+k,(\tilde{x}_{n},\tilde{y}_{k})}
 (y_{k},dy_{k+1})-P_{n+k,(\tilde{x}_{n},\tilde{y}_{k}')}
 (y_{k},dy_{k+1})\Big) \nonumber\\&&
 \qquad P_{n+k-1,(\tilde{x}_{n},\tilde{y}_{k-1})}
 (y_{k-1},dy_{k}) \dots
 \dots  P_{n,\tilde{x}_{n}}(x_{n},dy_{1}) \nonumber\\&
 \leq & R_{j-1-k} \int K_{1} \tau_{n} a_{k} V(y_{k}) \nonumber\\ &&
 \qquad P_{n+k-1,(\tilde{x}_{n},\tilde{y}_{k-1})}
 (y_{k-1},dy_{k}) \dots
 \dots  P_{n,\tilde{x}_{n}}(x_{n},dy_{1}) \nonumber\\ &
 \leq & r_{0} \tau_{n} a_{k}
 E_{x_{0}}\big(V(X_{n+k}) | \tilde{X}_{n}=\tilde{x}_{n}\big).
\end{eqnarray}}
Where the second inequality results from \emph{path-stability}
condition A.2 and $r_{0}$ is some finite constant, since $K_{1} <
\infty$ and $(R_{n}) \to 0 \textrm{ as } n \to \infty$.

Putting (\ref{right}) and (\ref{eta:V}) together in (\ref{split}),
we get:
 {\setlength\arraycolsep{2pt}
\begin{eqnarray} \label{bound:n} &&
\big| E_{x_{0}} \big(g_{n,\tilde{X}_{n}}(X_{n+j}) |
\mathcal{F}_{n} \big) \big| \leq \nonumber\\&&
 \qquad \qquad
 \leq R_{j}V(X_{n})+K_{2}\tau_{n} a_{j} + r_{0} \tau_{n} \sum_{k=1}^{j-1}
 a_{k} E_{x_{0}}\big(V(X_{n+k}) |
 \mathcal{F}_{n} \big).
\end{eqnarray}}
By Assumption A.3 we have
  {\setlength\arraycolsep{2pt}
\begin{eqnarray} \label{bound:n+j}
 \big| E_{x_{0}} \big(g_{n+j,\tilde{X}_{n+j}}(X_{n+j}) |
\mathcal{F}_{n} \big) \big| & \leq &
 \big| E_{x_{0}} \big(g_{n,\tilde{X}_{n}}(X_{n+j}) |
\mathcal{F}_{n} \big) \big| \nonumber\\&& {} +
 E_{x_{0}}\Big( \big| \pi_{n+j,\tilde{X}_{n+j}}(f) -
 \pi_{n,\tilde{X}_{n}}(f) \big| | \mathcal{F}_{n} \Big)
 \nonumber\\& \leq &
 \big| E_{x_{0}} \big(g_{n,\tilde{X}_{n}}(X_{n+j}) |
\mathcal{F}_{n} \big) \big| + K_{2} \tau_{n} a_{j}.
\end{eqnarray}}
We now combine (\ref{bound:n}) and (\ref{bound:n+j}) to obtain the
first inequality of the following bound:

{\setlength\arraycolsep{2pt}
\begin{eqnarray} \label{bound:n+j:2}
 \big\| E_{x_{0}} \big(g_{n+j,\tilde{X}_{n+j}}(X_{n+j}) |
\mathcal{F}_{n} \big) \big\|_{2} & \leq &
 R_{j}\|V(X_{n})\|_{2} + 2K_{2}\tau_{n} a_{j} + \nonumber\\ &&
 {}+ r_{0} \tau_{n} \sum_{k=1}^{j-1} a_{k} \|E_{x_{0}}\big(V(X_{n+k}) |
 \mathcal{F}_{n} \big)\|_{2} \nonumber\\ & \leq &
 R_{j}\|V(X_{n})\|_{2} + {} \nonumber\\ &&
 {} + \max\{r_{0}, 2K_{2}\} \tau_{n} \sum_{k=1}^{j} a_{k}
 \|V(X_{n+k})\|_{2} \nonumber\\ & \leq &
 R_{j}\sqrt{K_{3}}V(X_{0}) + {} \nonumber\\ &&
 {} + \max\{r_{0}, 2K_{2}\} \tau_{n}
 \sum_{k=1}^{j} a_{k} \sqrt{K_{3}}V(X_{0}) \nonumber\\ & \leq &
 V(x_{0}) (r_{3}R_{j} + r_{2}\tau_{n}\phi_{j}),
\end{eqnarray}}
where we use Assumption A.4 and apply {\setlength\arraycolsep{2pt}
\begin{eqnarray} \|E_{x_{0}}(V(X_{n+k}) |
 \mathcal{F}_{n})\|_{2} & = & \big\{E\big[\big(E_{x_{0}}(V(X_{n+k})
 |\mathcal{F}_{n})\big)^{2}\big]\big\}^{1/2} \nonumber\\ &
 \leq & \big\{E\big(E_{x_{0}}(V^{2}(X_{n+k})
 |\mathcal{F}_{n})\big)\big\}^{1/2} \nonumber\\ &
 = & \big\{EV^{2}(X_{n+k})\big\}^{1/2}=\|V(X_{n+k})\|_{2}
 \nonumber
\end{eqnarray}}
The constants in (\ref{bound:n+j:2}) are defined as
$r_{3}:=\sqrt{K_{3}}$, $r_{2}:=\max\{r_{0}, 2K_{2}\}\sqrt{K_{3}}$
and $\phi_{j}:=\sum_{k=1}^{j} a_{k}. $

Since $(\mathcal{F}_{n})_{n=-\infty}^{\infty}$ is a filtration,
$\mathcal{F}_{n} \subseteq \mathcal{F}_{n+j-k}$, for $k=1,\dots,j$
and therefore \[ E_{x_{0}} \big(g_{n+j,\tilde{X}_{n+j}}(X_{n+j}) |
\mathcal{F}_{n} \big) = E_{x_{0}} \Big( E_{x_{0}}
\big(g_{n+j,\tilde{X}_{n+j}}(X_{n+j}) | \mathcal{F}_{n+j-k} \big)
\big| \mathcal{F}_{n} \Big). \] This implies  \[ \Big\{E_{x_{0}}
\big(g_{n+j,\tilde{X}_{n+j}}(X_{n+j}) | \mathcal{F}_{n}
\big)\Big\}^{2} \leq E_{x_{0}} \Big( \big\{E_{x_{0}}
\big(g_{n+j,\tilde{X}_{n+j}}(X_{n+j}) | \mathcal{F}_{n+j-k} \big)
\big\}^{2} \big| \mathcal{F}_{n} \Big). \]
 And therefore
 \begin{equation} \label{war}
 \Big\|E_{x_{0}}
\big(g_{n+j,\tilde{X}_{n+j}}(X_{n+j}) | \mathcal{F}_{n}
\big)\Big\|_{2} \leq \Big\| E_{x_{0}}
\big(g_{n+j,\tilde{X}_{n+j}}(X_{n+j}) | \mathcal{F}_{n+j-k} \big)
\Big\|_{2}.
 \end{equation}
We now apply (\ref{bound:n+j:2}) to the right hand side of
(\ref{war}) and get:
 \begin{equation} \label{particular}
 \Big\|E_{x_{0}}
\big(g_{n+j,\tilde{X}_{n+j}}(X_{n+j}) | \mathcal{F}_{n}
\big)\Big\|_{2} \leq V(x_{0}) (r_{3}R_{k} +
r_{2}\tau_{n+j-k}\phi_{k}).
\end{equation}
Finally, since (\ref{particular}) holds for every $k=1,\dots,j$,
we can take the minimum:
 \begin{equation} \label{finally}
 \Big\|E_{x_{0}}
\big(g_{n+j,\tilde{X}_{n+j}}(X_{n+j}) | \mathcal{F}_{n}
\big)\Big\|_{2} \leq V(x_{0}) \min_{1\leq k \leq j}
\big\{r_{3}R_{k} + r_{2}\tau_{n+j-k}\phi_{k} \big\}.
\end{equation}
Obviously $V(x_{0}) \min_{1\leq k \leq j} \big\{r_{3}R_{k} +
r_{2}\tau_{n+j-k}\phi_{k} \big\} \leq V(x_{0})B(k_{1},k_{2},j)$
for some constants $k_{1}$ and $k_{2}$, which completes the proof
of the lemma.
\end{proof}
\bigskip
Hence the proof of Theorems (\ref{thm:conv}) and (\ref{thm:lln})
is complete as well.
%
%
%
%
%
%
%
%
\section{Appendix - Mixingales} \label{app:mix}
We present here a version of Strong Law of Large Numbers for
mixingales that is used to conclude the proof of Theorem
\ref{thm:lln}. Theorem \ref{thm:mix} presented here is a version
of Corollary 2.1 in \cite{deJong}. For an introduction to
mixingales see the books \cite{Hall} or \cite{David}.

Let $(Z_{n})_{n \geq 0}$ be a real-valued stochastic process on
some probability space $(\Omega, \mathcal{F}, P)$. Assume
$(Z_{n})$ is $L_{2}$-bounded, i.e. $\|Z_{n}\|_{2}=\big\{ \int
Z_{n}^{2}(\omega)dP(\omega)\big\}^{1/2} < \infty$ for all $n \geq
0$. Let $(\mathcal{F}_{n})_{n=-\infty}^{\infty}$ be a filtration.

\begin{defi} The process $(Z_{n})_{n \geq 0}$ is a
$L^{2}$-mixingale with respect to filtration
$(\mathcal{F}_{n})_{n=-\infty}^{\infty}$ if there exist real
number sequences $(c_{n})$ and $(\psi_{n})$, $\psi_{n} \to 0
\textrm{ as } j \to \infty$, such that for all $n \geq 0$ and all
$j \geq 0$,
\begin{equation} \label{mix:1} \big\| E(Z_{n} | \mathcal{F}_{n-j})
\big\|_{2} \leq c_{n}\psi_{j},
\end{equation} and
\begin{equation} \label{mix:2} \big\| Z_{n} - E(Z_{n} | \mathcal{F}_{n+j})
\big\|_{2} \leq c_{n}\psi_{j+1}.
\end{equation}

If for some $\lambda > 0$,
$\psi_{n}=\mathcal{O}(n^{-\lambda-\varepsilon})$ for some
$\varepsilon
> 0$, we say that mixingale $Z_{n}$ is of size $-\lambda$.
\end{defi}

\begin{thm} \label{thm:mix} Let $(Z_{n})$ be a $L^{2}$-mixingale
of size $-\lambda$. If $\frac{c_{n}}{n}=\mathcal{O}(n^{\alpha})$,
where $\alpha < \min\{-\frac{1}{2}, \lambda - 1\}$, then
$\frac{1}{n} \sum_{i=0}^{n-1} Z_{i} \to 0$ a.s.
\end{thm}

\end{document}